\documentclass[twocolumn,preprintnumbers,amsmath,amssymb]{revtex4}
\usepackage{epsfig}
\usepackage{graphicx}
\usepackage{dcolumn}
\usepackage{bm}
\usepackage{amsmath}
\usepackage{amsthm}
\usepackage{amsfonts}
\usepackage{color} 
\usepackage{caption}
\usepackage{subcaption}
\usepackage{siunitx,booktabs}
\usepackage{hyperref}
\usepackage{epstopdf}

\newcommand{\ket}[1]{\mbox{$ | #1 \rangle $}}
\newcommand{\bra}[1]{\mbox{$ \langle #1 | $}}

\newcommand{\braket}[2]{\ensuremath{\langle #1|#2\rangle}}

\begin{document}

\title{Long-distance device-independent quantum key distribution}
\author{V\'i­ctor Zapatero}
\email{vzapatero@com.uvigo.es}  
\author{Marcos Curty}
\affiliation{Escuela de Ingeniar\'ia de Telecomunicaci\'on, Department of Signal Theory and Communications, ­University of Vigo, Vigo E-36310, Spain}

\begin{abstract}
Besides being a beautiful idea, device-independent quantum key distribution (DIQKD) is probably the ultimate solution to defeat quantum hacking. To guarantee security, it requires, however, that the fair-sampling loophole is closed, which results in a very limited maximum achievable distance. To overcome this limitation, DIQKD must be furnished with fair-sampling devices like, for instance, qubit amplifiers.  These devices can herald the arrival of a photon to the receiver and thus decouple channel loss from the selection of the measurement settings. Consequently, one can safely post-select the heralded events and discard the rest, which results in a significant enhancement of the achievable distance. In this work, we investigate photonic-based DIQKD assisted by two main types of qubit amplifiers in the finite data block size scenario, and study the resources---particularly, the detection efficiency of the photodetectors and the quality of the entanglement sources---that would be necessary to achieve long-distance DIQKD within a reasonable time frame of signal transmission.
\end{abstract}

\maketitle

\section{Introduction}\label{Section1}

The use of quantum mechanics for cryptographic means was first proposed in the early 70's by Stephen Wiesner, aiming to create unfalsifiable banknotes~\cite{Wiesner}. Inspired by this seminal work, Charles Bennett and Gilles Brassard introduced a protocol to securely distribute cryptographic keys~\cite{Bennett}. Nowadays, intense theoretical and experimental research~\cite{Peev,Ribordy,Tamaki} has turned this latter task---called quantum key distribution (QKD)---into a feasible commercial solution~\cite{IdQuantique}. 

Despite such tremendous progress, a major flaw of QKD today is the existing big gap between the theory and the practice. This is so because security proofs of QKD typically rely on simple mathematical models to describe the behaviour of the different physical devices. As a result, any departure from these models might render real-life QKD implementations vulnerable to quantum hacking attacks~\cite{hack1,Zhao,Makarov,Vadim,Gerhardt,Weier,hack2}. 

To overcome this problem, the ultimate solution probably is device-independent QKD (DIQKD)~\cite{Mayers,Scarani,Masanes,Vazirani,Ekert,Miller}. Given that the users' devices are honest~\cite{honest1,HKLo}, DIQKD can guarantee security without characterizing the internal functioning of the apparatuses, thereby ruling out all hacking attacks against the physical implementation. It is based on a feature of some entangled states known as nonlocality~\cite{nonlocality}, which guarantees that two distant parties (say, Alice and Bob) sharing an ideal nonlocal quantum state observe perfectly correlated outcomes when performing adequate quantum measurements on their shares. Moreover, these correlations are monogamous, {\it i.e.}, the measurement outcomes are statistically independent of any pre-existing information held by a third party.  This property can be verified with a two-party Bell test~\cite{nonlocality,Bell,CHSH,Bell_exp1,Bell_exp2,Bell_exp3} known as the Clauser-Horne-Shimony-Holt (CHSH) test, which basically consists of repeatedly playing a two-party nonlocal game~\cite{Miller,Rotem}. The winning rate of the game indicates the  amount of monogamous correlations shared between Alice and Bob. 

The security of DIQKD has been rigorously established in different works, first against collective attacks~\cite{Scarani} in the asymptotic regime, then against coherent attacks~\cite{Vazirani} also in the asymptotic regime, and only recently in the practical scenario of finite data block sizes~\cite{Rotem} (see also~\cite{Wehner}). The security proof in~\cite{Rotem} relies on the so-called entropy accumulation theorem~\cite{Renner,entropy_ac2}, which effectively allows to prove the security of the full protocol from the security of a single round of the protocol by using a worst-case scenario. 

Security proofs require, however, that two fundamental loopholes are closed: the \textit{locality loophole}~\cite{nonlocality} and the \textit{fair-sampling loophole}~\cite{nonlocality,detection1,detection2}. The former is closed by enforcing a proper isolation of Alice's and Bob's devices. Closing the fair-sampling loophole is more tricky, specially if Alice and Bob wish to cover long distances. Indeed, if an adversary were able to correlate channel loss to Alice's or Bob's measurement settings, such an adversary could easily fake nonlocal correlations, and thus compromise the security of the distilled key. A simple solution is to assign a pre-established outcome value to each lost signal while running the CHSH test. The main drawback of this technique is however the limited achievable distance, because such mapping translates loss into errors. Indeed, with such an approach, even if the entanglement source could generate perfect Bell pairs and Alice and Bob could measure them with unit efficiency detectors, channel loss would limit the maximum DIQKD transmission distance to about only 3.5~km for a typical optical fiber. 

To enhance the distance, Alice and Bob need to use fair-sampling devices. These are devices that herald the arrival of a signal to the receiver, allowing a fair post-selection of the heralded events. Then, if Alice and Bob choose their measurement settings \textit{a posteriori}, {\it i.e.}, once their fair-sampling devices have declared the reception of a signal, they actually decouple channel loss from their measurement settings selection, thus paving the way for DIQKD over long distances. 

A fair-sampling device particularly suited for DIQKD is a qubit amplifier~\cite{Gisin,Pitkanen,Curty}, which basically consists of a teleportation gate. That is, a successful heralding corresponds to the teleportation of the state of the arriving signal to a signal at the output port of the qubit amplifier. DIQKD supported by qubit amplifiers has been analysed in~\cite{Gisin,Pitkanen,Curty,Takeoka,Acin2} in the asymptotic regime, {\it i.e.}, by considering an infinite number of signals. In this work, we focus on the practical finite data block size scenario. More precisely, we study finite-key DIQKD with the two different types of qubit amplifier architectures introduced in~\cite{Pitkanen} and~\cite{Curty}. We pay particular attention to the effect that typical device imperfections (specially, the finite detection efficiency of the photodetectors and the multi-photon pulses emitted by practical entanglement light sources) have on the performance of the system. In doing so, we determine the resources needed to achieve long-distance implementations of DIQKD within a reasonable time frame of signal transmission.  

The structure of the paper goes as follows. Sec.~\ref{Section2} introduces the mathematical models that we use to characterize the behaviour of the difference physical devices. Sec.~\ref{Section3} contains the description of the considered DIQKD protocol, and Sec.~\ref{Section4} introduces its secret key rate formula for the finite-key regime. Then, in Sec.~\ref{Section5} we present the main contributions of this paper. Here, we evaluate the performance of the DIQKD scheme as a function of the data block size and various device imperfections for two different qubit amplifier architectures. Finally, Sec.~\ref{Section6} summarizes the main conclusions. The paper includes a few appendices with complementary results and calculations.

\section{Toolbox}\label{Section2}

We begin by briefly introducing the mathematical models that we use in our analysis. They describe the main optical devices employed in a photonic implementation of a DIQKD setup, together with the behaviour of a typical lossy quantum channel. 

\subsection{Photonic sources}\label{A}

We will consider two types of light sources. The first one emits a coherent superposition of bipartite entangled states with the following form
\begin{equation}\label{1}
\ket{\psi}_{ab}=\sum_{n=0}^{\infty}\sqrt{p_{n}}\ket{\phi_{n}}_{ab},
\end{equation}
where $p_{n}$, with $\sum_{n=0}^{\infty}p_n=1$, stands for the probability of generating a $2n$-photon entangled state of the form
\begin{equation}\label{2}
\ket{\phi_{n}}_{ab}=\frac{1}{n!\sqrt{n+1}}(a_{\rm{h}}^{\dagger}b_{\rm{v}}^{\dagger}-a_{\rm{v}}^{\dagger}b_{\rm{h}}^{\dagger})^{n}\ket{0}_{ab}.
\end{equation}
In Eq.~(\ref{2}), $\ket{0}_{ab}$ is the vacuum state and $a_{\rm{h}}^{\dagger}$ and $a_{\rm{v}}^{\dagger}$ ($b_{\rm{h}}^{\dagger}$ and $b_{\rm{v}}^{\dagger}$) denote, respectively, the creation operators of horizontally and vertically polarized photons at the spatial mode $a$ ($b$). 

The case $p_{1}=1$ (and, thus, $p_n=0$ $\forall n\neq{}1$) in Eq.~(\ref{1}) corresponds to an ideal entanglement source generating singlet states, while the distribution
\begin{equation}\label{3}
p_{n}=\frac{(n+1)\lambda^{n}}{(1+\lambda)^{n+2}},
\end{equation}
corresponds to a type-II parametric down-conversion (PDC) source ~\cite{PDC,Xiongfeng}. The parameter $\lambda$ in Eq.~(\ref{3}) is one half of the expected number of photon pairs generated by the source each given time, {\it i.e.}, $\lambda=\frac{1}{2}\sum_{n=0}^\infty p_{n}n$.

The second type of light source that we will consider emits states of the form
\begin{equation}\label{4}
\ket{\Psi}_{ab}=\sum_{n=0}^{\infty}\sqrt{p_{n}}\ket{n,n}_{ab},
\end{equation}
where the $2n$-photon state $\ket{n,n}_{ab}=(a^{\dagger}b^{\dagger})^{n}/n!\ket{0}_{ab}$, $a^{\dagger}$ ($b^{\dagger}$) being the creation operator in the idler (signal) mode. As an example, a triggered PDC source can generate states like Eq.~(\ref{4}) with $p_{n}=\mu^{n}(1+\mu)^{-n-1}$, $\mu$ being the mean number of photon pairs per pulse~\cite{walls_book,trig_pdc}. 

This latter source can be used, for instance, to simulate practical single-photon sources. For example, upon the successful detection of one photon in the idler mode $a$ of $\ket{\Psi}_{ab}$, the normalized quantum state in the signal mode can be written as
\begin{equation}\label{5}
\rho_{\rm{single}}=\sum_{n=0}^{\infty}r_{n}\ket{n}_{b}\bra{n},
\end{equation}
for a certain probability distribution, $r_{n}$, which depends on the characteristics of the photodetector used to measure the idler mode. See Appendix~\ref{triggered_PDC} for the value of $r_{n}$ in the case of triggered PDC sources.

\subsection{Photodetectors}\label{B}

We shall consider that Alice and Bob have at their disposal photon-number-resolving (PNR) detectors, which are able to count the number of photons contained in each incoming optical pulse. In the relevant regime of low noise, they can be described by a positive operator valued measure (POVM) with the following elements:
\begin{equation}\label{6}
\Pi_{k} = \left\{
\begin{array}{ll}
(1-p_{\rm{d}})\tilde{\Pi}_{0}      & \mathrm{if\ } k=0, \\
(1-p_{\rm{d}})\tilde{\Pi}_{k}+p_{\rm{d}}\tilde{\Pi}_{k-1} & \mathrm{if\ } k\geq{1}, \\
\end{array}
\right.
\end{equation}
where the quantity $p_{\rm{d}}$ stands for the dark count rate of the photodetectors, which is, to a good approximation, independent of the incoming signals. On the other hand, the operators $\tilde{\Pi}_{k}$ that appear in Eq.~(\ref{6}), with $k\in\mathbb{N}$, are given by
\begin{equation}\label{7}
\tilde{\Pi}_{k}=\sum_{j=k}^{\infty}\binom{j}{k}\eta_{\rm{d}}^{k}(1-\eta_{\rm{d}})^{j-k}\ket{j}\bra{j},
\end{equation}
with $\eta_{\rm{d}}$ denoting the detection efficiency of the detectors, and where $\ket{j}$ stands for a Fock state with $j$ photons. 

We remark that the mathematical model given by Eq.~(\ref{6}) assumes, for simplicity, that dark counts can only increment by one unit the number of photons observed in a given pulse. That is, if an optical pulse contains, say, $k$ photons, we assume that the measurement outcome is at most $k+1$ photons due to the dark counts, but not greater than this. This is a fair approximation given that $p_{\rm{d}}$ is sufficiently low, which indeed is typically the case in practice.

\subsection{Heralded qubit amplifiers}\label{D} 

As already discussed above, to achieve long-distance DIQKD, we will consider that Alice and Bob use heralded qubit amplifiers~\cite{Gisin,Pitkanen,Curty} to notify them the arrival of a photon before they select their measurement settings. That is, only after the qubit amplifier confirms that a photon has arrived, Alice (Bob) selects the measurement and measures the photon. If no successful heralding takes place, the optical pulse is simply discarded. 
 
\begin{figure}
	\centering 
	\includegraphics[width=0.96\columnwidth]{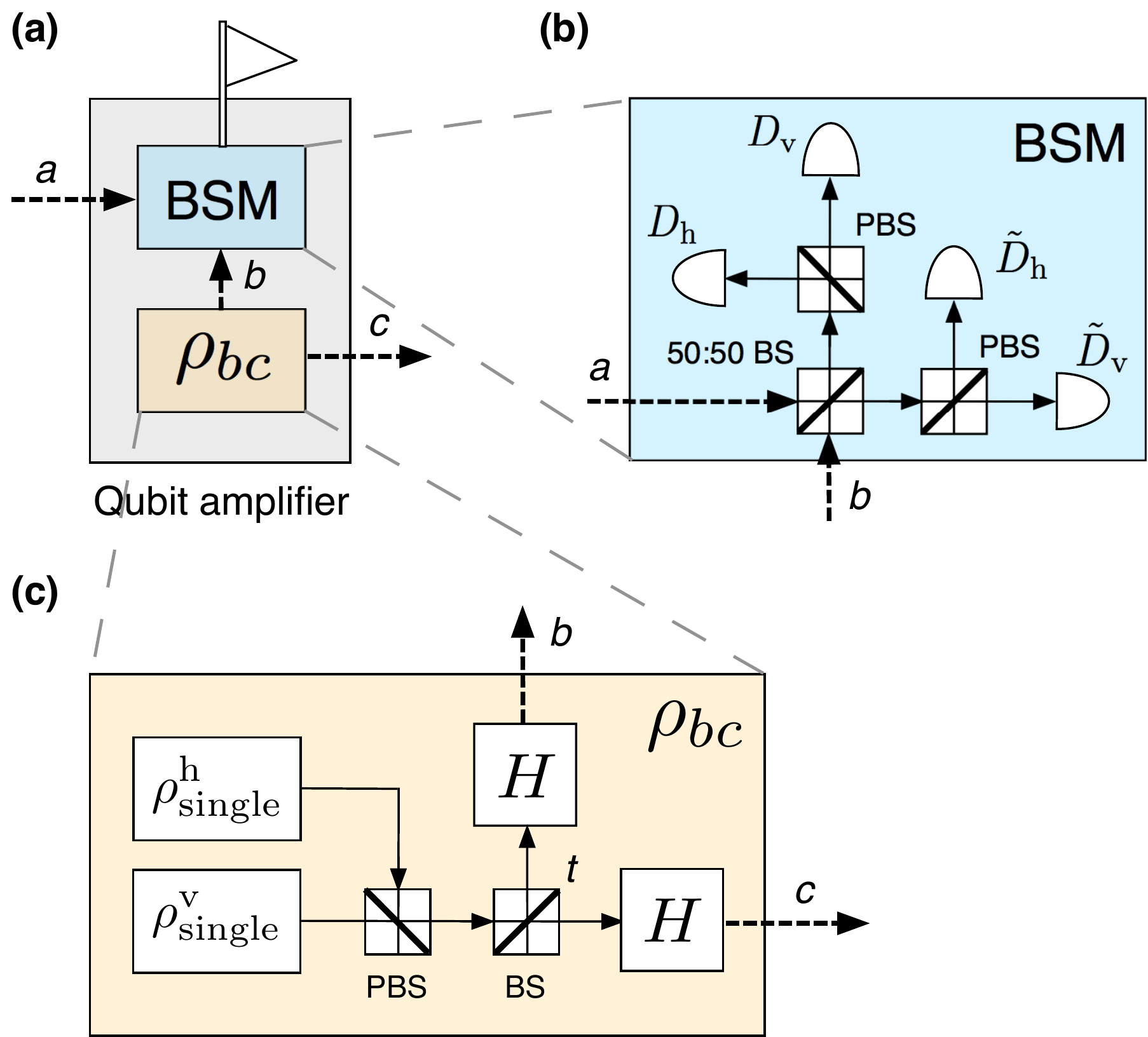}
	\caption{(a) Working principle of an heralded qubit amplifier based on teleportation~\cite{Gisin,Pitkanen,Curty}. A successful heralding is indicated with a flag. It notifies that a photon at the input port, $a$, of the qubit amplifier was teleported to a photon at its output port, $c$. For this, the qubit amplifier first generates a bipartite entangled state, ${\rho}_{bc}$, and then measures the signals in modes $a$ and $b$ with a Bell state measurement (BSM). In doing so, the state of the photon at mode $a$ is teleported to that at mode $c$ up to a unitary rotation. The main difference between the qubit amplifiers proposed in~\cite{Gisin,Pitkanen,Curty} is the mechanism to generate the entangled states ${\rho}_{bc}$. See the main text for further details. (b) Linear-optics BSM. The input states in modes $a$ and $b$ interfere at a 50:50 beamsplitter (BS). A polarizing BS (PBS) located at each output port of the BS separates vertically and horizontally polarized photons. Here we shall assume that all detectors are PNR detectors. A successful BSM corresponds to detecting two photons with orthogonal polarizations, {\it i.e.}, only when exactly two detectors detect one input photon each for any of the following photodetector pairs: $(D_{\rm{h}},D_{\rm{v}})$, $(D_{\rm{h}},\tilde{D}_{\rm{v}})$, $(\tilde{D}_{\rm{h}},D_{\rm{v}})$ or $(\tilde{D}_{\rm{h}},\tilde{D}_{\rm{v}})$. (c) Scheme introduced in~\cite{Pitkanen} to generate $\rho_{bc}$. A light source emits horizontally (vertically) polarized single-photons $\rho^{\rm{h}}_{\rm{single}}$ ($\rho^{\rm{v}}_{\rm{single}}$), which interfere at a PBS and then go through a BS of tunable transmittance $t$. Two Hadamard gates, denoted by $H$ in the figure, are used to avoid (if one disregards noise effects) that input vacuum signals at mode $a$ can produce a successful heralding flag when the BSM is that given by Fig.~\ref{fig:1}(b).
	}
	\label{fig:1}
\end{figure}

 Typical qubit amplifiers consist in a teleportation gate~\cite{teleportation}. That is, a successful heralding occurs when the state of the arriving photon is teleported to a photon at the output port of the qubit amplifier. The general mechanism is depicted in Fig.~\ref{fig:1}(a), while Fig.~\ref{fig:1}(b) shows the standard linear-optics Bell state measurement (BSM) used by the qubit amplifiers introduced in~\cite{Gisin,Pitkanen,Curty} to teleport the input photon. More efficient BSMs exist~\cite{efficient_BSM1,efficient_BSM2} and could be used here as well, although they require complicated entangled ancilla states.
Depending on the mechanism used to generate the bipartite entangled states, $\rho_{bc}$, illustrated in Fig.~\ref{fig:1}(a), one can distinguish two types of qubit amplifiers: polarization qubit amplifiers (PQAs)~\cite{Gisin,Pitkanen} and entanglement swapping relays (ESRs)~\cite{Curty}.

PQAs, first introduced in~\cite{Gisin} based on the seminal work reported in~\cite{Ralph}, typically employ practical single-photon sources to generate $\rho_{bc}$. For instance, the PQA proposed in~\cite{Pitkanen} uses the linear-optics circuit shown in Fig.~\ref{fig:1}(c) for this purpose, where $\rho^{\rm{h}}_{\rm{single}}$ ($\rho^{\rm{v}}_{\rm{single}}$) represents the state of a single-photon pulse prepared in horizontal (vertical) polarization. In the ideal case of perfect single-photon sources and unit detection and coupling efficiencies, it is straightforward to show that the circuit given by Fig.~\ref{fig:1}(c) generates states $\rho_{bc}=\ket{\phi}_{bc}\bra{\phi}$ with
\begin{equation}\label{9}
\ket{\phi}_{bc}=(1-t)\ket{\psi}_{bb}+t\ket{\varphi}_{cc}-\sqrt{2t(1-t)}\ket{\chi}_{bc},
\end{equation}
where the parameter $t$ is the transmittance of the tunable beamsplitter (BS) illustrated in Fig.~\ref{fig:1}(c), and the states $\ket{\psi}_{bb}$, $\ket{\varphi}_{cc}$, and $\ket{\chi}_{bc}$ have the form
\begin{eqnarray}\label{10}
\ket{\psi}_{bb}&=&\frac{1}{2}{({b_{\rm{h}}^{\dagger}}^{2}-{b_{\rm{v}}^{\dagger}}^{2})}\ket{0}_{b}, \nonumber \\
\ket{\varphi}_{cc}&=&\frac{1}{2}{({c_{\rm{h}}^{\dagger}}^{2}-{c_{\rm{v}}^{\dagger}}^{2})}\ket{0}_{c}, \nonumber \\
\ket{\chi}_{bc}&=&\frac{1}{\sqrt{2}}{(b_{\rm{h}}^{\dagger}c_{\rm{h}}^{\dagger}-b_{\rm{v}}^{\dagger}c_{\rm{v}}^{\dagger})}\ket{0}_{bc}.
\end{eqnarray}
In Eq.~(\ref{10}), the states $\ket{0}_{b}$, $\ket{0}_{c}$ and $\ket{0}_{bc}$ denote the vacuum states of the corresponding modes. The expression for the output states of the PQA in the practical scenario with non-ideal sources and non-unit detector and coupling efficiencies can be found in Appendix~\ref{PQA}. 

Let us continue assuming, for simplicity and for the moment, an ideal scenario where the BSM within the qubit amplifier uses perfect PNR detectors ({\it i.e.}, $p_{\rm{d}}=0$ and $\eta_{\rm{d}}=1$ in Eqs.~(\ref{6})-(\ref{7})) and lossless BSs and PBSs. Then, from Eq.~(\ref{9}), it can be shown that whenever a single-photon pulse prepared in, say, the pure state $\ket{\varphi_{\rm in}}_{a}=({\alpha}a_{\rm{h}}^{\dagger}+{\beta}a_{\rm{v}}^{\dagger})\ket{0}_a$ (with $|\alpha|^2+|\beta|^2=1$) arrives at the input port $a$ of the qubit amplifier, a successful BSM occurs with probability $t(1-t)$. Also, in the case of a successful result, the state of the output photon at mode $c$ (after applying an appropriate unitary transformation) is equal to $\ket{\varphi_{\rm out}}_{c}=({\alpha}c_{\rm{h}}^{\dagger}+{\beta}c_{\rm{v}}^{\dagger})\ket{0}_c$. That is, the state $\ket{\varphi_{\rm in}}_{a}$ of the input photon is successfully teleported to an output photon. On the other hand, if a vacuum pulse, $\ket{\varphi_{\rm in}}_{a}=\ket{0}_a$, arrives at the input port $a$ of the qubit amplifier, this state can never lead to a spurious heralding event, at least in the ideal scenario. This is so because, when the BSM uses perfect PNR detectors with no dark counts, the state $\ket{0}_a\ket{\phi}_{bc}$, with $\ket{\phi}_{bc}$ given by Eq.~(\ref{9}), cannot produce two detection clicks associated to orthogonal polarizations if it is measured with the BSM shown in Fig.~\ref{fig:1}(b). 

Finally, qubit amplifiers based on ESRs~\cite{Curty} directly prepare the state $\rho_{bc}$ with practical entanglement light sources like, for example, PDC sources. Indeed, in contrast to the arguments presented in~\cite{Gisin,Gisin2}, it was shown in~\cite{Curty} that when this type of qubit amplifier is used in DIQKD it can provide higher secret key rates than those achievable with the PQA introduced in~\cite{Gisin} when using PDC sources. 

\subsection{Optical couplers}\label{E}

In our analysis in Sec.~\ref{Section5}, we will consider a fiber-based implementation of DIQKD. Thus, we will model the coupling of the photons generated by the light sources into the optical fibers by means of a BS of transmittance $\eta_{\rm{c}}$. One input to the BS is the quantum signal, while the other input is a vacuum state. Similarly, one of the outputs of the BS is the optical fiber, while we assume that the other output is not accessible and represents the loss. For further details about this model we refer the reader to Appendix~\ref{honest}.

\subsection{Quantum channel}\label{C}

For simplicity, we will suppose that the quantum channel mainly introduces loss. That is, we shall disregard any noise effect due for example to polarization or phase misalignment. 

The channel loss is modeled with a BS of transmittance $\eta_{\rm ch}=10^{-\Lambda/10}$, where the parameter $\Lambda$ (dB) is related to the transmission distance $L$ (km) by an attenuation coefficient $\alpha$ (dB/km) via the expression $\Lambda={\alpha}L$. The specific value of $\alpha$ depends on the considered channel. For instance, a typical value for $\alpha$ in the case of single-mode optical fibers in the telecom wavelength is $\alpha=0.2$ dB/km.

\section{DIQKD PROTOCOL}\label{Section3}

We consider the DIQKD protocol introduced in~\cite{Rotem}. It is based on a CHSH test~\cite{CHSH}, and is equivalent to a certain two-party nonlocal game known as the CHSH game. 

Before presenting the steps of the protocol in detail, we introduce some notation first. Alice's measurement setting in the $i$-th successfully heralded round of the protocol is denoted by $X_{i}\in\{0,1\}$, where $X_{i}=0$ and $X_{i}=1$ tag the measurements described by the two following Pauli operators
\begin{equation}\label{11}
\sigma_{z}=
\begin{pmatrix}
1 & \hspace{.2cm}0 \\
0 & -1
\end{pmatrix}\hspace{.2cm}\rm{and}\hspace{.2cm}
\sigma_{x}=
\begin{pmatrix}
0 & \hspace{.2cm}1 \\
1 & \hspace{.2cm}0
\end{pmatrix},
\end{equation}
respectively. On the other hand, Bob's measurement setting in the $i$-th successfully heralded round of the protocol is denoted by $Y_{i}\in\{0,1,2\}$, where $Y_{i}=0$ tags the measurement $\sigma_{+}=(\sigma_{z}+\sigma_{x})/\sqrt{2}$, $Y_{i}=1$ indicates the measurement $\sigma_{-}=(\sigma_{z}-\sigma_{x})/\sqrt{2}$ and $Y_{i}=2$ refers to the measurement $\sigma_{z}$, with $\sigma_{z}$ and $\sigma_{x}$ again given by Eq.~(\ref{11}).
Similarly, Alice's (Bob's) outcome in the $i$-th successfully heralded round is denoted by $A_{i}\in\{0,1\}$ ($B_{i}\in\{0,1\}$). 

Next, we present the different steps of the  protocol. A schematic is shown in Fig.~\ref{fig:4}. For simplicity and for the moment, we shall assume that only Bob holds a qubit amplifier to compensate channel loss, while Alice has the entanglement source, $\rho_{ab}$, in her lab. The case where both Alice and Bob hold a qubit amplifier and the entanglement source is located in the middle of the channel between them is analyzed in Appendix~\ref{two_QA}.  
\begin{figure}[!htbp]
	\centering 
	\includegraphics[width=8.6cm]{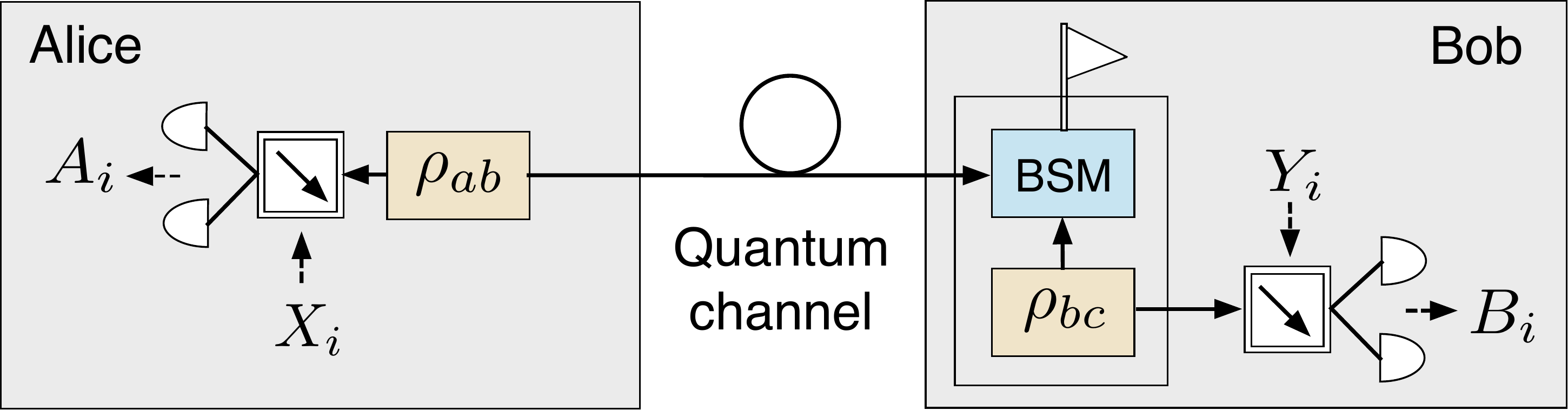} 
	\caption{Schematic of the considered DIQKD protocol. Alice holds an entanglement source, $\rho_{ab}$, in her lab, and Bob holds a qubit amplifier to mitigate the effect of channel loss. In every round of the protocol in which a successful heralding takes place at Bob's qubit amplifier, he randomly chooses a bit value $T_i\in\{0,1\}$. If $T_i=0$, Alice (Bob) chooses as measurement setting $X_{i}=\sigma_{z}$ ($Y_{i}=\sigma_{z}$). If $T_i=1$, Alice chooses at random her measurement setting $X_{i}\in\left\{\sigma_{z},\sigma_{x}\right\}$, with $\sigma_{z}$ and $\sigma_{x}$ being the Pauli matrices given by  Eq.~(\ref{11}). Similarly, in this latter case, Bob chooses at random his measurement setting $Y_{i}\in\left\{\sigma_{+},\sigma_{-}\right\}$, where $\sigma_{\pm}=(\sigma_{z}\pm{\sigma_{x}})/\sqrt{2}$. Their respective outcomes are recorded as $A_{i},B_{i}\in\left\{0,1\right\}$, where $A_{i}$ ($B_{i}$) indicates which of Alice's (Bob's) two photodetectors registered a single-photon pulse. If, say, Alice obtains an inconclusive result ({\it i.e.}, no photons or multiple photons are observed), she deterministically selects $A_{i}=1$, and similarly for Bob. The reader is referred to the main text for further details.}
	\label{fig:2}
\end{figure}
\\

\noindent\textbf{Protocol steps}\\

\noindent 1. \textit{Initialization.} Bob sets the counter $i$ of successfully heralded rounds to 0.\\

While $i<{n_{\rm{SH}}}$ for a certain prefixed value ${n_{\rm{SH}}}$, steps $2$ and $3$ below are repeated.\\

\noindent 2. \textit{Distribution.} Alice prepares a bipartite entangled state, $\rho_{{ab}}$, and sends system $b$ to Bob through the quantum channel. If no successful heralding takes place at Bob's qubit amplifier, the signal is discarded and step 2 is repeated. Otherwise, Bob updates the counter $i$ to $i+1$. Then, he randomly chooses a bit value $T_{i}\in{\left\{0,1\right\}}$ with probabilities ${\rm P}(T_{i}=0)=1-\gamma$ and ${\rm P}(T_{i}=1)=\gamma$, respectively, and sends it to Alice through an authenticated classical channel. We denote by $T=(T_{1},T_{2},..., T_{n_{\rm{SH}}})$ the string of all bit values $T_{i}$.\\

\noindent 3. \textit{Measurement.} If $T_{i}=0$, the $i$-th successfully heralded round is considered to be a key generation round, and Alice and Bob choose the settings $(X_{i},Y_{i})=(0,2)$. If $T_{i}=1$, such round is considered to be a test round ({\it i.e.}, a CHSH game round), and they independently select $X_{i},Y_{i}\in{\left\{0,1\right\}}$ uniformly at random. 

Alice and Bob record their measurement outcomes as $A_{i}$, ${B}_{i}\in\{0,1\}$, respectively. If, say, Alice (Bob) obtains an inconclusive result (\textit{i.e.}, no photon or multiple photons are observed in the detectors) she (he) deterministically assigns $A_{i}=1$ ($B_{i}=1$) to keep the fair-sampling loophole closed. Finally, Alice (Bob) publicly announces the measurement settings $X_{i}$ ($Y_{i}$). In what follows, we will denote by $X$ ($Y$) the bit string $X_{1},X_{2},...,X_{n_{\rm{SH}}}$ ($Y_{1},Y_{2},...,Y_{n_{\rm{SH}}}$) of Alice's (Bob's) measurement settings for the successfully heralded rounds. Similarly, $A$ ($B$) will denote the string of measurement outcomes $A_{1},A_{2},...,A_{n_{\rm{SH}}}$ ($B_{1},B_{2},...,B_{n_{\rm{SH}}}$).\\

\noindent 4. \textit{Information reconciliation.} Alice and Bob use an error correction protocol to obtain from $A$ and $B$ two identical bit strings, $Z_{\rm{A}}$ and $Z_{\rm{B}}$, respectively. For this, Alice sends Bob $leak_{\rm{IR}}$ bits of syndrome information and Bob obtains an estimate, $Z_{\rm{B}}$, of $A$. Next, they perform an error verification step (using two-universal hash functions) that leaks at most $\lceil{\log_{2}{1/\epsilon_{\rm{IR}}}}\rceil$ bits of information to Eve, for a certain prefixed parameter $\epsilon_{\rm IR}$. If this last step is successful, it is guaranteed that Alice's and Bob's bit strings $Z_{\rm{A}}=A$ and $Z_{\rm{B}}$ satisfy ${\rm P}(Z_{\rm{A}}\neq{Z_{\rm{B}}})\leq{\epsilon_{\rm{IR}}}$. Otherwise, the protocol aborts.\\

\noindent 5. \textit{Parameter estimation.} Bob sets the parameter $C_{i}=\perp$ for the key generation rounds ({\it i.e.}, when $T_{i}=0$) and $C_{i}=\omega_{\rm{CHSH}}(Z_{{\rm B}i},B_{i},X_{i},Y_{i})$ for the test rounds ({\it i.e.}, when $T_{i}=1$), with $i=1,2,...,n_{\rm{SH}}$, and where $Z_{{\rm B}i}$ denotes the $i$-th bit of the string $Z_{\rm{B}}$ and the function $\omega_{\rm{CHSH}}$ is defined as
\[   \omega_{\rm{CHSH}}(a,b,x,y)=\left\{
\begin{array}{ll}
1 & \rm{if}\hspace{.2cm}a\oplus{b}={\it x}\cdot{\it y}, \\
0 & \rm{otherwise,} \\
\end{array} 
\right. \]
with $\oplus$ denoting bit addition modulo~$2$ and $\cdot$ denoting bit multiplication. The overall number of test rounds in which $C_{i}=1$ (and thus the parties win the CHSH game) is denoted by $C_{\rm{SH}}={\sum_{\left\{i:T_{i}=1\right\}}C_{i}}$. This quantity determines a lower bound on the number of secret bits that can be extracted from $Z_{\rm{A}}$ and $Z_{\rm{B}}$ using privacy amplification~\cite{Rotem}. 

Bob aborts the protocol if the fraction of wins lies below a certain prefixed threshold value, {\it i.e.}, when $C_{\rm{SH}}/{n_{\rm{SH}}}<\omega|_{\rm{SH}}\gamma-\delta_{\rm{est}}$, where $\omega|_{\rm{SH}}$ is the expected winning rate of the CHSH game in the test rounds for the considered channel model~\cite{comment0}, $\gamma$ is again the probability that Bob uses a successfully heralded round as a test round, and $\delta_{\rm{est}}$ is the confidence interval that defines the abortion threshold. That is, $\delta_{\rm{est}}$ is the maximum difference between the expected and the actual winning rates of the CHSH game that Bob accepts without aborting.\\

\noindent 6. \textit{Privacy amplification.} 
Alice and Bob apply a privacy amplification protocol to their bit strings $Z_{\rm{A}}$ and $Z_{\rm{B}}$ to obtain the final keys, $K_{\rm A}$ and $K_{\rm B}$, of length $l$. This protocol uses a randomness extractor that succeeds except with error probability $\epsilon_{\rm{PA}}$. \\

The main protocol arguments are summarized in~Table~\ref{table:1}.
\begin{table}
	\centering
	\begin{tabular}{|l|l|}
		\hline
		\multicolumn{2}{|c|}{Protocol arguments} \\ \hline
		$n_{\rm{SH}}$ & Post-processing block size \\ \hline
		$\gamma$ & Probability of a test round\\ \hline
		$\delta_{\rm{est}}$ & Confidence interval for the CHSH game winning rate \\ \hline
		$\epsilon_{\rm{IR}}$ & Error probability of information reconciliation \\ \hline
		$\epsilon_{\rm{PA}}$ & Error probability of privacy amplification \\ \hline
		$l$ & Length of the final keys $K_{\rm A}$ and $K_{\rm B}$ \\ \hline
	\end{tabular}
	\caption{List containing the main protocol arguments.}
	\label{table:1}
\end{table}

We remark that in the distribution step of the protocol, Alice and Bob need to store their signals until they choose their measurement settings and measure the signals in the third step of the protocol. For simplicity, below we will optimistically assume that, for this purpose, both of them hold noiseless and lossless quantum memories in their labs. Alternatively, they could also decide which rounds are key generation rounds and which ones are test rounds {\it a posteriori} by using the typical sifting step in QKD, though this approach results in a slightly less efficient solution. This is so because the data associated to Alice measuring $\sigma_{\rm x}$ and Bob measuring $\sigma_z$ is not used in the protocol. 

Also, we note that in a photonic implementation of the DIQKD scheme, the measurements $X_{i}$ and $Y_{i}$ can be realized by means of a polarization modulator that rotates the polarization state of the incoming signals, together with a PBS that separates vertical and horizontal polarization modes, followed by two photodetectors. For example, the rotation angles of the polarization modulator associated to the measurements $\sigma_{z}$, $\sigma_{x}$, $\sigma_{+}$ and $\sigma_{-}$ are $0$, $\pi/4$, $\pi/8$ and $-\pi/8$ radians, respectively. The observation of one single-photon in, say, horizontal (vertical) polarization is recorded by Alice as $A_{i}=0$ ($A_{i}=1$), and the same applies to Bob. 

\section{Security analysis}\label{Section4}

We use the security analysis introduced in~\cite{Rotem}, which is valid against coherent attacks. Prior to the execution of the protocol, Alice and Bob agree on three parameters that tag the security of the final keys, $K_{\rm A}$ and $K_{\rm B}$. These parameters are the secrecy parameter, $\epsilon_{\rm{sec}}$, the correctness parameter, $\epsilon_{\rm{cor}}${,} and the robustness parameter, $\epsilon_{\rm{rob}}$. 

In particular, a protocol is said to be $\epsilon_{\rm{sec}}$-secret, when implemented using a device D, if it satisfies
\begin{equation}\label{12}
[1-{\rm P(abort)}]||\rho_{K_{\rm A}{\rm E}}-\rho_{{\rm U}_{l}}\otimes{\rho_{\rm{E}}}||_{1}\leq{\epsilon_{\rm{sec}}},
\end{equation} 
where $\rm{P(abort)}$ is the abortion probability of the protocol,  $||\rho||_{1}={\sqrt{\rho\rho^{\dagger}}}$ stands for the trace norm, E is a quantum register held by the eavesdropper that may be initially correlated with D, $\rho_{K_{\rm A}{\rm E}}$ is the output state of the DIQKD protocol describing Alice's key string $K_{\rm A}$ and the quantum register E conditioned on not aborting, $\rho_{{\rm U}_{l}}=\sum_z \frac{1}{2^l}\ket{z}_{\rm A}\bra{z}$ is the uniform mixture of all possible values of a $l$-bit string $K_{\rm A}$, and $\rho_{{\rm U}_{l}}\otimes{\rho_{\rm{E}}}$ is the perfectly secret output state.

The parameter $\epsilon_{\rm{sec}}$ is upper bounded by
\begin{equation}\label{13}
\epsilon_{\rm{sec}}\leq{\epsilon_{\rm{PA}}+\epsilon_{\rm{s}}}+\epsilon_{\rm{EA}}, 
\end{equation}
where $\epsilon_{\rm{PA}}+\epsilon_{\rm{s}}$ is the total failure probability associated to the privacy amplification step, $\epsilon_{\rm{PA}}$ being an upper bound on the error probability of the randomness extractor and $\epsilon_{\rm{s}}$ being the smoothing parameter of the $\epsilon_{\rm{s}}$-smooth min-entropy~\cite{Rotem}. The term $\epsilon_{\rm{EA}}$ is the failure probability associated to the entropy accumulation theorem~\cite{Renner}, which only guarantees that a certain lower bound on the $\epsilon_{\rm{s}}$-smooth min-entropy holds with a probability larger than $1-\epsilon_{\rm{EA}}$.

The correctness parameter, $\epsilon_{\rm{cor}}$, quantifies the probability that the final keys, $K_{\rm A}$ and $K_{\rm B}$, are not equal. More precisely, a protocol is said to be $\epsilon_{\rm{cor}}$-correct if ${\rm P}[K_{\rm A}\neq K_{\rm B}]\leq{\epsilon_{\rm{cor}}}$. According to the protocol definition given in the previous section, we have that $\epsilon_{\rm{cor}}={\epsilon_{\rm{IR}}}$.

Finally, a protocol is said to be $\epsilon_{\rm{rob}}$-robust for a specific honest implementation ({\it i.e.}, for a particular implementation where the eavesdropper does not intervene) if it aborts with a probability smaller than $\epsilon_{\rm{rob}}$. The protocol described above can only abort in two steps: the information reconciliation step, and the parameter estimation step. Therefore, we have that $\epsilon_{\rm{rob}}$ satisfies
\begin{equation}\label{14}
\epsilon_{\rm{rob}}\leq{\epsilon_{\rm{rob}}^{\rm{IR}}+\epsilon_{\rm{rob}}^{\rm{PE}}},
\end{equation}
where $\epsilon_{\rm{rob}}^{\rm{IR}}$ ($\epsilon_{\rm{rob}}^{\rm{PE}}$) is the probability of aborting at the information reconciliation (parameter estimation) step for the considered honest implementation. Moreover, we have that the quantity $\epsilon_{\rm{rob}}^{\rm{PE}}$ verifies
\begin{equation}\label{15} 
\epsilon_{\rm{rob}}^{\rm{PE}}\leq{\epsilon_{\rm{rob}}^{\rm{EA}}+\epsilon_{\rm{IR}}},
\end{equation}
where $\epsilon_{\rm{rob}}^{\rm{EA}}$ is the probability of the fraction of CHSH wins, $C_{\rm{SH}}/{n_{\rm{SH}}}$, being lower than the threshold $\omega|_{\rm{SH}}\gamma-\delta_{\rm{est}}$. That is,
\begin{equation}\label{16}
\epsilon_{\rm{rob}}^{\rm{EA}}={\rm{P}\left(\omega|_{\rm{SH}}\gamma-\frac{C_{\rm{SH}}}{n_{\rm{SH}}}>\delta_{\rm{est}}\right)}.
\end{equation}
Note that for fixed values of $\epsilon_{\rm{rob}}^{\rm{EA}}$ and $n_{\rm{SH}}$, the minimum value of $\delta_{\rm{est}}$, such that Eq.~(\ref{16}) holds, satisfies~\cite{Hoeffding}
\begin{equation}\label{17}
\delta_{\rm{est}}\geq{\sqrt{\frac{1}{2n_{\rm{SH}}}\ln{\frac{1}{\epsilon_{\rm{rob}}^{\rm{EA}}}}}}.
\end{equation}
Also, we note that the parameter $\epsilon_{\rm{IR}}$ contributes to $\epsilon_{\rm{rob}}^{\rm{PE}}$ in Eq.~(\ref{15}) because, conditioned on not aborting the protocol in the error verification step, Bob performs the parameter estimation step by using his original bit string of outcomes, $B$, and his estimate, $Z_{\rm B}$, of Alice's string, which is equal to $Z_{\rm A}$ except with probability $\epsilon_{\rm{IR}}$.

A list with the main parameters related to the security of the protocol is provided in Table~\ref{table:2}.
\begin{table}
	\centering
	\begin{tabular}{|l|l|}
		\hline
		\multicolumn{2}{|c|}{Finite-key security parameters} \\ \hline
		$\epsilon_{\rm{sec}}$ & Secrecy parameter \\ \hline
		$\epsilon_{\rm{cor}}$ & Correctness parameter \\ \hline
		$\epsilon_{\rm{rob}}$ & Robustness parameter \\ \hline
		$\epsilon_{\rm{s}}$ & Smoothing parameter of the min-entropy\\ \hline
		$\epsilon_{\rm{rob}}^{\rm{PE}}$ & Abort probability of parameter estimation \\ \hline
		$\epsilon_{\rm{rob}}^{\rm{IR}}$ & Abort probability of information reconciliation \\ \hline
		$\epsilon_{\rm{rob}}^{\rm{EA}}$ & Abort probability of entropy accumulation \\ \hline
		$\epsilon_{\rm{EA}}$ & Error probability of entropy accumulation bound \\ \hline
	\end{tabular}
	\caption{List containing the main finite-key security parameters.}
	\label{table:2}
\end{table}\rm

Then, it turns out that, conditioned on not aborting, the DIQKD protocol presented in the previous section delivers $\epsilon_{\rm{cor}}$-correct and $\epsilon_{\rm{sec}}$-secret output keys, $K_{\rm A}$ and $K_{\rm B}$, whose length, $l$, is given by
\begin{eqnarray}\label{18}
l&=&n_{\rm SH}(\eta_{\rm opt}-\gamma)-2{\log{7}}\sqrt{1-2\log{\left[\frac{\epsilon_{\rm s}}{4}(\epsilon_{\rm{EA}}+\epsilon_{\rm IR})\right]}}\nonumber \\
&\times&\sqrt{n_{\rm SH}}-leak_{\rm IR}-3\log{\left[1-\sqrt{1-\left(\frac{\epsilon_{\rm{s}}}{4}\right)^2}\right]}\nonumber \\
&-&2\log{\frac{1}{\epsilon_{\rm PA}}}. 
\end{eqnarray}
Here, the term $\eta_{\rm{opt}}$ represents a lower bound on the entropy generation rate of the CHSH game (see~Appendix~\ref{ap2} for further details on $\eta_{\rm{opt}}$ and $leak_{\rm IR}$).
Importantly, the {\it conditional} secret key rate, {\it i.e.}, the number of secret key bits per successful heralding event, reads $K|_{\rm SH}=l/n_{\rm SH}$, and the secret key rate is defined as
\begin{equation}\label{19}
K={}\frac{l}{N},
\end{equation}
where $N$ is the total number of transmitted signals. Below, we will use the term ``success rate'' to refer to the number of successful heralding events per transmitted signal, {\it i.e.}, $r_{\rm SH}=n_{\rm SH}/N$. 

We remark that, in the limit where $n_{\rm SH}\to{\infty}$, Eq.~(\ref{19}) matches the asymptotic secret key rate against general attacks reported in~\cite{Vazirani}, which we will denote by $K_{\infty}$. To see this, note that the quantum communication phase of the protocol described in the previous section only ends when $n_{\rm SH}$ heralded events are observed. This means that the number of transmitted signals, $N$, is a negative binomial random variable. Let $\rm{Var}\left(N/n_{\rm SH}\right)$ be the variance of the related random variable $N/n_{\rm SH}$. Then, we have that $\lim_{n_{\rm SH}\rightarrow{\infty}}{\rm{Var}\left(N/n_{\rm SH}\right)}=\lim_{n_{\rm SH}\rightarrow{\infty}}{(1-{P_{\rm SH}})/({n_{\rm SH}}{P_{\rm SH}}^{2})}=0$. This means that $\lim_{n_{\rm SH}\rightarrow{\infty}}{N/n_{\rm SH}}=1/{P_{\rm SH}}$, where $P_{\rm{SH}}$ is the successful heralding probability of the qubit amplifier. By using this result, it is straightforward to show that $K_{\infty}$ satisfies
\begin{eqnarray}\label{20}
K_{\infty}&=&\lim_{n_{\rm SH}\to\infty}\frac{n_{\rm SH}\eta_{\rm opt}-leak_{\rm IR}}{N} \nonumber \\
&=&P_{\rm SH}\bigg\{1-h\bigg[\frac{1}{2}+\frac{1}{2}\sqrt{16\omega|_{\rm SH}(\omega|_{\rm SH}-1)+3}\bigg] \nonumber \\
&-&h(Q|_{\rm SH})\bigg\},
\end{eqnarray}
which is the result provided in~\cite{Vazirani}. In this equation, $h(x)=-x\log_2{x}-(1-x)\log_2{(1-x)}$ is the binary entropy function, and $Q|_{\rm SH}$ represents the quantum bit error rate of the key generation rounds. In the asymptotic regime, the conditional secret key rate reduces to $K|_{\rm SH}=K_{\infty}/P_{\rm SH}$, while for the success rate we have $r_{\rm SH}=P_{\rm SH}$.

In the next section, we evaluate the effect that device imperfections---particularly, finite detection and coupling efficiencies and the multi-photon pulses emitted by practical sources---have on the performance of DIQKD in terms of the resulting secret key rate for various honest optical implementations.

\section{Evaluation}\label{Section5}

Here, we use the device models introduced in Sec.~\ref{Section2} to evaluate the performance of the DIQKD protocol presented in Sec.~\ref{Section3}, by using the security analysis given in the previous section. The goal is to determine the resources needed to implement DIQKD over long distances. All the relevant calculations to reproduce the results presented in this section are included in Appendix~\ref{honest}.

For simplicity, in the simulations below we assume that the coupling efficiency of the light sources is equal to the detection efficiency of the photodetectors, {\it i.e.}, we set $\eta_{\rm c}=\eta_{\rm d}=\eta_{\rm c,d}$. This decision is motivated because DIQKD requires very high values of $\eta_{\rm c,d}$, so the effect of this simplification is negligible, and it reduces the number of experimental parameters to consider. Also, unless otherwise stated, we fix the dark count rate of the photodetectors to $p_{\rm d}=10^{-7}$, which is a reasonable value with current technology~\cite{det1,det2}. 

Furthermore, for illustration purposes, we consider two examples of security parameter sets $(\epsilon_{\rm sec}, \epsilon_{\rm cor}, \epsilon_{\rm rob})$, which we denote by $S_{1}$ and $S_{2}$. These sets are given in Table~\ref{table:3}.
\begin{table}
	\centering
	\begin{tabular}{|c|c|c|c|}
		\hline
		\quad & $\epsilon_{\rm sec}$ & $\epsilon_{\rm cor}$ & $\epsilon_{\rm rob}$ \\ \hline
		$S_{1}$ & $10^{-5}$ & $10^{-10}$ & $10^{-2}$  \\ \hline
		$S_{2}$ & $10^{-9}$ & $10^{-15}$ & $10^{-3}$  \\ \hline
	\end{tabular}
	\caption{Sets of security parameters $\epsilon_{\rm sec}$, $\epsilon_{\rm cor}$ and $\epsilon_{\rm rob}$ considered in the performance evaluation of DIQKD. The set $S_{1}$ provides a lower level of security than the set $S_{2}$.}
	\label{table:3}
\end{table}
Also, to simplify the numerics, we fix the value of the failure probability of the entropy accumulation theorem to $\epsilon_{\rm EA}=10^{-6}$ ($\epsilon_{\rm EA}=10^{-10}$) for the set $S_{1}$ ($S_{2}$). We remark, however, that according to our simulations the loss of generality that results from fixing the value of $\epsilon_{\rm EA }$ in advance is very small. The secret key rate is then maximized over the remaining parameters, which include the quantities $\epsilon_{\rm PA}$, $\epsilon_{\rm s}$, $\epsilon_{\rm rob}^{\rm IR}$, $\epsilon_{\rm rob}^{\rm EA}$, the value of the transmittance $t$ in the case of a PQA (see Fig.~\ref{fig:3}(c)), as well as the intensities of the different light sources. 

In the absence of real experimental data, in our simulations we set the denominator of Eq.~(\ref{19}) to the expected number of transmitted signals until $n_{\rm SH}$ successful heralding detection events occur at the qubit amplifier, which is given by
\begin{equation}\label{mean_tx_signals}
\left\langle{N}\right\rangle=\frac{n_{\rm SH}}{P_{\rm SH}}.
\end{equation}
This also amounts to setting the success rate of the qubit amplifier to its expected value, {\it i.e.}, $\left\langle{r_{\rm SH}}\right\rangle=P_{\rm SH}$. 

Finally, in all the plots below we assume a threshold value for the secret key rate as low as $10^{-10}$. That is, whenever the resulting secret key rate is smaller than this threshold value, it is considered to be impractical and we neglect it, as this value is already probably too low to have any practical relevance.  

\subsection{Ideal sources}
We start by analyzing the ideal scenario where Alice and Bob hold perfect photon sources. Obviously, this case provides the best possible performance, and thus it can be used as a reference about the minimum resources (say, {\it e.g.}, the minimum value of the detection and coupling efficiency, $\eta_{\rm c,d}$, and the minimum block size, $n_{\rm SH}$) that are required to achieve a certain secret key rate.

More precisely, we consider here that the entanglement source, $\rho_{ab}$, at Alice's lab generates perfect polarization Bell pairs, {\it i.e.}, $\rho_{ab}=\ket{\phi_{1}}_{ab}\bra{\phi_{1}}$ with $\ket{\phi_{1}}_{ab}$ given by Eq.~(\ref{2}). Also, in the case of a PQA, we set $\rho^{\rm{h}}_{\rm{single}}$ ($\rho^{\rm{v}}_{\rm{single}}$) to be a perfect single-photon source generating horizontally (vertically) polarized single photons. That is, $\rho^{\rm{h}}_{\rm{single}}=\ket{1,0}\bra{1,0}$ ($\rho^{\rm{v}}_{\rm{single}}=\ket{0,1}\bra{0,1}$) with $\ket{1,0}=a^\dagger_{\rm h}\ket{0}$ ($\ket{0,1}=a^\dagger_{\rm v}\ket{0}$), being $\ket{0}$ the vacuum state. Similarly, in the case of an ESR, we set ${\rho}_{bc}=\ket{\phi_{1}}_{bc}\bra{\phi_{1}}$. 

\subsubsection{No channel loss}

To begin with, we compare the achievable performance when using PQAs and ESRs in the absence of channel loss, {\it i.e.}, we set the distance at $L=0$~km. This scenario allows us to determine the minimum value of $n_{\rm SH}$ as a function of $\eta_{\rm c,d}$.

As we will show below, it turns out that $\eta_{\rm c,d}$ is quite high even in this ideal scenario. This means that the probability that any of these two amplifiers provides a spurious success at Bob's side due to the dark counts of the PNR detectors within the BSM is negligible compared to that of a genuine success triggered by a single photon from Alice. Therefore, for simplicity, in this subsection we set the dark count rate $p_{\rm d}$ equal to zero. With this approximation, it is straightforward to derive simple analytical expressions for the quantities $P_{\rm SH}$, $\omega|_{\rm SH}$ and $Q|_{\rm SH}$. In the case of an ESR, we obtain
\begin{eqnarray}\label{21}
P_{\rm SH}^{\rm ESR}&=&\frac{\xi^{2}}{2}, \nonumber \\
\omega|_{\rm SH}^{\rm ESR}&=&{\frac{2+\sqrt{2}}{4}}\xi^{2}+{\frac{3}{4}}{(1-\xi)^{2}}+\xi{(1-\xi)}, \nonumber \\
Q|_{\rm SH}^{\rm ESR}&=&\xi{(1-\xi)},
\end{eqnarray}
where the parameter $\xi={\eta_{\rm c,d}^{2}}$. Similarly, it can be shown that in the case of a PQA we have
\begin{eqnarray}\label{22}
P_{\rm SH}^{\rm PQA}&=&(1-t)\xi^{2}\left[1-\xi(1-t)\right], \nonumber \\
\omega|_{\rm SH}^{\rm PQA}&=& \frac{1}{1-\xi(1-t)}\bigg[\frac{2+\sqrt{2}}{4}{t}{\xi^2}+\frac{3}{4}{(1-\xi)^2} \nonumber \\
&+& \frac{(1+t)}{2}\xi(1-\xi)\bigg], \nonumber \\
Q|_{\rm SH}^{\rm PQA}&=&\frac{(1+t){\xi}(1-\xi)}{2\left[1-\xi(1-t)\right]},
\end{eqnarray}
where the parameter $t$ corresponds to the transmittance of the BS within the amplifier. See Appendix~\ref{honest} for details of the calculations to derive Eqs.~(\ref{21}) and (\ref{22}).

From Eq.~(\ref{22}), it is evident that there is a trade-off on the coefficient $t$. The terms $\omega|_{\rm SH}^{\rm PQA}$ and $Q|_{\rm SH}^{\rm PQA}$ favor $t\approx 1$, and thus the conditional secret key rate $K|_{\rm SH}$, which depends on these parameters but not on $P_{\rm SH}^{\rm PQA}$, also favors $t\approx 1$. Indeed, in the limit $t\rightarrow1$ we have that $\omega|_{\rm SH}^{\rm PQA}=\omega|_{\rm SH}^{\rm ESR}$ and $Q|_{\rm SH}^{\rm PQA}=Q|_{\rm SH}^{\rm ESR}$. On the other hand, the average success rate of the qubit amplifier, $P_{\rm SH}^{\rm PQA}$, is maximized when $t=1-(2\xi)^{-1}$, and it actually vanishes when $t=1$. This behaviour can be easily understood by examining the states $\rho_{bc}=\ket{\phi}_{bc}\bra{\phi}$ generated within the PQA and which are given by Eq.~(\ref{9}). By setting $t$ close to 1 we have that whenever a successful heralding takes place at the qubit amplifier, with a high probability this event is due to the entangled pair $\ket{\chi}_{bc}$ (see Eq.~(\ref{10})), and thus $K|_{\rm SH}$ is maximized. On the contrary, the lower the transmittance $t$ is, the more likely is that a successful heralding comes from the spurious term $\ket{\psi}_{bb}$ and, consequently, $K|_{\rm SH}$ is minimized. This might happen when only one photon from $\ket{\psi}_{bb}$ is detected due to the finite detection efficiency of the detectors, and the remaining necessary detection is caused by a single photon coming from Alice. In our simulations, we numerically optimize the value of $t$ so that the overall secret key rate $K$ is maximized.

Also, we remark that from Eqs.~(\ref{20})-(\ref{21})-(\ref{22}) it is easy to show that a positive secret key rate requires a very high value of $\xi\gtrapprox{92.3\%}$ (or, equivalently, $\eta_{\rm c,d}=\sqrt{\xi}\gtrapprox{96.1\%}$) for both types of qubit amplifiers. Similarly, from Eqs.~(\ref{21}) and (\ref{22}), it can be shown that, irrespectively of the value of $t$, whenever $\xi\geq 50\%$ we have that $P_{\rm SH}^{\rm ESR}>{P_{\rm SH}^{\rm PQA}}$, $\omega|_{\rm SH}^{\rm ESR}\geq{\omega|_{\rm SH}^{\rm PQA}}$ and $Q|_{\rm SH}^{\rm ESR}\leq{Q|_{\rm SH}^{\rm PQA}}$. That is, an ESR-based qubit amplifier always outperforms a PQA in the absence of channel loss, if perfect sources are assumed. 

\begin{figure}
	\includegraphics[width=10.5cm,height=12cm]{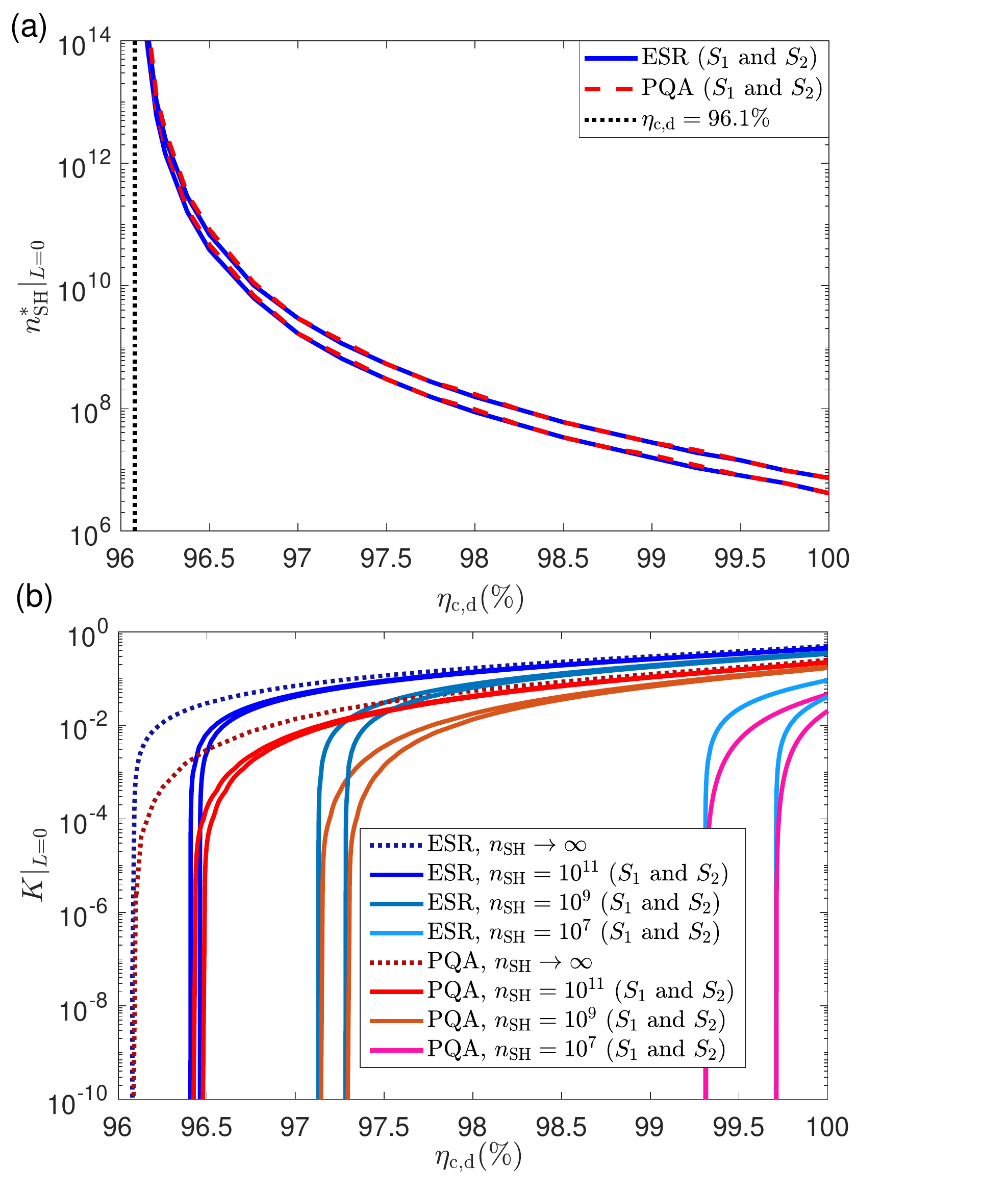}
	\caption{Performance evaluation of DIQKD with ideal photon sources at $L=0$ km. Bluish (reddish) lines are used for the ESR (PQA) architecture.
		(a) Minimum value of the detection and coupling efficiency, $\eta_{\rm c,d}$, and minimum value of the block size, $n_{\rm SH}$, required to obtain a zero-distance secret key rate $K|_{L=0}\geq{10^{-10}}$. Both sets of security requirements, $S_1$ and $S_2$, are compared for each qubit amplifier. Any combination of parameters $\eta_{\rm c,d}$ and $n_{\rm SH}$ must be above the lower (upper) lines to achieve a secret key rate above the threshold value with the security requirements given by the sets $S_{1}$ ($S_{2}$). The dotted black vertical line indicates the (asymptotic) minimum efficiency, $\eta_{\rm c,d}\approx{96.1\%}$, which is the smallest detection efficiency that delivers a zero-distance asymptotic secret key rate $K_{\infty}|_{L=0}\geq{10^{-10}}$.
		(b) Zero-distance secret key rate, $K|_{L=0}$, as a function of $\eta_{\rm c,d}$ for various values of the block size $n_{\rm SH}$. For each qubit amplifier, four different block sizes are considered: $n_{\rm SH}\rightarrow{\infty}$, $n_{\rm SH}=10^{11}$, $n_{\rm SH}=10^{9}$ and $n_{\rm SH}=10^{7}$. The finite secret key rates appear in pairs of solid lines, one for the security set $S_{1}$ (upper line) and another one for the security set $S_{2}$ (lower line). The asymptotic secret key rates corresponding to $n_{\rm SH}\rightarrow{\infty}$ are illustrated with dotted lines.}
	\label{fig:3}
\end{figure}
In Fig.~\ref{fig:3}(a) we plot the minimum block size, $n_{\rm SH}$, and the minimum detection efficiency, $\eta_{\rm c,d}$, that are needed to obtain a secret key rate above the threshold value of $10^{-10}$ at $L=0$~km. We denote this secret key rate by $K|_{L=0}$, and the minimum block size is denoted by $n_{\rm SH}^{*}|_{L=0}$. The value of $n_{\rm SH}^{*}|_{L=0}$ is obtained for each value of $\eta_{\rm c,d}$ via exhaustive numerical search over all the free parameters contained in the finite-key rate formula given by Eq.~(\ref{19}). The solid (dashed) bluish (reddish) lines correspond to the ESR (PQA) architecture, and in each case the lower (upper) line uses the set of security requirements $S_{1}$ ($S_{2}$); see Table~\ref{table:3}. Remarkably, despite the security parameters of $S_{2}$ being significantly more demanding than those of $S_{1}$, it turns out that the set $S_{2}$ does not require much larger block sizes than $S_1$. 

Also, Fig.~\ref{fig:3}(a) indicates that both qubit amplifiers require a similar minimum block size, $n_{\rm SH}^{*}|_{L=0}$, to deliver a secret key rate above the threshold value. Indeed, it is easy to show that if the threshold value for the secret key rate were zero (instead of $10^{-10}$) then the value of $n_{\rm SH}^{*}|_{L=0}$ would be equal for both qubit amplifiers. However, since we use a threshold value greater than zero implies that $n_{\rm SH}^{*}|_{L=0}$ is always slightly lower for the ESR than for the PQA. This is so because, even though the latter can mimic the conditional secret key rate of the former for any efficiency $\eta_{\rm c,d}$, the ESR has a higher success probability $P_{\rm SH}^{\rm ESR}$, thus leading to a higher overall secret key rate. Nevertheless, this effect cannot be fully appreciated with the resolution of Fig.~\ref{fig:3}(a).

Finally, the dotted black vertical line illustrated in Fig.~\ref{fig:3}(a) corresponds to the (asymptotic) minimum efficiency, $\eta_{\rm c,d}\approx{96.1\%}$, required to obtain $K_{\infty}|_{L=0}\geq{10^{-10}}$. That is, no secret key rate above such threshold value is possible when $\eta_{\rm c,d}\lessapprox{96.1\%}$, no matter how much we increase the block size.

In what follows, we will refer to the lines in Fig.~\ref{fig:3}(a) as the \textit{critical lines}, since every pair $(\eta_{\rm c,d},n_{\rm SH})$ lying below these lines delivers a negligible secret key rate with the corresponding security requirements.

Fig.~\ref{fig:3}(b) shows the zero-distance secret key rate, $K|_{L=0}$, as a function of the detection and coupling efficiency, $\eta_{\rm c,d}$, for different values of the block size, $n_{\rm SH}$. As already discussed above, the ESR architecture always leads to larger secret key rates for all values of $\eta_{\rm c,d}$, while the minimum efficiencies required to have a secret key rate larger than the threshold value are roughly equal for both qubit amplifiers. Again, the small mismatch between the minimum efficiencies required by both amplifiers occurs because the selected threshold is greater than zero. Otherwise, the minimum efficiencies would match. For illustrative purposes, Fig.~\ref{fig:3}(b) considers four different block sizes: $n_{\rm SH}\rightarrow{\infty}$, $n_{\rm SH}=10^{11}$, $n_{\rm SH}=10^{9}$ and $n_{\rm SH}=10^{7}$. As already shown in Fig.~\ref{fig:3}(a), the smaller the block size is, the larger the value of the minimum efficiency $\eta_{\rm c,d}$ that is required. For instance, for a block size as large as, say, $n_{\rm SH}=10^{11}$, and if one considers the weaker set of security requirements $S_{1}$, the minimum efficiency is at least $\eta_{\rm c,d}\approx{96.4\%}$. Also, for any given value of $n_{\rm SH}$, the greater the detection efficiency considered (with respect to its minimum value), the closer the resulting secret key rates corresponding to the security settings $S_{1}$ and $S_{2}$ become. This is so because, in this situation, the effect of finite statistics is less prominent. Note that in the limit given by the asymptotic regime, the secret key rate $K_{\infty}$ does not depend on the security sets $S_{1}$ and $S_{2}$, but these sets are only relevant in the finite-key regime.

\subsubsection{Channel loss}\label{channel_loss_ideal}

\begin{figure}
	\includegraphics[width=10.5cm,height=12cm]{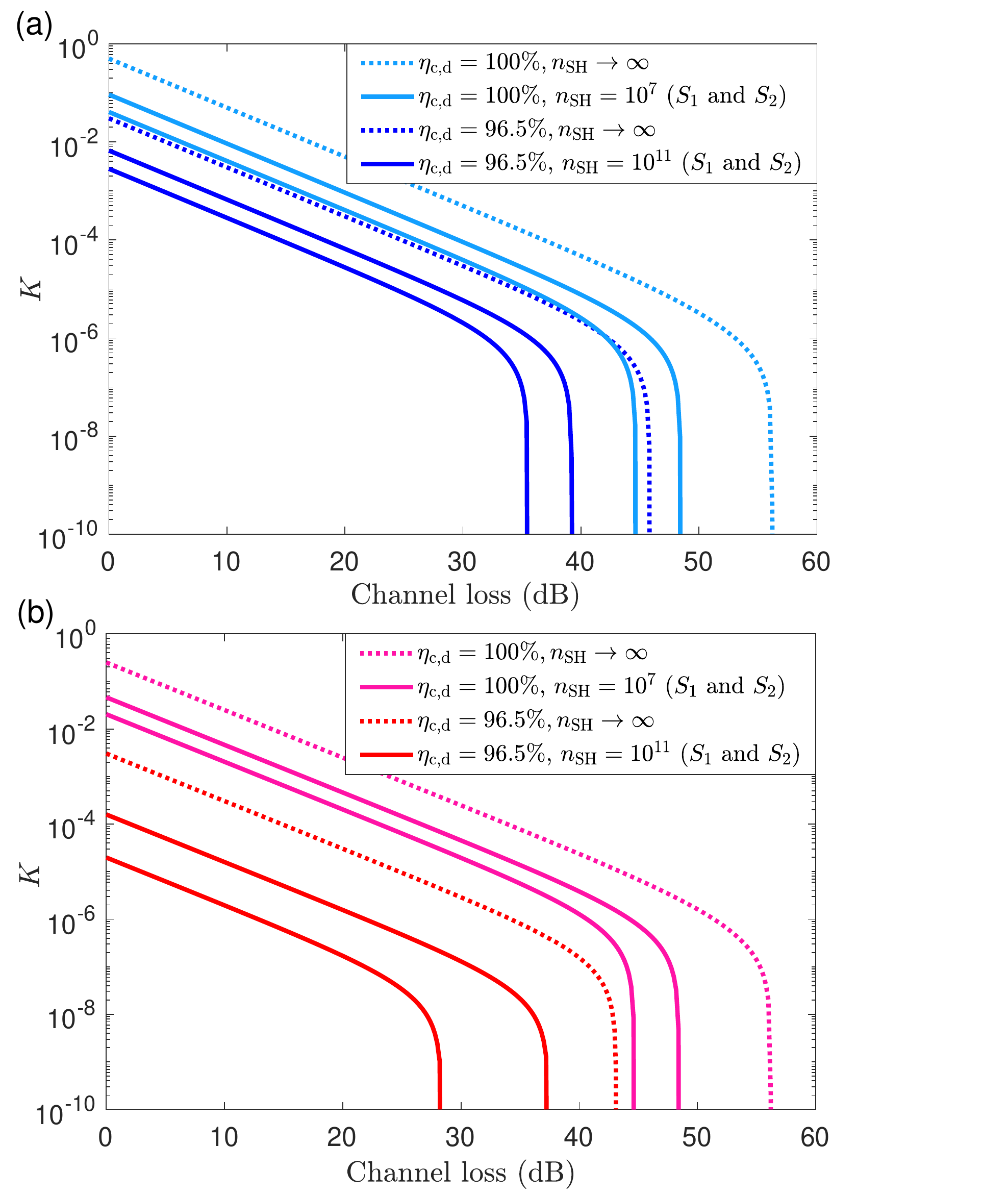}
	\caption{Secret key rate $K$ as a function of the overall channel loss $\Lambda=10\log_{10}{(1/\eta_{\rm ch})}$ measured in dB for the case of ideal photon sources. Figure~(a) corresponds to the ESR architecture and figure~(b) to the PQA architecture. In both figures, we use two different detection and coupling efficiencies, $\eta_{\rm c,d}=100\%$ and $\eta_{\rm c,d}=96.5\%$, each of them tagged with a different color. For each value of the efficiency, we plot the asymptotic secret key rate $K_\infty$ (dotted line),	together with two finite-key rates for different values of $n_{\rm SH}$ (solid lines). Each finite-key rate is plotted twice, one line corresponds to the security settings $S_1$ and the other line to the security settings $S_2$. Both finite-key rates assume a common block size $n_{\rm SH}$ close to the critical one (see Fig.~\ref{fig:3}(a)). More precisely, we take $n_{\rm SH}=10^{7}$ when $\eta_{\rm c,d}=100\%$ and $n_{\rm SH}=10^{11}$ when $\eta_{\rm c,d}=96.5\%$. By increasing the value of $\eta_{\rm c,d}$ and/or $n_{\rm SH}$ the finite-key rates approach those of the optimal scenario, which corresponds to $K_\infty$ assuming $\eta_{\rm c,d}=100\%$.}
	\label{fig:4}
\end{figure}
In this subsection, we consider the effect of channel loss, which is parametrized by the quantity $\Lambda={\alpha}L$ measured in dB (see Sec.~\ref{C}). Also, we set here the dark count rate of the detectors to $p_{\rm d}=10^{-7}$, as the effect of dark counts becomes relevant in this scenario. 

Fig.~\ref{fig:4} plots the secret key rate $K$ as a function of $\Lambda$ for various values of $\eta_{\rm c,d}$ and $n_{\rm SH}$, and for the two considered qubit amplifier architectures. More precisely, we use two values for $\eta_{\rm c,d}$: the ideal one, {\it i.e.}, $\eta_{\rm c,d}=100\%$, and one close to the threshold value of $96.1\%$ discussed above, say, $\eta_{\rm c,d}=96.5\%$. Moreover, for each of these values of the efficiency $\eta_{\rm c,d}$, we plot three different secret key rates: the asymptotic one $K_\infty$, and two finite-key rates, one for the security settings $S_{1}$ and another one for the security settings $S_{2}$. In both finite-key cases, we use a common block size $n_{\rm SH}$ close to the critical value obtained from Fig.~\ref{fig:3}.(a). Specifically, we set $n_{\rm SH}=10^{7}$ when $\eta_{\rm c,d}=100\%$, and  $n_{\rm SH}=10^{11}$ when $\eta_{\rm c,d}=96.5\%$. In doing so, and for the considered security analysis, we are simultaneously providing upper bounds (given by $K_\infty$) and lower bounds (given by the finite-key rates) to the finite-key performance that could be achieved with the chosen detection and coupling efficiencies, and the security requirements. By increasing the value of $n_{\rm SH}$, the finite-key rates approach the asymptotic scenario. Also, $K_\infty$ with $\eta_{\rm c,d}=100\%$ provides a clear upper bound for the achievable secret key rate with the security analysis introduced in Sec.~\ref{Section4}. 

Figs.~\ref{fig:4}.(a) and (b) further show, as expected, that in the case of ideal sources the ESR architecture outperforms the PQA architecture also in the presence of channel loss. 
 
 As a final remark, we note that if Bob did not use a qubit amplifier, then the maximum possible value of $\Lambda$ would be very limited. Indeed, it can be shown that in the case of ideal sources, and even if one sets $\eta_{\rm c,d}=100\%$ and $n_{\rm SH}\to\infty$, the maximum value of $\Lambda$ is as low as $\Lambda\lessapprox 0.7$ dB. See Appendix~\ref{unassisted_scenario} for further details.
 
\subsubsection{Time constraints}

In the discussion so far, we have not considered the duration of a DIQKD session, which is another crucial experimental parameter. Indeed, this parameter imposes strong restrictions on the loss that DIQKD can tolerate. We study it in this section. 

In the protocol described in Sec.~\ref{Section3}, we have that the post-processing block size, $n_{\rm SH}$, is fixed a priori. This means, in particular, that the number of transmitted signals, $N$, which is needed to achieve $n_{\rm SH}$ successful heralding events, and thus the duration of the distribution step of the protocol, which we shall denote by $\tau$, are random variables. Their mean values are given by Eq.~(\ref{mean_tx_signals}) and $\left\langle{\tau}\right\rangle={\nu}\left\langle{N}\right\rangle$, respectively, where $\nu$ represents the clock rate of system. 

\begin{figure}
	\includegraphics[width=10cm,height=6.5cm]{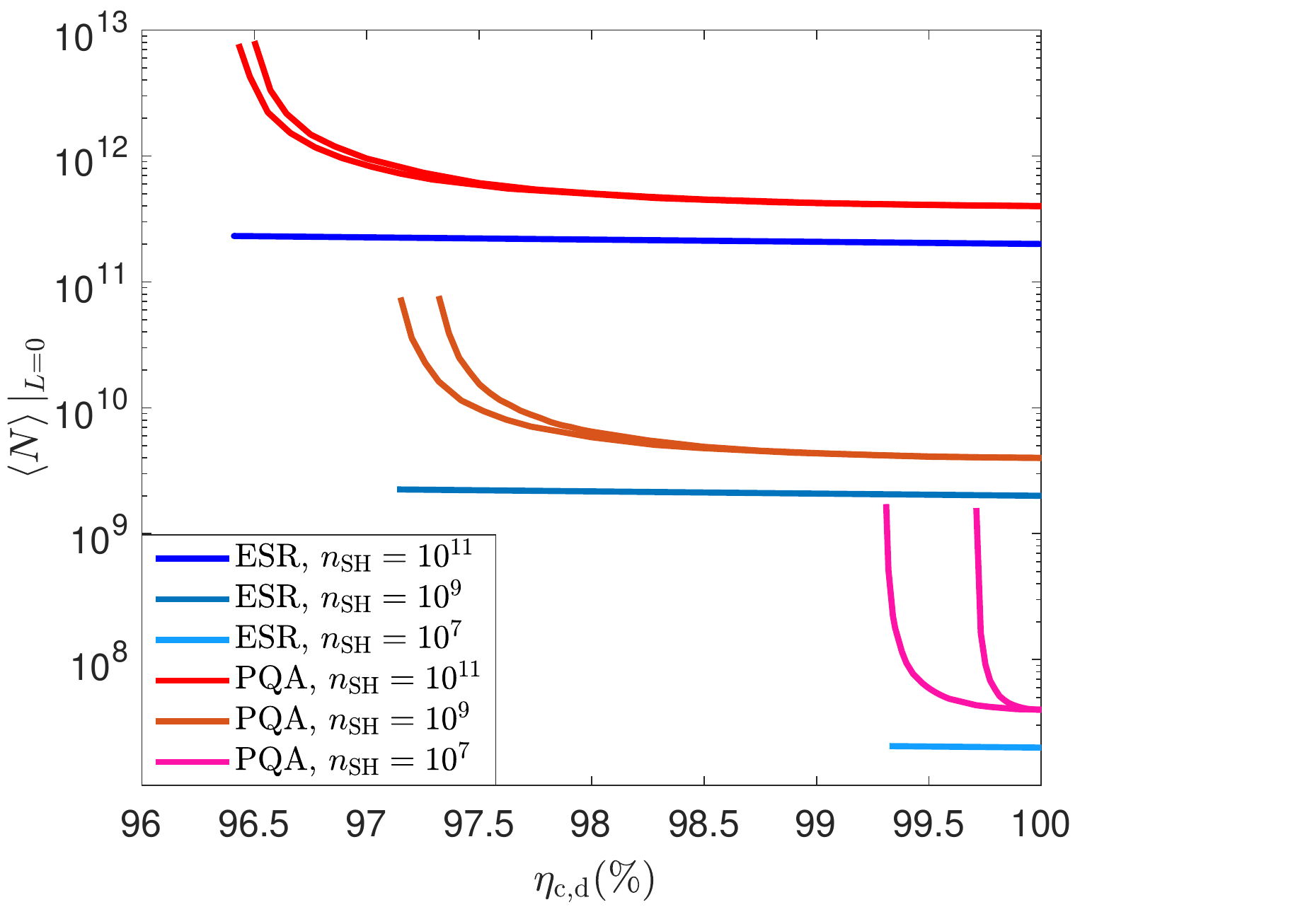}
	\caption{Average number of signals, $\left\langle{N}\right\rangle|_{L=0}$, that Alice needs to send Bob to collect a data block size equal to $n_{\rm SH}$ when using ideal photon sources, as a function of the detection and coupling efficiency $\eta_{\rm c,d}$ at $L=0$~km. As in Eqs.~(\ref{21})-(\ref{22}), in this figure we disregard dark counts because their effect at $L=0$~km is negligible. Also, we set the free experimental and security parameters to those values that optimize the secret key rate given by Fig.~\ref{fig:3}(b). The figure considers three different data block sizes, {\it i.e.}, $n_{\rm SH}=10^{7}$, $n_{\rm SH}=10^{9}$ and $n_{\rm SH}=10^{11}$. All the plots are cut at the value of $\eta_{\rm c,d}$ for which the resulting secret key rate is below the threshold value of $10^{-10}$. We note that, since in the case of the ESR the value of $\left\langle{N}\right\rangle|_{L=0}$ does not depend on any parameter to be optimized, the cases $S_1$ and $S_2$ only differ in the minimum $\eta_{\rm c,d}$ that still provides $K\geq{10^{-10}}$, which can be extracted from Fig.~\ref{fig:3}. In the case of the PQA, $\left\langle{N}\right\rangle|_{L=0}$ depends on the transmittance $t$ to be optimized, and therefore the cases $S_1$ and $S_2$ differ more from each other.}
	\label{fig:5}
\end{figure}
From Eq.~(\ref{mean_tx_signals}) we have that, for given $n_{\rm SH}$, the value of $\left\langle N \right\rangle$ increases when the success probability of the qubit amplifier decreases, for instance, due to channel and/or detection loss. Indeed, according to Eqs.~(\ref{21})-(\ref{22}) we find that $\left\langle N \right\rangle$ at $L=0$~km, which we will denote by $\left\langle{N}\right\rangle|_{L=0}$, is, in the case of an ESR, equal to 
\begin{equation}
\left\langle{N}\right\rangle|_{L=0}=\frac{2n_{\rm SH}}{\eta_{\rm c,d}^{2}}, 
\end{equation}
while in the case of a PQA it satisfies 
\begin{eqnarray}
\left\langle{N}\right\rangle|_{L=0}&=&\frac{n_{\rm SH}}{(1-t)\eta_{\rm c,d}^{4}[1-\eta_{\rm c,d}^{2}(1-t)]} \nonumber \\
&\approx& \frac{n_{\rm SH}}{(1-t)\eta_{\rm c,d}^{4}}. 
\end{eqnarray}
This is illustrated in Fig.~\ref{fig:5}, which shows $\left\langle{N}\right\rangle|_{L=0}$ as a function of $\eta_{\rm c,d}$ when $n_{\rm SH}=\{10^{7},10^{9},10^{11}\}$. From Fig.~\ref{fig:5} we find that the value of $\left\langle{N}\right\rangle|_{L=0}$ associated to the PQA presents a much steeper slope than that of the ESR architecture when $\eta_{\rm c,d}$ decreases. This is because the optimal transmittance $t$ of the PQA approaches $1$ in that regime.

In the scenario where $L>0$~km, the success probability of the qubit amplifier decreases exponentially with the channel loss. In particular, we find that the value of $\left\langle{N}\right\rangle|_{L\geq0}$ in this case is given by
\begin{eqnarray}\label{N_loss_ESR}
\left\langle{N}\right\rangle|_{L\geq0}&=&\frac{2n_{\rm SH}}{\eta_{\rm{c,d}}^2}\bigg\{(1-4p_{\rm d})\eta_{\rm ch}{\eta_{\rm c,d}}^{2} \nonumber \\
&+&4{p_{\rm d}}\left[1+\eta_{\rm ch}(1-2{\eta_{\rm{c,d}}}^{2})\right]\bigg\}^{-1},
\end{eqnarray}
for the ESR architecture, and
\begin{eqnarray}\label{N_loss_PQA}
\left\langle{N}\right\rangle|_{L\geq0}&=&\frac{n_{\rm SH}}{\eta_{\rm c,d}^{2}{[1-\eta_{\rm c,d}^{2}(1-t)]}}\bigg\{(1-10p_{\rm d})(1-t)\nonumber \\
&\times&\eta_{\rm ch}\eta_{\rm c,d}^{2}+4p_{\rm d}(1-t+\eta_{\rm ch}/2)\bigg\}^{-1}, \nonumber \\
\end{eqnarray}
for the PQA. In these two equations, for simplicity, the success probability $P_{\rm SH}$ is computed to the first order in $p_{\rm d}$.
\begin{table}
	\centering
	\begin{tabular}{|c|c|c|c|c|c|}
		\hline
		\quad $S_{1}\hspace{.2cm}$ & $\eta_{\rm c,d}$ & $n_{\rm SH}$ & $\Lambda_{\rm cutoff}$ & $K_{\rm cutoff}$ & $\left\langle{N}\right\rangle$ \\ \hline
		ESR & $100\%$ & $10^{7}$ & $48$ dB & $1.3\times{10^{-7}}$ & $1.2\times{10^{12}}$ \\ \hline
		ESR & $96.5\%$ & $10^{11}$ & $39$ dB & $4.2\times{10^{-8}}$ & $1.8\times{10^{15}}$ \\ \hline
		PQA & $100\%$ & $10^{7}$ & $48$ dB & $6.3\times{10^{-8}}$ & $2.5\times{10^{12}}$ \\ \hline
		PQA & $96.5\%$ & $10^{11}$ & $37$ dB & $2.7\times{10^{-9}}$ & $3.2\times{10^{16}}$ \\ \hline
	\end{tabular}
    \begin{tabular}{|c|c|c|c|c|c|}
	    \hline
	    \quad $S_{2}\hspace{.2cm}$ & $\eta_{\rm c,d}$ & $n_{\rm SH}$ & $\Lambda_{\rm cutoff}$ & $K_{\rm cutoff}$ & $\left\langle{N}\right\rangle$ \\ \hline
	    ESR & $100\%$ & $10^{7}$ & $44$ dB & $2.1\times{10^{-7}}$ & $5.0\times{10^{11}}$ \\ \hline
	    ESR & $96.5\%$ & $10^{11}$ & $35$ dB & $9.8\times{10^{-8}}$ & $7.2\times{10^{14}}$ \\ \hline
	    PQA & $100\%$ & $10^{7}$ & $44$ dB & $1.0\times{10^{-7}}$ & $9.9\times{10^{11}}$ \\ \hline
	    PQA & $96.5\%$ & $10^{11}$ & $28$ dB & $2.4\times{10^{-9}}$ & $4.0\times{10^{15}}$ \\ \hline
    \end{tabular}
	\caption{Average number of signals, $\left\langle{N}\right\rangle$, that Alice needs to send Bob to collect a data block size equal to $n_{\rm SH}$, when using ideal photon sources. The dark count rate of the photodetectors is set to $p_{\rm d}=10^{-7}$, and the detection and coupling efficiency is $\eta_{\rm c,d}$. For each combination $(\eta_{\rm c,d},n_{\rm SH})$, the considered value of $\Lambda$ is approximately equal to the cutoff loss for which the secret key rate starts dropping down to zero in Fig.~\ref{fig:4}. Also, the values of $\eta_{\rm c,d}$ and $n_{\rm SH}$ correspond to the cases illustrated in Fig.~\ref{fig:4}, and both sets of security settings, $S_{1}$ and $S_{2}$, are considered.}
	\label{table:4}
\end{table}
Table~\ref{table:4} provides the value of $\left\langle{N}\right\rangle|_{L\geq0}$ for a couple of detection and coupling efficiencies $\eta_{\rm c,d}$ and data block sizes $n_{\rm SH}$. This table shows that the average number of signals that has to be transmitted increases significantly with $\Lambda$, both for the ESR and the PQA. 

Indeed, if one considers, for example, that the clock rate of the system is, say, $10$ GHz, we find that when $\eta_{\rm c,d}=96.5\%$ it would take about $2.1$ ($37.0$) days to establish a secret key of length $7.56\times{10^{7}}$ ($8.64\times{10^{7}}$) bits---out of a block size $n_{\rm SH}=10^{11}$---over a channel loss of 39 dB (37 dB) when using the ESR (PQA) architecture and the security settings given by $S_1$. Of course, the result improves when $\eta_{\rm c,d}$ increases. For instance, if $\eta_{\rm c,d}=100\%$ then it would take of the order of $120$ ($250$) seconds to establish a secret key of length $1.56\times{10^{5}}$ ($1.58\times{10^{5}}$)---out of a block size $n_{\rm SH}=10^{7}$---over 48 dB when using the ESR (PQA) architecture and the security settings given by $S_1$.

\subsection{PDC sources}\label{pdc_sources_main}

In this subsection we consider now the more practical case where Alice and Bob use PDC sources instead of ideal sources. 

That is, here we suppose that the quantum state, $\rho_{ab}$, emitted by Alice's entanglement source in Fig.~\ref{fig:2} is given by Eq.~(\ref{1}) with the statistics $p_n$ given by Eq.~(\ref{3}). The quantum state $\rho_{bc}$ generated by the entanglement source within the qubit amplifier depends on the architecture considered. In the case of a PQA, we will assume that the single-photon states, $\rho^{\rm h}_{\rm single}$ and $\rho^{\rm v}_{\rm single}$, have the form given by Eq.~(\ref{5}) with the statistics $r_n$ given by Eq.~(\ref{weights2}), {\it i.e.}, they are generated with a triggered single-photon PDC source in combination with PNR detectors. In the case of a ESR architecture, we will asume that the state $\rho_{bc}$ is directly generated with a PDC source, {\it i.e.}, it has the form given by Eq.~(\ref{1}) with $p_n$ given by Eq.~(\ref{3}). 

To simplify the numerics, in our simulations below we consider a contribution of up to three photon pairs per source. That is, we set $p_{n}=0$ in Eq.~(\ref{1}) for all $n\geq{4}$ and we choose $p_{3}=1-p_{0}-p_{1}-p_{2}$. Likewise, we do the same with the statistics $r_n$ in Eq.~(\ref{5}). This is a reasonable approximation when the optimal intensities of the light sources are sufficiently small, which, indeed, is what we expect and observe in our simulations. A full mode analysis for this scenario is included in Appendix~\ref{honest}, which in principle could be use to evaluate the contribution of up to any desired number of photon pairs per source.

\subsubsection{No channel loss}

Like in the case of ideal sources, we start our analysis by evaluating the critical lines corresponding to the security setting sets $S_{1}$ and $S_{2}$ when the channel loss and the dark count rate of the detectors is set to zero. 

\begin{figure}
	\includegraphics[width=10.5cm,height=12cm]{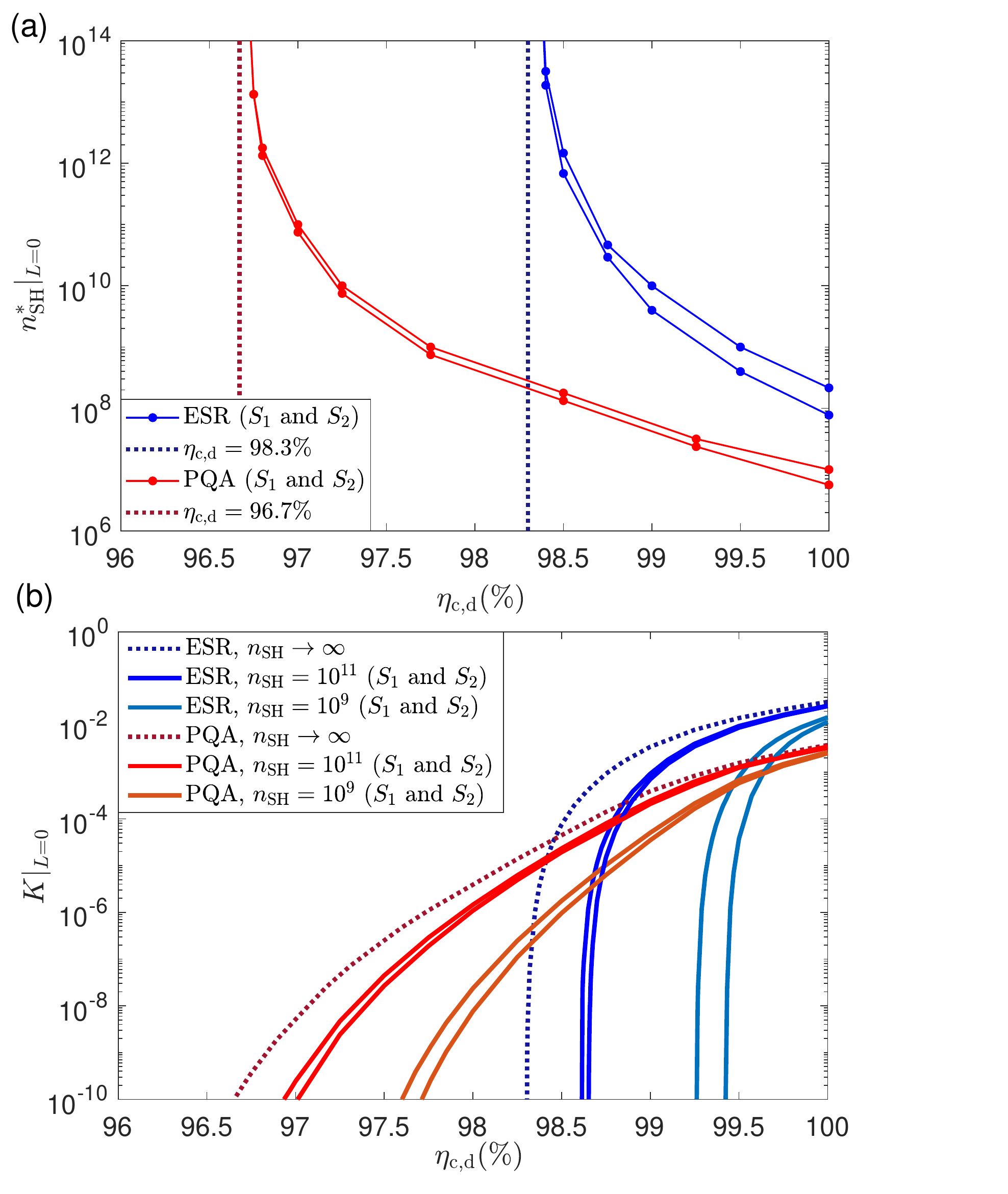}
	\caption{Performance evaluation of DIQKD with PDC sources. Bluish (reddish) lines are used for the ESR (PQA) architecture. (a) Minimum value of the detection and coupling efficiency, $\eta_{\rm c,d}$, and minimum value of the data block size, $n_{\rm SH}$, required to obtain a zero-distance secret key rate $K|_{L=0}\geq{10^{-10}}$. Both sets of security requirements, $S_{1}$ and $S_{1}$, are compared for each qubit amplifier. Any combination of parameters $\eta_{\rm c,d}$ and $n_{\rm SH}$ must be above the lower (upper) lines to achieve a secret key rate above the threshold value with the security requirements given by the sets $S_1$ ($S_2$). The dotted blue (red) vertical line indicates the (asymptotic) minimum efficiency, $\eta_{\rm c,d}\approx 98.3\%$ ($\eta_{\rm c,d}\approx 96.7\%$), which is the smallest detection efficiency that delivers a zero-distance asymptotic secret key rate $K_\infty|_{L=0}\geq{10^{-10}}$ when using the ESR (PQA) architecture. (b) Zero-distance secret key rate, $K|_{L=0}$, as a function of $\eta_{\rm c,d}$ for various values of the data block size $n_{\rm SH}$. For each qubit amplifier, three different block sizes are considered: $n_{\rm SH}\rightarrow\infty$, $n_{\rm SH}=10^{11}$ and $n_{\rm SH}=10^{9}$. The finite secret key rates appear in pairs of solid lines, one for the security set $S_1$ (upper line) and another one for the security set $S_2$ (lower line). The asymptotic secret key rates corresponding to $n_{\rm SH}\rightarrow\infty$ are illustrated with dotted lines. }
	\label{fig:6}
\end{figure}
The results are illustrated in Fig.~\ref{fig:6}(a), which shows the minimum data block size, $n^*_{\rm SH}|_{L=0}$, and the minimum detection and coupling efficiency, $\eta_{\rm c,d}$, that are needed to obtain $K|_{L=0}\geq{10^{-10}}$. By definition, any combination of $\eta_{\rm c,d}$ and $n_{\rm SH}$ lying below the critical lines illustrated in the figure leads to a minuscule (if not zero) secret key rate below $10^{-10}$. The blue (red) solid lines correspond to the ESR (PQA) architecture, and in each case the lower (upper) line uses the security requirements $S_1$ ($S_2$). Fig.~\ref{fig:6}(a) shows clearly that the ESR architecture demands much larger data block sizes and detection and coupling efficiencies than the PQA architecture in this case. This can be understood if one compares $P_{\rm SH}$ and $K|_{\rm SH}$ separately for both qubit amplifiers: while the ESR has a considerably larger success probability in the relevant efficiency regime, its conditional secret key rate is lower. This suggests that the PQA performs better at filtering genuine entanglement, and thus it is more robust to efficiency decrease and statistical fluctuations (partly due to the availability of an extra tunable parameter $t$). Also, as already observed in the case of ideal sources, we note that using the security set $S_{2}$ (instead of that given by $S_{1}$) does not affect much the minimum data block size, despite the significant difference that exists between both sets in terms of security requirements (see Table~\ref{table:3}). 

The dotted blue (red) vertical line shown in Fig.~\ref{fig:6}(a) corresponds to the (asymptotic) minimum efficiency, $\eta_{\rm c,d}\approx{98.3\%}$ ($\eta_{\rm c,d}\approx{96.7\%}$), that is necessary to obtain a secret key rate above the threshold value with the ESR-based qubit amplifier (PQA) when the block size $n_{\rm SH}$ tends to infinity. As expected, these values are much higher than those required for the case of ideal sources, specially for the ESR. These results are also in accordance with the results reported in~\cite{Curty}, as that work also considers a similar setup and device models like here, though the analysis of~\cite{Curty} is restricted to the asymptotic key rate scenario. From Fig.~\ref{fig:6}(a) we observe that when $n_{\rm SH}$ decreases, the minimum value of $\eta_{\rm c,d}$ increases even further as expected. For example, if $n_{\rm SH}=10^{11}$ and we focus on, say, the weaker set of security requirements, $S_{1}$, we find that the minimum values of $\eta_{\rm c,d}$ are about $98.6\%$ and $97\%$ for the ESR and PQA, respectively.

Fig.~\ref{fig:6}(b) illustrates the zero-distance secret key rate, $ K|_{L=0}$, as a function of $\eta_{\rm c,d}$ for varios values of the block size $n_{\rm SH}$. This figure shows that, in the absence of channel loss, an ESR-based qubit amplifier outperforms a PQA in the regime of very high detection and coupling efficiencies, while in principle the PQA can tolerate slightly lower values of $\eta_{\rm c,d}$, as we have already seen in Fig.~\ref{fig:6}(a). In any case, the minimum value of $\eta_{\rm c,d}$, specially for the ESR, seems to be already probably too high to have practical relevance. The main reasons for this behaviour are twofold. First, the vacuum and multiple photon pairs emitted by the light sources significantly reduce the probability to have a successful heralding event in the amplifier. And, second, multiple photon pairs are also responsable for spurious heralding events which increase (decrease) $Q|_{\rm SH}$ ($\omega|_{\rm SH}$) and thus decrease the resulting $R$. The higher $\eta_{\rm c,d}$ is, the higher is the number of multiple photon pairs which can be filtered out by Alice and Bob's PNR detectors and, therefore, the better the resulting performance. As already mentioned above, in our simulations we optimize over the intensities of the different light sources and, as expected, the optimal intensities decrease when $\eta_{\rm c,d}$ decreases in order to reduce the number of multiple photon pairs generated.

\subsubsection{Channel loss}

Next, we consider the effect of the channel loss and, as in Sec.~\ref{channel_loss_ideal}, we set $p_{\rm d}=10^{-7}$.

\begin{figure}
	\includegraphics[width=10.5cm,height=12cm]{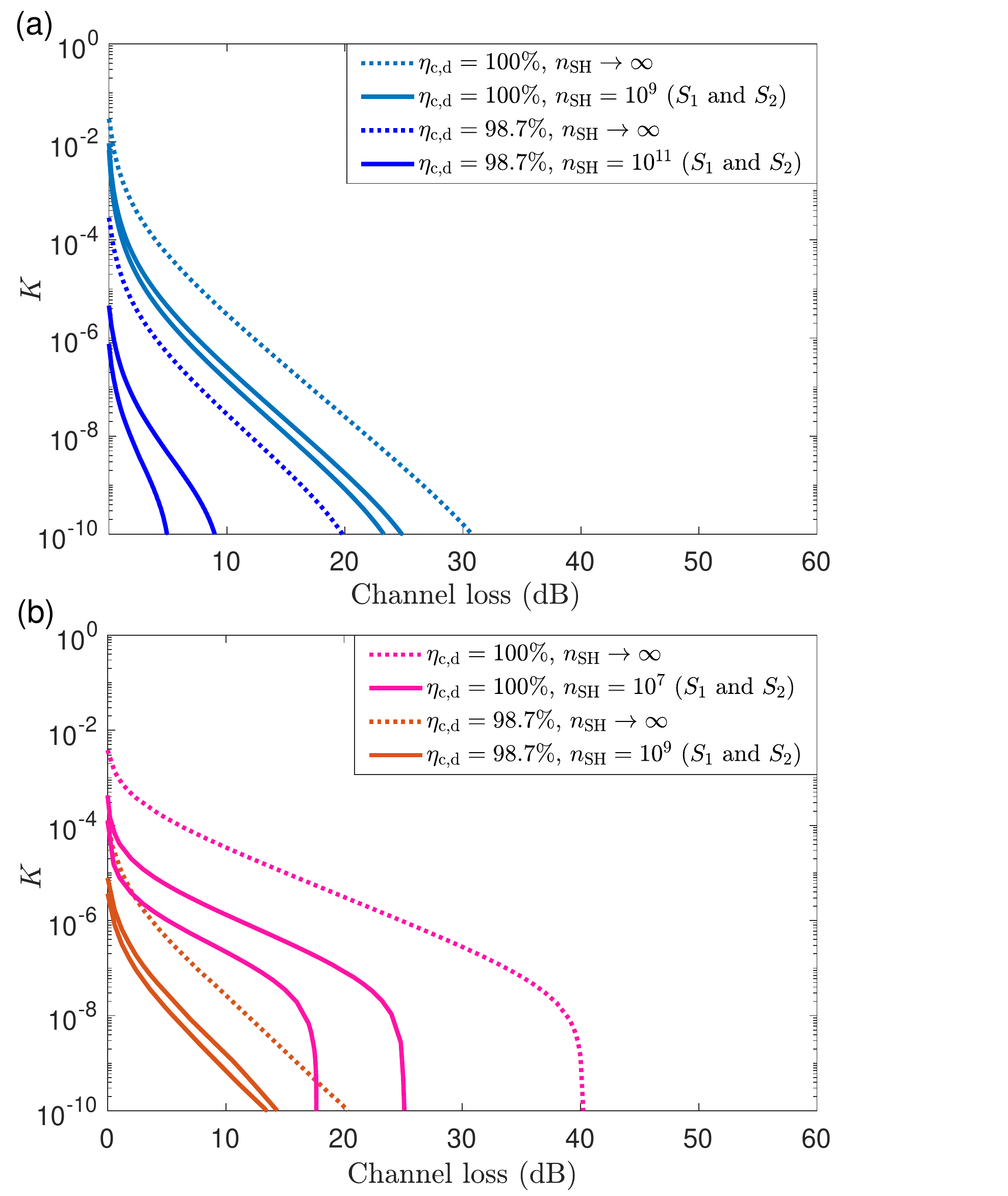}
	\caption{Secret key rate $K$ as a function of the overall channel loss $\Lambda$ measured in dB for the case of PDC sources. Figure (a) corresponds to the ESR architecture and figure (b) to the PQA architecture. For each qubit amplifier, we use two different detection and coupling efficiencies, $\eta_{\rm c,d}=100\%$ and $\eta_{\rm c,d}=98.7\%$, each of them tagged with a different color. For each value of the efficiency, we plot the asymptotic secret key rate $K_\infty$ (dotted line), together with two finite-key rates for different values of $n_{\rm SH}$ (solid lines). Each finite-key rate is plotted twice, one line corresponds to the security settings $S_1$ and the other line to the security settings $S_2$. We take $n_{\rm SH}\in\{10^{9}, 10^{11}\}$ for the ESR-based qubit amplifier and $n_{\rm SH}\in\{10^{7}, 10^{9}\}$ for the PQA. By increasing the value of $\eta_{\rm c,d}$ and/or $n_{\rm SH}$ the finite-key rates approach those of the optimal scenario, which corresponds to $K_\infty$ assuming $\eta_{\rm c,d}=100\%$.} 
	\label{fig:7}
\end{figure}

The results are illustrated in~Fig.~\ref{fig:7}, which shows the secret key rate $K$ as a function of $\Lambda$ for various values of  $\eta_{\rm c,d}$ and $n_{\rm SH}$. More precisely, Figs.~\ref{fig:7}(a) and~\ref{fig:7}(b) are respectively devoted to the ESR-architecture and to the PQA-architecture. For each case, we assume two values of $\eta_{\rm c,d}$: the ideal one $\eta_{\rm c,d}=100\%$, and another one close to the threshold value of the ESR, say, $\eta_{\rm c,d}=98.7\%$. Regarding the block size $n_{\rm SH}$, we set it to a value near the critical line in each case. That is, we consider the pairs $(\eta_{\rm c,d},n_{\rm SH})\in\{(100\%,10^{9}), (98.7\%,10^{11})\}$ for the ESR, and $(\eta_{\rm c,d},n_{\rm SH})\in\{(100\%,10^{7}), (98.7\%,10^{9})\}$ for the PQA. Also, Fig.~\ref{fig:7} includes the results for the sets $S_1$ and $S_2$ of security requirements, as well as the asymptotic curves corresponding to $K_{\infty}$. These latter curves serve as upper bounds to the attainable finite-key rates for each $\eta_{\rm c,d}$.

Generally speaking, we observe that, in the presence of channel loss, the performance of DIQKD with PDC sources is again significantly worse than that achievable with ideal sources. The reason for this, as already explained above, is due to the presence of vacuum and multiple photon pairs, which require that the intensities of the sources are quite low to palliate their negative effect. Indeed, in the case of an ESR, we find that even when $n_{\rm SH}=10^{11}$, the detection and coupling efficiency is as high as $98.7\%$, and the weaker set of security requirements is considered ($S_{1}$), the resulting secret key rate is already as low as $K\approx{10^{-10}}$ for a channel loss of only $9$~dB. Similarly, for the same value of $\eta_{\rm c,d}$ and a block size $n_{\rm SH}=10^{9}$, we find that the PQA can only tolerate about $14$ dB channel loss. In this regard, we remark that setting $n_{\rm SH}$ to a different value for each qubit amplifier does not necessarily lead to an unfair comparison between them, as the average number of signals, $\left\langle{N}\right\rangle$, required to gather a particular block size is different in both cases. This is discussed in more detail in the following subsection. 

Finally, from Fig.~\ref{fig:7} we observe again that increasing the security requirements from $S_{1}$ to $S_{2}$ does not affect the system performance significantly. Also, we note that if Bob did not use a qubit amplifier, the maximum possible value of $\Lambda$ in this scenario would be as low as $\Lambda\lessapprox 0.4$ dB, even if one sets $\eta_{\rm c,d}=100\%$ and $n_{\rm SH}\to\infty$. See Appendix~\ref{unassisted_scenario} for further details.

\subsubsection{Time constraints}

To conclude this part, we now consider the duration of a DIQKD session with PDC sources. As we show below, in this case the time requirements are much more demanding than in the scenario with ideal sources. As discussed above, this happens because the optimal intensity of Alice's source, $\rho_{ab}$, is quite low in the high loss regime. As a result, the average number, $\left\langle{N}\right\rangle$, of signals that Alice has to send Bob to achieve $n_{\rm SH}$ successful heralding events turns out to be quite high. 

\begin{figure}
	\includegraphics[width=10cm,height=6.5cm]{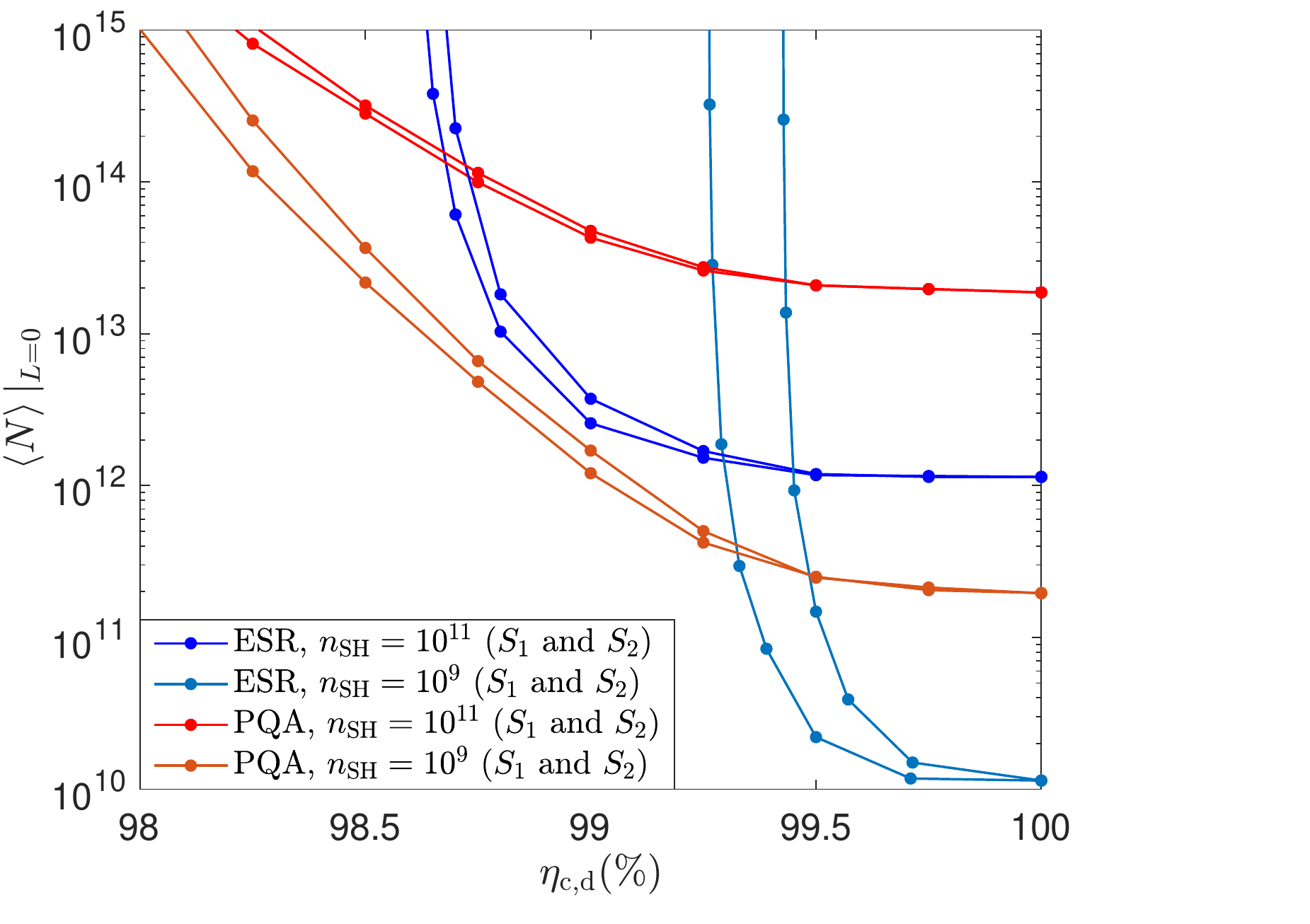}
	\caption{Average number of transmitted signals, $\left\langle{N}\right\rangle|_{L=0}$, that Alice needs to send Bob to collect a data block size equal to $n_{\rm SH}$ when using PDC sources, as a function of the detection and coupling efficiency $\eta_{\rm c,d}$ at $L=0$ km. The free experimental and security parameters are set to the values that optimize the secret key rate given by Fig.~\ref{fig:6}(b). The figure considers two different data block sizes, {\it i.e.}, $n_{\rm SH}=10^{9}$ and $n_{\rm SH}=10^{11}$. All the plots are cut at the value of $\eta_{\rm c,d}$ for which the resulting secret key rate is below the threshold value of $10^{-10}$. Also, we consider both sets of security requirements ($S_{1}$ and $S_{2}$) and, like in Fig.~\ref{fig:5}, here we disregard dark counts because their effect at $L=0$ km is negligible.}
	\label{fig:8}
\end{figure}
This is illustrated in Fig.~\ref{fig:8}, which shows the value of $\left\langle{N}\right\rangle$ as a function of $\eta_{\rm c,d}$ at $L=0$ km. As in Fig.~\ref{fig:5}, dark counts are disregarded here because their effect is negligible. Also, we consider both sets of security requirements ($S_{1}$ and $S_{2}$) and two different block sizes for each qubit amplifier: $n_{\rm SH}=10^{11}$ and $n_{\rm SH}=10^{9}$. We note that the case $\left\langle{N}\right\rangle|_{L=0}\geq10^{15}$ leads to DIQKD sessions that would take longer than one day even with $10$ GHz PDC sources. Actually, if the conditions $\left\langle{N}\right\rangle|_{L=0}<10^{15}$ (see Fig.~\ref{fig:8}) and $K\geq{10^{-10}}$ (see Fig.~\ref{fig:6}) are imposed, one finds that the detection efficiency must satisfy $\eta_{\rm c,d}\gtrapprox{98\%}$, irrespectively of the qubit amplifier and the data block size. This means that the examples shown in Fig.~\ref{fig:6} for the PQA where $K\geq{10^{-10}}$ is possible for $\eta_{\rm c,d}<98\%$, are probably not too practical, as they require too long DIQKD sessions.
\begin{table}
	\centering
	\begin{tabular}{|c|c|c|c|c|c|}
		\hline
		\quad $S_{1}\hspace{.2cm}$ & $\eta_{\rm c,d}$ & $n_{\rm SH}$ & $\Lambda$ & $K$ & $\left\langle{N}\right\rangle$ \\ \hline
		ESR & $100\%$ & $10^{9}$ & $13$ dB & $5.7\times{10^{-8}}$ & $10^{15}$  \\ \hline
		ESR & $98.7\%$ & $10^{11}$ & $1$ dB & $5.9\times{10^{-7}}$ & $10^{15}$  \\ \hline
		PQA & $100\%$ & $10^{7}$ & $24$ dB & $1.1\times{10^{-8}}$ &  $1.7\times{10^{13}}$   \\ \hline
		PQA & $98.7\%$ & $10^{9}$ & $4$ dB & $5.5\times{10^{-8}}$ & $10^{15}$   \\ \hline
	\end{tabular}
    \begin{tabular}{|c|c|c|c|c|c|}
	    \hline
	    \quad $S_{2}\hspace{.2cm}$ & $\eta_{\rm c,d}$ & $n_{\rm SH}$ & $\Lambda$ & $K$ & $\left\langle{N}\right\rangle$ \\ \hline
	    ESR & $100\%$ & $10^{9}$ & $12$ dB & $4.9\times{10^{-8}}$ & $10^{15}$  \\ \hline
	    ESR & $98.7\%$ & $10^{11}$ & $0$ dB & $2.9\times{10^{-7}}$ & $10^{15}$  \\ \hline
	    PQA & $100\%$ & $10^{7}$ & $17$ dB & $6.9\times{10^{-9}}$ & $8.6\times{10^{12}}$   \\ \hline
	    PQA & $98.7\%$ & $10^{9}$ & $3$ dB & $5.7\times{10^{-8}}$ & $10^{15}$   \\ \hline
    \end{tabular}
	\caption{Maximum value of the channel loss, $\Lambda$, before either $\left\langle{N}\right\rangle\geq10^{15}$ or the secret key rate $K$ starts dropping down to zero, depending on the pair $(\eta_{\rm c,d},n_{\rm SH})$ and on the qubit amplifier. The considered detection and coupling efficiencies, as well as the data block sizes, correspond to the finite-key rates illustrated in Fig.~\ref{fig:7} with security settings $S_{1}$ and $S_{2}$. As shown by the table, in the case of the PQA with $\eta_{\rm c,d}=100\%$ and $n_{\rm SH}=10^{7}$, the secret key rate drops down to zero before $\left\langle{N}\right\rangle$ exceeds $10^{15}$ signals.
    To be precise, the cutoff for $S_{1}$ ($S_{2}$) roughly occurs at $\Lambda=24$ dB ($17$ dB), and the corresponding $\left\langle{N}\right\rangle$ is still $1.7\times{10^{13}}$ ($8.6\times{10^{12}}$).}
	\label{table:5}
\end{table}

Obviously, if one considers the case of nonzero channel loss, the time constraints become sharper. This is illustrated in Table~\ref{table:5}, which shows the maximum value of the channel loss, $\Lambda$, for which $\left\langle{N}\right\rangle\leq10^{15}$ for various pairs $(\eta_{\rm c,d},n_{\rm SH})$ previously evaluated in Fig.~\ref{fig:7}. For instance, when $\eta_{\rm c,d}=98.7\%$, and assuming the ESR (PQA) architecture, the maximum $\Lambda$ decreases from $9$ ($14$) dB as shown in Fig.~\ref{fig:7} to roughly $1$ ($4$) dB for $n_{\rm SH}=10^{11}$ ($10^{9}$).

Indeed, in Appendix~\ref{two_QA} we show that, due to similar time constraints, locating the entanglement source $\rho_{ab}$ in the middle of the channel between Alice and Bob, and furnishing both Alice and Bob with a qubit amplifier, does not seem to improve the performance that can be obtained when Alice holds the source and only Bob holds a qubit amplifier, at least in the case of PDC sources.

\subsection{Generic sources}\label{general}

\begin{figure*}
	\includegraphics[width=\textwidth,height=9cm]{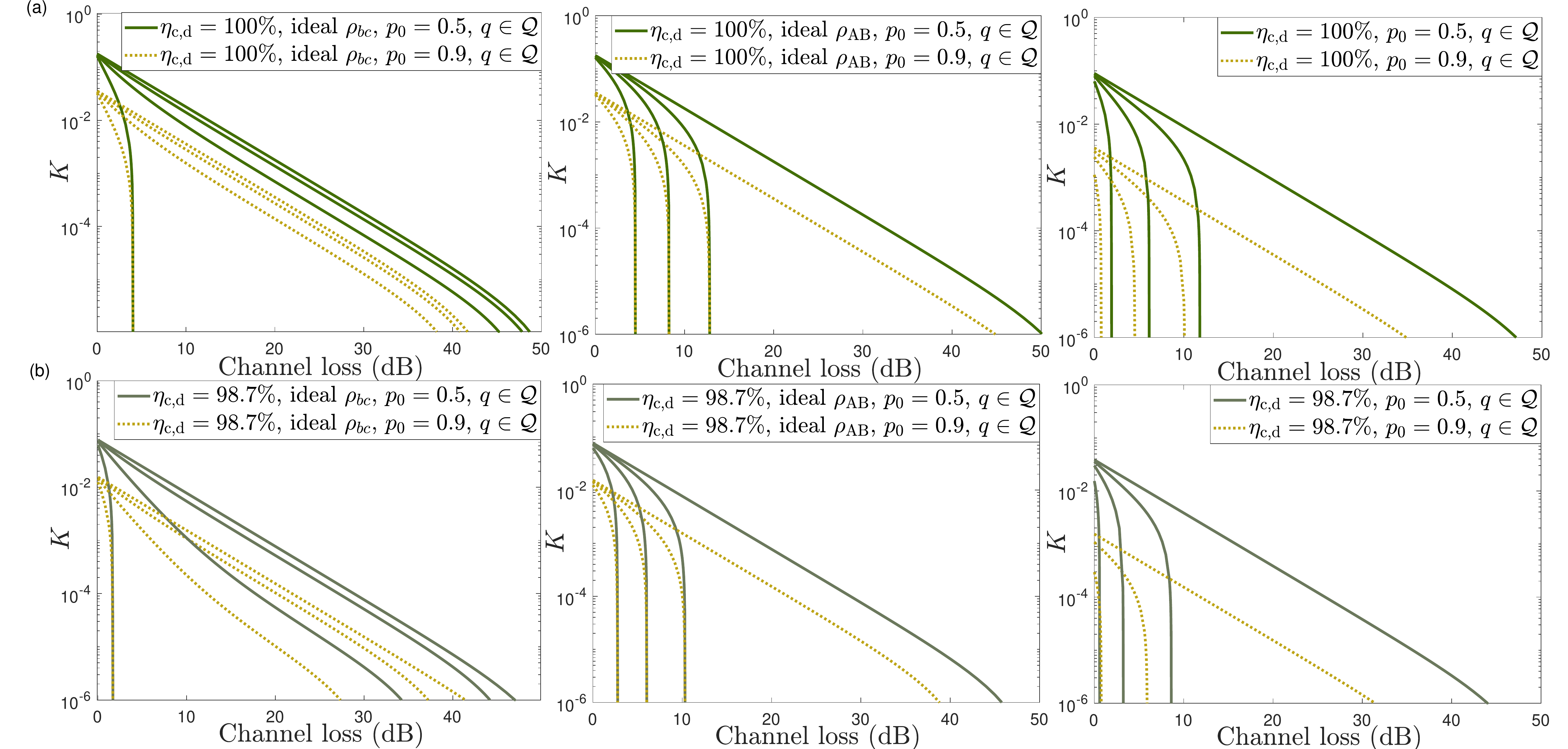} 
	\caption{Secret key rate $K$ as a function of the overall channel loss $\Lambda$ measured in dB for generic photonic sources and assuming the ESR architecture. Figure (a) considers a detection and coupling efficiency $\eta_{\rm c,d}=100\%$ and figure (b) considers $\eta_{\rm c,d}=98.7\%$. Each figure evaluates three different cases. The first (second) case, assumes that the entanglement source $\rho_{bc}$ ($\rho_{ab}$) is an ideal entanglement source, while $\rho_{ab}$ ($\rho_{bc}$) is characterized by the parameters $p_0$ and $q=p_2/p_1$. The third case considers that both $\rho_{bc}$ and $\rho_{ab}$ are characterized by the parameters $p_0$ and $q=p_2/p_1$. All figures consider two possible values for $p_0$, {\it i.e.}, $p_{0}=0.5$ (solid lines) and $p_{0}=0.9$ (dotted lines), and four different values for the parameter $q\in\mathcal{Q}=\{0,10^{-2},10^{-1.5},10^{-1}\}$. Also, for concreteness, in all cases we set $n_{\rm SH}=10^{9}$ and choose the security settings $S_{1}$.}
	\label{fig:9}
\end{figure*}
Finally, in this section we further investigate the effect that vacuum pulses and multiple photon pairs, generated by practical entanglement sources, has on the performance of DIQKD. For concreteness, we focus on the ESR architecture and we consider entanglement sources generating signals of the form given by Eqs.~(\ref{1}) and (\ref{2}). Also, for simplicity, we set $p_{n}=0$ for $n\geq{3}$ with $p_2=1-p_0-p_1$, the underlying assumption being that the effect of multiple photon pairs is properly encompassed by the effect of double photon pairs, which is supported by our numerical simulations. In doing so, we can characterize the photon-number statistics of the entanglement sources by means of only two parameters: the probability $p_{0}$ of emitting vacuum, and the ratio $q={p_{2}/p_{1}}$ between the probability of emitting a double photon pair and that of emitting a single photon pair. 

Of course, if one considers a practical entanglement source, the photon-number statistics $p_{n}$ cannot be controlled separately, but they typically depend on an intensity parameter. For instance, in the case of PDC sources, we have that $p_{n}$ is fixed for all $n$ once we set the value of the probability $p_{0}$ (or, equivalently, the value of $\lambda$). In this type of sources, by using Eq.~(\ref{3}), we have that $q={p_{2}/p_{1}}=(1-p_0-p_1)/p_1=(p_{0}^{-1/2}-1)(p_{0}^{-1/2}+2)/2={}q_{\rm PDC}$. The case of ideal sources, on the other hand, corresponds to $p_{0}=p_2=0$ and thus $q=0$. Since both scenarios have already been evaluated above, and the case $q>q_{\rm PDC}$ delivers worse results than those evaluated in Sec.~\ref{pdc_sources_main}, below we consider combinations of $(p_{0},q)$ that  satisfy $0\leq{}q<q_{\rm PDC}$ for the corresponding $p_{0}>0$. In doing so, we investigate an intermediate scenario between the photon number statistics of ideal sources and those of PDC sources. 

We remark, however, that Appendix~\ref{honest} includes a full mode analysis, both for the ESR and PQA architectures, which allows the evaluation of any desired photon number distribution for the different light sources, including the contribution of up to any wanted number of photon pairs per source. That is, such analysis could be used to investigate any generic source of the form given by Eqs.~(\ref{1}) and (\ref{2}).

The results for the simplified scenario discussed above are illustrated in Fig.~\ref{fig:9}, which shows the secret key rate as a function of the channel loss $\Lambda$ for $n_{\rm SH}=10^{9}$ and  the security requirements given by $S_{1}$. To simplify the comparison between this case and the one based on PDC sources, we use the values of $\eta_{\rm c,d}$ employed in Fig.~\ref{fig:7}, {\it i.e.}, $\eta_{\rm c,d}\in\{100\%, 98.7\%\}$. Moreover, for each value of $\eta_{\rm c,d}$, we plot three different cases. The first (second) case, assumes that the entanglement source $\rho_{bc}$ ($\rho_{ab}$) is an ideal entanglement source, while $\rho_{ab}$ ($\rho_{bc}$) is characterized by the parameters $(p_{0},q)$. The third case considers that both $\rho_{bc}$ and $\rho_{ab}$ are characterized by the parameters $(p_{0},q)$, which, for simplicity, we assume are the same for both sources, though, in general, the optimal intensity for each source will depend of the value of the channel loss. In each figure, we evaluate two possible values for $p_{0}$: $p_{0}=0.5$ (solid lines) and $p_{0}=0.9$ (dotted lines). Also, we consider four different values for the parameter $q\in\mathcal{Q}=\{0,10^{-2},10^{-1.5},10^{-1}\}$.

In all the plots within Fig.~\ref{fig:9}, if one compares the solid lines with the dotted lines, we observe that reducing the value of the probability $p_0$ for a fixed value of $q$ basically leads to a rigid increase of the secret key rate. This is so because vacuum signals rarely lead to false heralding flags in the qubit amplifier: if $\rho_{ab}$ ($\rho_{bc}$) emits a vacuum state, it is necessary that either $\rho_{bc}$ ($\rho_{ab}$) emits more than one photon pair or that at least one dark count takes place at the detectors within the qubit amplifier in order to have a (spurious) successful heralding event. As a consequence, to a good extent, $p_{0}$ affects mainly the pre-factor $P_{\rm SH}$, but not the conditional secret key rate, $K|_{\rm SH}$. The greater the value of $p_0$, the smaller the value of $P_{\rm SH}$, and thus the rigid decrease of the secret key rate.

On the other hand, for a fixed value of $p_{0}$, increasing $q$ significantly affects $K|_{\rm SH}$, so that multiple photon pairs are responsible for the changing slope of the secret key rate as well as for the position of the cutoff point where the secret key rate starts dropping down to zero, as shown in Fig.~\ref{fig:9}. This is so because multiple photon pairs lead to spurious heralding events that limit the utility of the qubit amplifier, and, as expected, this effect is amplified when the detection and coupling efficiency $\eta_{\rm c,d}$ decreases. In this regard, we also note that the performance of DIQKD seems to be more robust to the presence of multiple photon pairs in $\rho_{ab}$ than in $\rho_{bc}$. The reason goes as follows. Multi-photons arising from $\rho_{ab}$ need to undergo a lossy channel, but multi-photons from $\rho_{bc}$ do not. Therefore, the latter are more likely to trigger a spurious success at the qubit amplifier when the input from the channel is a vacuum signal. Actually, from Fig.~\ref{fig:9} we observe that the curves with an ideal source $\rho_{bc}$, and $\rho_{ab}$ characterized with $q=10^{-2}$ or $10^{-1.5}$ are relatively close to the curve corresponding to $q=0$. This suggests that the cutoff points of these curves (at the high loss regime) are probably mainly due to the dark counts of the detectors at the qubit amplifier, as in the case $q=0$, rather than due to the presence of multiple photon pairs in $\rho_{ab}$. On the contrary, all the curves with an ideal source $\rho_{ab}$, and $\rho_{bc}$ characterized with a nonzero $q$, show an early cutoff point induced by the multiple photon pairs in $\rho_{bc}$.

Furthermore, we note that the cutoff points match for $p_{0}=0.5$ and $p_{0}=0.9$ if they are caused by the presence of multiple photon pairs, but they do not match if they are caused by the dark counts of the detectors. This is so because, in the former case, the cutoff point is roughly determined by the double-to-single photon pair ratio (\textit{i.e.}, by the  parameter $q$), while in the latter case it is determined by the dark count to single photon pair ratio, which is different for each curve. Either way, Fig.~\ref{fig:9} suggests that the noise induced by multiple photon pairs generated by the sources, particularly those generated by the sources within the qubit amplifier, seems to be the major challenge to achieve long-distance DIQKD with the considered setup. 

\begin{figure}
	\includegraphics[width=10cm,height=6.5cm]{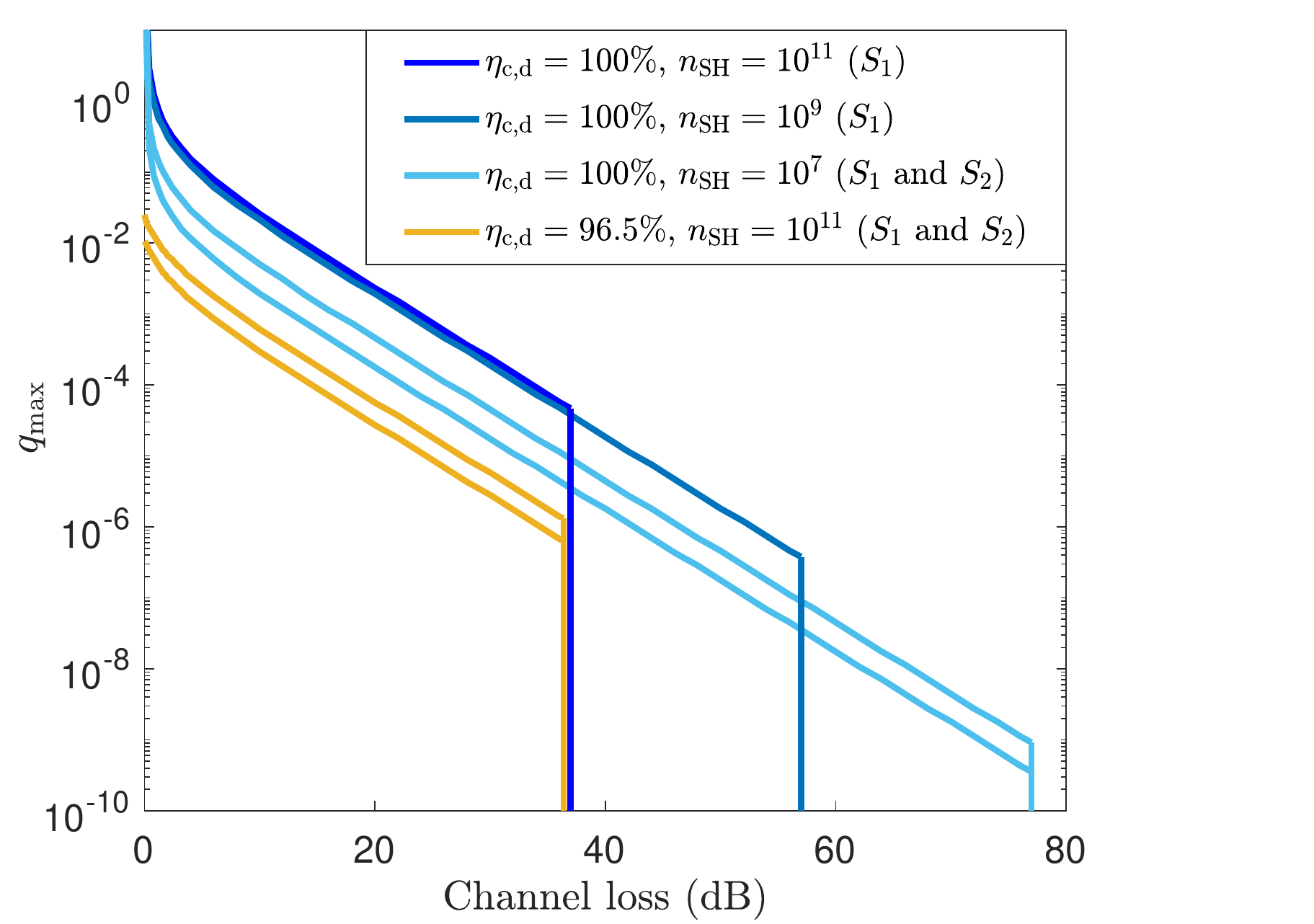}
	\caption{Upper bound on the maximum double-to-single photon pair ratio $q_{\rm max}$ of the entanglement source $\rho_{{bc}}$ required to achieve $K\geq{0}$ with $\left\langle{N}\right\rangle\leq10^{15}$, as a function of the channel loss $\Lambda$. Here, we set $p_{\rm d}=0$, so that the multiphotons generated in the qubit amplifier are the only source of noise in the system. The bluish lines use coupling and detection efficiency $\eta_{\rm c,d}=100\%$ and they include the block sizes $n_{\rm SH}=10^{11}$, $10^{9}$ and $10^{7}$, while the yellow lines use $\eta_{\rm c,d}=96.5\%$ and they only include the case $n_{\rm SH}=10^{11}$. This is so because $n_{\rm SH}=10^{9}$ and $n_{\rm SH}=10^{7}$ do not deliver a positive secret key rate for $\eta_{\rm c,d}=96.5\%$. Also, for each pair $(\eta_{\rm c,d},n_{\rm SH})$, the graphs corresponding to both sets of security settings, $S_{1}$ and $S_{2}$, are included whenever they are significantly different. Otherwise, we only plot that of $S_{1}$ for simplicity. The vertical cutoffs in the graphs indicate the points where $\left\langle{N}\right\rangle\approx{10^{15}}$. As expected, when $\eta_{\rm c,d}$ decreases the value of $q_{\rm max}$ decreases as well.}
	\label{fig:10}
\end{figure}

This effect is investigated further in Fig.~\ref{fig:10}, where we plot an upper bound on the maximum value of the parameter $q$, which we denote by $q_{\rm max}$, to achieve $K\geq{0}$ with $\left\langle{N}\right\rangle\leq10^{15}$, as a function of the channel loss $\Lambda$. For this, we assume that the source $\rho_{ab}$ is an ideal source and we parametrize the source $\rho_{bc}$ with the quantity $q$. Note that since here we use the condition that the secret key rate is strictly greater than zero, we can set the parameter $p_{0}$ corresponding to $\rho_{bc}$ to zero. This is so because setting $p_{0}>0$ simply translates into a rigid decrease of the secret key rate, thus not affecting the value of $q_{\rm max}$. That is, we define $q_{\rm max}=\min_{p_1}\left\{(1-p_{1})/p_{1}|K\geq{0}\right\}$. In addition, and in order to investigate the limitations imposed by the noise due to multiple photon pairs alone, we set $p_{\rm d}=0$. In this scenario, Fig.~\ref{fig:10} suggests that, irrespectively of the block size $n_{\rm SH}$, the value of the detector efficiency $\eta_{\rm c,d}$ and the security settings, the double-to-single photon pair ratio $q$ severely restricts the maximum distance that is achievable with DIQKD. Moreover, note that in a realistic situation with a non-ideal $\rho_{ab}$, $q_{\rm max}$ would be lower than the value shown in Fig.~\ref{fig:10}. The vertical cutoffs in the graphs indicate the points where $\left\langle{N}\right\rangle\approx{10^{15}}$, as this value is already probably too large for a QKD session today. Since $q_{max}$ is very small at the cutoff points, the corresponding values of $\Lambda$ are very close (indistinguishable to our numerical precision) to those of the case $q=0$, which are given by
\begin{equation}
\Lambda=150-10\log_{10}\left(\frac{2n_{\rm SH}}{\eta_{\rm c,d}^{4}}\right)
\end{equation}
for the different pairs $(\eta_{\rm c,d},n_{\rm SH})$. This expression is directly obtained from Eq.~(\ref{N_loss_ESR}) assuming $p_{\rm d}=0$.

As a final remark, note that one might achieve a source whose parameter $q<q_{\rm max}$ for a given distance by simply decreasing the intensity of the source. Indeed, this is the case, for example, of PDC sources, where one can reduce $\lambda$ and thus $q$ at the price of significantly increasing the probability $p_0$ of emitting vacuum. While this might provide a positive key rate according to Fig.~\ref{fig:10} (by assuming still that $\rho_{ab}$ is an ideal source), the resulting secret key rate might be probably too low to be practical because the probability of having a successful heralding event would be very low. The situation gets worse in the presence of dark counts.

\section{Conclusions}\label{Section6}

Device independence is a desirable feature for quantum key distribution (QKD) to ultimately defeat quantum hacking. However, it comes at a high price, in terms of achievable performance and required resources. Indeed, long distance device-independent QKD (DIQKD) requires the use of fair-sampling devices, like for instance qubit amplifiers, which can herald the arrival of a photon and thus decouple channel loss from the measurement settings selection. 

In this work, we have investigated all-photonic DIQKD assisted by two general types of qubit amplifiers---entanglement swapping relays and polarization qubit amplifiers---in the finite-key regime. In doing so, we have quantified some crucial experimental parameters that are essential to achieve DIQKD over practical distances and within a reasonable time frame of signal transmission. This includes, for example, the minimum value of the detection efficiency of the photodetectors and the quality of the entanglement light sources, in terms of their vacuum and multi-photon contributions. In this regard, we have shown that, even if perfect entanglement sources and photon-number-resolving detectors were available, the ability to achieve large enough violations of a loophole-free CHSH test, within a DIQKD session of a reasonable time duration, already imposes very strong restrictions on the minimum detection efficiency ($\gtrsim 96,5\%$), which further increases quickly with the length of the transmission link. Similarly, we have shown that multi-photon pulses emitted by practical entanglement sources have a severe effect on the performance of DIQKD assisted by qubit amplifiers, as multiple photon pairs lead to spurious heading successes that strongly decrease the conditional secret key rate.

Altogether, our results suggest that the possibility of implementing DIQKD over long distances is probably quite far-off, as it seems to require a significant improvement of our current experimental capabilities. 

\section*{Acknowledgements}

We thank Rotem Arnon-Friedman for very useful discussions related to the security analysis in~\cite{Rotem}. We thank the Spanish Ministry of Economy and Competitiveness (MINECO), the Fondo Europeo de Desarrollo Regional (FEDER) through grants TEC2014-54898-R and TEC2017-88243-R, and the European Union's Horizon 2020 research and innovation programme under the Marie Sk\l{}odowska-Curie grant agreement No 675662 (project QCALL) for financial support. VZ gratefully acknowledges support from a FPU scholarship from the Spanish Ministry of Education.

\appendix

\section{Honest implementations}\label{honest}
Here, we present the main calculations needed to reproduce our simulation results. For this, we use the mathematical models introduced in the main text. As already mentioned in Sec.~\ref{Section2}, we model channel, coupling and detection loss with a beamsplitter (BS), whose unitary transformation is of the form
\begin{align}\label{loss}
&a_{\rm h,v}^{\dagger}\xrightarrow{}\sqrt{\eta}c_{\rm h,v}^{\dagger}+\sqrt{1-\eta}d_{\rm h,v}^{\dagger}, \nonumber \\
&b_{\rm h,v}^{\dagger}\xrightarrow{}\sqrt{\eta}d_{\rm h,v}^{\dagger}-\sqrt{1-\eta}c_{\rm h,v}^{\dagger},
\end{align}
where the input mode $a$ is the quantum signal, the input mode $b$ is a vacuum state, the output mode $c$ is the optical fiber, and the output mode $d$ represents loss. In Eq.~(\ref{loss}), the subscript ``h, v'' indicates again horizontal and vertical polarization respectively. That is, the BS transformation given by Eq.~(\ref{loss}) applies to both polarizations. 

Similarly, polarization modulators are simply described by a rotation that transforms the input modes $a_{\rm h}^{\dagger}$ and $a_{\rm v}^{\dagger}$ as follows
\begin{align}\label{rotation}
&a_{\rm h}^{\dagger}\xrightarrow{}\cos{\theta}b_{\rm h}^{\dagger}+\sin{\theta}b_{\rm v}^{\dagger}, \nonumber \\
&a_{\rm v}^{\dagger}\xrightarrow{}\cos{\theta}b_{\rm v}^{\dagger}-\sin{\theta}b_{\rm h}^{\dagger},
\end{align}
where $b_{\rm h}^{\dagger}$ and $b_{\rm v}^{\dagger}$ denote the output modes. The case $\theta=\pi/4$ corresponds to the Hadamard transformation.

\subsection{Entanglement swapping relay}\label{ESR}
\begin{figure*}[!htbp]
	\includegraphics[width=\textwidth,height=12cm]{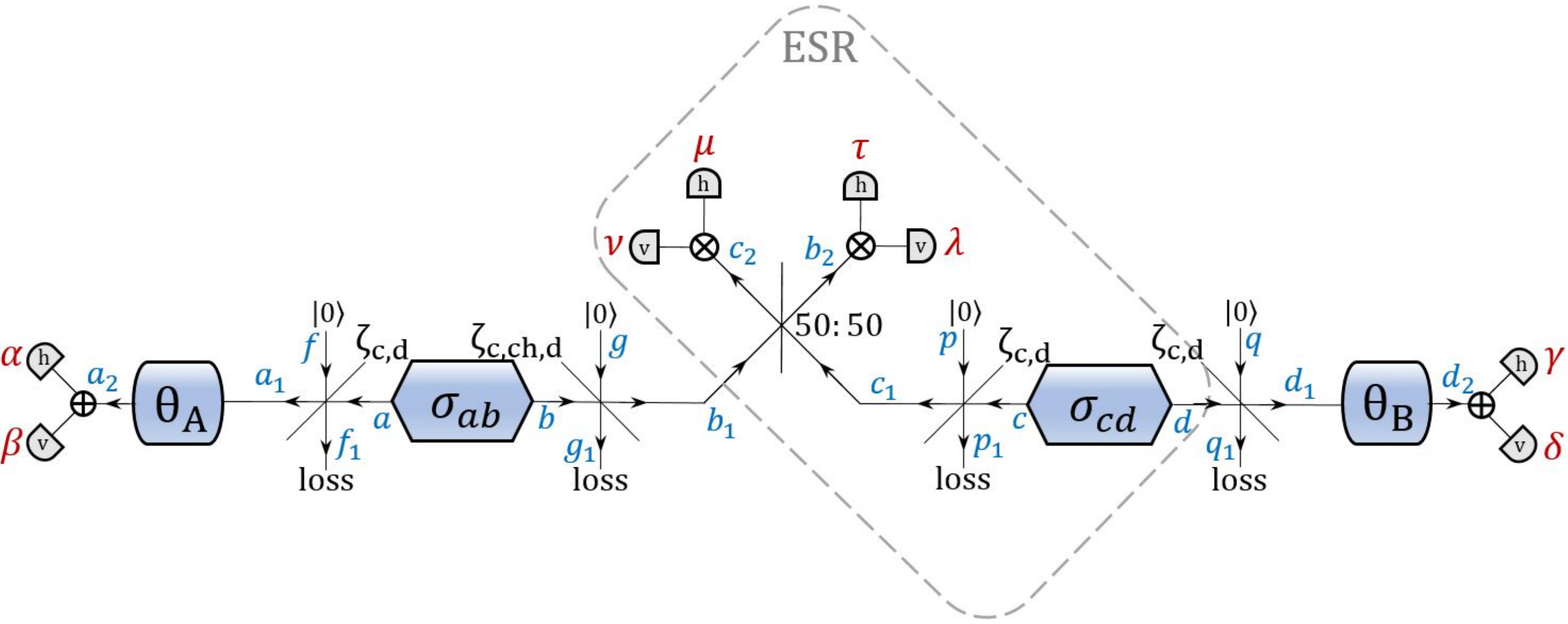} 
	\caption{Schematic of the ESR-based DIQKD setup matching the mathematical models presented in the main text: $\sigma_{ab}$ and $\sigma_{cd}$ stand for the entanglement sources, $\theta_{\rm A}$ ($\theta_{\rm B}$) denotes the rotation angle of Alice's (Bob's) measurement settings and $\zeta_{\rm c,d}$ and $\zeta_{\rm c,ch,d}$ tag the effective efficiency parameters, $\zeta_{\rm c,d}=\eta_{\rm c}\eta_{\rm d}$ and $\zeta_{\rm c,ch,d}=\eta_{\rm c}\eta_{\rm ch}\eta_{\rm d}$, where $\eta_{\rm c}$, $\eta_{\rm d}$ and $\eta_{\rm ch}$ denote, respectively, the transmittance of the BSs modeling the coupling loss, the detection inefficiency of the detectors and the channel loss. The symbol ``$\oplus$'' is used to denote the PBSs that precede the photodetectors. The latin letters in blue color indicate the different modes, while the greek letters in red color are used to tag the number of photons observed at each of the detectors in any given detection event. The output modes $f_{1}$, $g_{1}$, $p_{1}$ and $q_{1}$ correspond to the losses, $\ket{0}$ is the vacuum state, and a dashed grey rectangle identifies the ESR.}
	\label{fig:ESR_setup}
\end{figure*}
Here, we calculate the parameters $P_{\rm SH}$, $Q|_{\rm SH}$ and $\omega|_{\rm SH}$ for an honest implementation of the DIQKD protocol assisted by an ESR at Bob's side. A schematic of the mathematical model that describes the optical setup is given in Fig.~(\ref{fig:ESR_setup}). This figure includes all the parameters and the relevant notation that shall be used in what follows.

\subsubsection{Click pattern distribution}\label{ESR_1}
Since all photodetectors are described with a POVM whose elements are diagonal in the Fock basis (see Eq.~(\ref{7})), for convenience we will consider that Alice's source emits mixed states of the form 
\begin{equation}\label{PDC_1}
\sigma_{ab}=\sum_{n=0}^{\infty}{p_{n}}\ket{\phi_{n}}_{ab}\bra{\phi_{n}},
\end{equation}
instead of pure states $\ket{\psi}_{ab}=\sum_{n=0}^{\infty}\sqrt{p_{n}}\ket{\phi_{n}}_{ab}$ (see Eq.~(\ref{2}) for the explicit expression of $\ket{\phi_{n}}_{ab}$). Both states deliver exactly the same output statistics when measured in the Fock basis, and thus we can use the state $\sigma_{ab}$ for the calculations below.

Similarly, we will consider that the state emitted by the entanglement source within the qubit amplifier is of the form
\begin{equation}\label{PDC_2}
\sigma_{cd}=\sum_{n=0}^{\infty}{p'_{n}}\ket{\phi_{n}}_{cd}\bra{\phi_{n}},
\end{equation}
where the apostrophe in ${p'_n}$ indicates that the statistics of $\sigma_{cd}$ are generally different from those of $\sigma_{ab}$. For instance, $\sigma_{ab}$ and $\sigma_{cd}$ could originate from PDC sources with different intensity parameters. 

The starting point for our calculations is then the quantum state
\begin{align}\label{state_0}
&\rho_{0}=\sigma_{ab}\otimes\sigma_{cd}= \nonumber \\
&\sum_{n=0}^{\infty}\sum_{n'=0}^{\infty}{p_{n}p'_{n'}}\ket{\phi_{n}}_{ab}\bra{\phi_{n}}\otimes\ket{\phi_{n'}}_{cd}\bra{\phi_{n'}}.
\end{align}
Prior to the interference in the linear optics BSM, the states $\sigma_{ab}$ and $\sigma_{cd}$ evolve separately. In particular, let us focus on the evolution of $\sigma_{ab}$ first. We have that the states $\ket{\phi_{n}}_{ab}$ can be written as
\begin{equation}\label{singlet_newton}
\ket{\phi_{n}}_{ab}=\frac{1}{n!\sqrt{n+1}}\sum_{i=0}^{n}{{n}\choose{i}}(-1)^{i}(a_{\rm{v}}^{\dagger}b_{\rm{h}}^{\dagger})^{i}(a_{\rm{h}}^{\dagger}b_{\rm{v}}^{\dagger})^{n-i}\ket{0}.
\end{equation}
Here and in what follows, we shall simply use $\ket{0}$ to denote the vacuum
state in all spatial modes. Since all detectors are assumed to have the same detection efficiency, $\eta_{\rm d}$, we model them by means of a BS of transmittance $\eta_{\rm d}$ together with lossless PNR detectors, whose POVM elements are simply given by projectors onto Fock states: $\Pi_{j}=\ket{j}\bra{j}$, $j\in{\mathbb{N}}$. The effect of dark counts is incorporated a posteriori. This model is equivalent to that given by Eq.~(\ref{6}).  Moreover, we combine the effect of finite detection efficiency and coupling efficiency (as well as channel loss) in one BS of transmittance $\zeta_{\rm c,d}={\eta_{\rm c}\eta_{\rm d}}$ ($\zeta_{\rm c,ch,d}={\eta_{\rm c}\eta_{\rm ch}}\eta_{\rm d}$, where $\eta_{\rm ch}=10^{-\Lambda/10}$ is the transmission efficiency of the fiber link connecting Alice's and Bob's labs, which depends on the channel loss $\Lambda$). In doing so, we find that the quantum states $\ket{\phi_{n}}_{ab}$ evolve to
\begin{widetext}
	\begin{eqnarray}\label{first_branch}
	\ket{\phi_{n}}_{a_{1}b_{1};f_{1}g_{1}}&=&\frac{1}{n!\sqrt{n+1}}\sum_{i=0}^{n}\sum_{j=0}^{i}\sum_{k=0}^{i}\sum_{l=0}^{n-i}\sum_{m=0}^{n-i}(-1)^{i}{{n}\choose{i}}{{i}\choose{j}}{{i}\choose{k}}{{n-i}\choose{l}}{{n-i}\choose{m}}T_{\rm c,d}^{j+l}T_{\rm c,ch,d}^{k+m}R_{\rm c,d}^{n-j-l}R_{\rm c,ch,d}^{n-k-m} \nonumber \\
	&\times&{a_{1,\rm{h}}^{\dagger}}^{l}{a_{1,\rm{v}}^{\dagger}}^{j}{b_{1,\rm{h}}^{\dagger}}^{k}{b_{1,\rm{v}}^{\dagger}}^{m}{f_{1,\rm{h}}^{\dagger}}^{n-i-l}{f_{1,\rm{v}}^{\dagger}}^{i-j}{g_{1,\rm{h}}^{\dagger}}^{i-k}{g_{1,\rm{v}}^{\dagger}}^{n-i-m}\ket{0},
	\end{eqnarray}
\end{widetext}
where the different modes are illustrated in Fig.~(\ref{fig:ESR_setup}). In Eq.~(\ref{first_branch}) we define $T_{\rm c,d}={\zeta_{\rm c,d}}^{1/2}$, $T_{\rm c,ch,d}={\zeta_{\rm c,ch,d}}^{1/2}$, $R_{\rm c,d}=(1-\zeta_{\rm c,d})^{1/2}$ and $R_{\rm c,ch,d}=(1-\zeta_{\rm c,ch,d})^{1/2}$ for readability. 

Similarly, the state $\sigma_{cd}$ within the ESR undergoes a similar transformation, so that the pure states $\ket{\phi_{n'}}_{cd}$ evolve to $\ket{\phi_{n'}}_{c_{1}d_{1};p_{1}q_{1}}$, whose form is identical to that of $\ket{\phi_{n}}_{a_{1}b_{1};f_{1}g_{1}}$ but taking $T_{\rm c,ch,d}\to{T_{\rm c,d}}$ and $R_{\rm c,ch,d}\to{R_{\rm c,d}}$, and now obviously referred to the modes $c_{1}, d_{1}, p_{1}$ and $q_{1}$ (instead of $a_{1}, b_{1}, f_{1}$ and $g_{1}$).

Putting it all together, the overall quantum state prior to the interference in the ESR is given by
\begin{eqnarray}\label{BSM}
\rho_{\rm BSM}&=&\sum_{n=0}^{\infty}\sum_{n'=0}^{\infty}{p_{n}p'_{n'}}\ket{\phi_{n}}_{a_{1}b_{1};f_{1}g_{1}}\bra{\phi_{n}} \nonumber \\
&\otimes&\ket{\phi_{n'}}_{c_{1}d_{1};p_{1}q_{1}}\bra{\phi_{n'}},
\end{eqnarray}
where the pure states $\ket{\phi_{n}}_{a_{1}b_{1};f_{1}g_{1}}\otimes{\ket{\phi_{n'}}_{c_{1}d_{1};p_{1}q_{1}}}$ are written as
\begin{widetext}
	\begin{align}\label{input_ESR}
	&\ket{\phi_{n}}_{a_{1}b_{1};f_{1}g_{1}}\otimes{\ket{\phi_{n'}}_{c_{1}d_{1};p_{1}q_{1}}}=\frac{1}{n!n'!\sqrt{(n+1)(n'+1)}}\sum_{i=0}^{n}\sum_{x=0}^{n'}\sum_{j=0}^{i}\sum_{y=0}^{x}\sum_{k=0}^{i}\sum_{z=0}^{x}\sum_{l=0}^{n-i}\sum_{t=0}^{n'-x}\sum_{m=0}^{n-i}\sum_{w=0}^{n'-x}(-1)^{i+x}{{n}\choose{i}}{{n'}\choose{x}}{{i}\choose{j}} \nonumber \\
	&\times{{x}\choose{y}}{{i}\choose{k}}{{x}\choose{z}}{{n-i}\choose{l}}{{n'-x}\choose{t}}{{n-i}\choose{m}}{{n'-x}\choose{w}}T_{\rm c,d}^{j+l+y+z+t+w}T_{\rm c,ch,d}^{k+m}R_{\rm c,d}^{n+2n'-j-l-y-z-t-w}R_{\rm c,ch,d}^{n-k-m}{a_{1,\rm{h}}^{\dagger}}^{l}{a_{1,\rm{v}}^{\dagger}}^{j} \nonumber \\
	&\times{}{b_{1,\rm{h}}^{\dagger}}^{k}{b_{1,\rm{v}}^{\dagger}}^{m}{c_{1,\rm{h}}^{\dagger}}^{t}{c_{1,\rm{v}}^{\dagger}}^{y}{d_{1,\rm{h}}^{\dagger}}^{z}{d_{1,\rm{v}}^{\dagger}}^{w}{f_{1,\rm{h}}^{\dagger}}^{n-i-l}{f_{1,\rm{v}}^{\dagger}}^{i-j}{g_{1,\rm{h}}^{\dagger}}^{i-k}{g_{1,\rm{v}}^{\dagger}}^{n-i-m}{p_{1,\rm{h}}^{\dagger}}^{n'-x-t}{p_{1,\rm{v}}^{\dagger}}^{x-y}{q_{1,\rm{h}}^{\dagger}}^{x-z}{q_{1,\rm{v}}^{\dagger}}^{n'-x-w}\ket{0}.
	\end{align}
\end{widetext}
Next, modes $b_{1}$ and $c_{1}$ interfere at a 50:50 BS within the ESR. The corresponding transformation is given by
\begin{align}\label{interference}
&{b_{1,\rm{h}}^{\dagger}}^{k}{b_{1,\rm{v}}^{\dagger}}^{m}{c_{1,\rm{h}}^{\dagger}}^{t}{c_{1,\rm{v}}^{\dagger}}^{y}\xrightarrow{50:50}\sum_{h=0}^{k}\sum_{r=0}^{m}\sum_{u=0}^{t}\sum_{v=0}^{y}\frac{(-1)^{u+v}}{{\sqrt{2}}^{k+m+t+y}} \nonumber \\
&\times{{k}\choose{h}}{{m}\choose{r}}{{t}\choose{u}}{{y}\choose{v}}{b_{2,\rm{h}}^{\dagger}}^{h+u}{b_{2,\rm{v}}^{\dagger}}^{r+v}{c_{2,\rm{h}}^{\dagger}}^{k+t-h-u}\nonumber \\
&\times{c_{2,\rm{v}}^{\dagger}}^{m+y-r-v}.
\end{align}
If we define the sum variables $s=h+u$ and $o=r+v$, and we use the fact that a rectangle $\{0\leq{a}\leq{A}, 0\leq{b}\leq{B}\}$ can be equivalently characterized by $\{0\leq{s}\leq{A+B}, \max(0,s-A)\leq{b}\leq{\min(s,B)}\}$, we have that the RHS of Eq.~(\ref{interference}) can be written as
\begin{align}\label{interference2}
&\sum_{s=0}^{k+t}\sum_{o=0}^{m+y}\sum_{u=\max\{0,s-k\}}^{\min\{s,t\}}\sum_{v=\max\{0,o-m\}}^{\min\{o,y\}}\frac{(-1)^{u+v}}{{\sqrt{2}}^{k+m+t+y}} \nonumber \\
&\times{{k}\choose{s-u}}{{m}\choose{o-v}}{{t}\choose{u}}{{y}\choose{v}}{b_{2,\rm{h}}^{\dagger}}^{s}{b_{2,\rm{v}}^{\dagger}}^{o}{c_{2,\rm{h}}^{\dagger}}^{k+t-s}\nonumber \\
&\times{c_{2,\rm{v}}^{\dagger}}^{m+y-o}. 
\end{align}
This means that the state $\ket{\phi_{n}}_{a_{1}b_{1};f_{1}g_{1}}\otimes{\ket{\phi_{n'}}_{c_{1}d_{1};p_{1}q_{1}}}$ is transformed into $\ket{\phi_{n,n'}}_{a_{1}b_{2}c_{2}d_{1};f_{1}g_{1}p_{1}q_{1}}$ given by
\begin{widetext}
	\begin{align}\label{after_the_BSM}
	&\ket{\phi_{n,n'}}_{a_{1}b_{2}c_{2}d_{1};f_{1}g_{1}p_{1}q_{1}}=\frac{1}{n!n'!\sqrt{(n+1)(n'+1)}}\sum_{i=0}^{n}\sum_{x=0}^{n'}\sum_{j=0}^{i}\sum_{y=0}^{x}\sum_{k=0}^{i}\sum_{z=0}^{x}\sum_{l=0}^{n-i}\sum_{t=0}^{n'-x}\sum_{m=0}^{n-i}\sum_{w=0}^{n'-x}\sum_{s=0}^{k+t}\sum_{o=0}^{m+y}\sum_{u=\max\{0,s-k\}}^{\min\{s,t\}} \nonumber \\
	&\times\sum_{v=\max\{0,o-m\}}^{\min\{o,y\}}\frac{(-1)^{i+x+u+v}}{{\sqrt{2}}^{k+m+t+y}}{{n}\choose{i}}{{n'}\choose{x}}{{i}\choose{j}}{{x}\choose{y}}{{i}\choose{k}}{{x}\choose{z}}{{n-i}\choose{l}}{{n'-x}\choose{t}}{{n-i}\choose{m}}{{n'-x}\choose{w}}{{k}\choose{s-u}}{{m}\choose{o-v}}{{t}\choose{u}} \nonumber \\
	&\times{}{{y}\choose{v}}T_{\rm c,d}^{j+l+y+z+t+w}T_{\rm c,ch,d}^{k+m}R_{\rm c,d}^{n+2n'-j-l-y-z-t-w}R_{\rm c,ch,d}^{n-k-m}{a_{1,\rm{h}}^{\dagger}}^{l}{a_{1,\rm{v}}^{\dagger}}^{j}{d_{1,\rm{h}}^{\dagger}}^{z}{d_{1,\rm{v}}^{\dagger}}^{w}{b_{2,\rm{h}}^{\dagger}}^{s}{b_{2,\rm{v}}^{\dagger}}^{o}{c_{2,\rm{h}}^{\dagger}}^{k+t-s}{c_{2,\rm{v}}^{\dagger}}^{m+y-o} \nonumber \\
	&\times{f_{1,\rm{h}}^{\dagger}}^{n-i-l}{f_{1,\rm{v}}^{\dagger}}^{i-j}{g_{1,\rm{h}}^{\dagger}}^{i-k}{g_{1,\rm{v}}^{\dagger}}^{n-i-m}{p_{1,\rm{h}}^{\dagger}}^{n'-x-t}{p_{1,\rm{v}}^{\dagger}}^{x-y}{q_{1,\rm{h}}^{\dagger}}^{x-z}{q_{1,\rm{v}}^{\dagger}}^{n'-x-w}\ket{0}.
	\end{align}
\end{widetext}
Now, we incorporate Alice's and Bob's measurement settings by transforming the operators of the affected modes, $a_{1}$ and $d_{1}$, with rotation angles $\theta_{\rm{A}}$ and $\theta_{\rm{B}}$, respectively. We denote the corresponding output pure state by $\ket{\phi_{n,n'}}_{a_{2}b_{2}c_{2}d_{2};f_{1}g_{1}p_{1}q_{1}}^{\theta_{\rm A},\theta_{\rm B}}$, and it is given by
\begin{widetext}
	\begin{align}\label{detectors}
	&\ket{\phi_{n,n'}}_{a_{2}b_{2}c_{2}d_{2};f_{1}g_{1}p_{1}q_{1}}^{\theta_{\rm A},\theta_{\rm B}}=\frac{1}{n!n'!\sqrt{(n+1)(n'+1)}}\sum_{i=0}^{n}\sum_{x=0}^{n'}\sum_{j=0}^{i}\sum_{y=0}^{x}\sum_{k=0}^{i}\sum_{z=0}^{x}\sum_{l=0}^{n-i}\sum_{t=0}^{n'-x}\sum_{m=0}^{n-i}\sum_{w=0}^{n'-x}\sum_{s=0}^{k+t}\sum_{o=0}^{m+y}\sum_{u=\max\{0,s-k\}}^{\min\{s,t\}} \nonumber \\
	&\times\sum_{v=\max\{0,o-m\}}^{\min\{o,y\}}\sum_{e=0}^{l+j}\sum_{r=0}^{z+w}\sum_{h=\max\{0,e-l\}}^{\min\{e,j\}}\sum_{\tilde{n}=\max\{0,r-z\}}^{\min\{r,w\}}{{n}\choose{i}}{{n'}\choose{x}}{{i}\choose{j}}{{x}\choose{y}}{{i}\choose{k}}{{x}\choose{z}}{{n-i}\choose{l}}{{n'-x}\choose{t}}{{n-i}\choose{m}}{{n'-x}\choose{w}} \nonumber \\
	&\times{{k}\choose{s-u}}{{m}\choose{o-v}}{{t}\choose{u}}{{y}\choose{v}}{{l}\choose{e-h}}{{j}\choose{h}}{{z}\choose{r-\tilde{n}}}{{w}\choose{\tilde{n}}}\frac{(-1)^{i+x+u+v+h+\tilde{n}}}{{\sqrt{2}}^{k+m+t+y}}T_{\rm c,d}^{j+l+y+z+t+w}T_{\rm c,ch,d}^{k+m}R_{\rm c,d}^{n+2n'-j-l-y-z-t-w} \nonumber \\
	&\times{}R_{\rm c,ch,d}^{n-k-m}\cos{\theta_{\rm A}}^{e+j-2h}\sin{\theta_{\rm A}}^{l+2h-e}\cos{\theta_{\rm B}}^{r+w-2\tilde{n}}\sin{\theta_{\rm B}}^{z+2\tilde{n}-r}{a_{2,\rm{h}}^{\dagger}}^{e}{a_{2,\rm{v}}^{\dagger}}^{l+j-e}{d_{2,\rm{h}}^{\dagger}}^{r}{d_{2,\rm{v}}^{\dagger}}^{z+w-r}{b_{2,\rm{h}}^{\dagger}}^{s}{b_{2,\rm{v}}^{\dagger}}^{o}{c_{2,\rm{h}}^{\dagger}}^{k+t-s} \nonumber \\
	&\times{c_{2,\rm{v}}^{\dagger}}^{m+y-o}{f_{1,\rm{h}}^{\dagger}}^{n-i-l}{f_{1,\rm{v}}^{\dagger}}^{i-j}{g_{1,\rm{h}}^{\dagger}}^{i-k}{g_{1,\rm{v}}^{\dagger}}^{n-i-m}{p_{1,\rm{h}}^{\dagger}}^{n'-x-t}{p_{1,\rm{v}}^{\dagger}}^{x-y}{q_{1,\rm{h}}^{\dagger}}^{x-z}{q_{1,\rm{v}}^{\dagger}}^{n'-x-w}\ket{0}.
	\end{align}
\end{widetext}
That is, Eq.~(\ref{detectors}) describes the quantum state immediately prior to the detectors, conditioned on the photon pair numbers $n$ and $n'$. 

Importantly, as the detection efficiencies of all detectors were already accounted for in the effective BS models, projective measurements onto Fock states must be considered now, as explained above. More precisely, we are interested in the conditional probability, ${{\rm{P}}\left(\vec{\alpha}|n,n'\right)}_{\theta_{\rm A},\theta_{\rm B}}$, of observing a detection pattern $\vec{\alpha}$ given the quantum state $\ket{\phi_{n,n'}}_{a_{2}b_{2}c_{2}d_{2};f_{1}g_{1}p_{1}q_{1}}^{\theta_{\rm A},\theta_{\rm B}}$. Here, we introduced the vector notation $\vec{\alpha}=(\alpha,\beta,\gamma,\delta,\mu,\nu,\tau,\lambda)$, where each element of the vector denotes the number of photons in a certain output mode: $\alpha$ refers to mode $a_{2,\rm{h}}$, $\beta$ to mode $a_{2,\rm{v}}$, $\gamma$ to mode $d_{2,\rm{h}}$, $\delta$ to mode $d_{2,\rm{v}}$, $\mu$ to mode $c_{2,\rm{h}}$, $\nu$ to mode $c_{2,\rm{v}}$, $\tau$ to mode $b_{2,\rm{h}}$ and $\lambda$ to mode $b_{2,\rm{v}}$. With this notation, and if we disregard for the moment the effect of dark counts, we have that ${{\rm{P}}\left(\vec{\alpha}|n,n'\right)}_{\theta_{\rm A},\theta_{\rm B}}$ is given by
\begin{equation}\label{qwe}
{{\rm P}\left(\vec{\alpha}|n,n'\right)}_{\theta_{\rm A},\theta_{\rm B}}=\left\lVert{\ket{\tilde{\phi}_{n,n'}}_{\vec{\alpha};f_{1}g_{1}p_{1}q_{1}}^{\theta_{\rm A},\theta_{\rm B}}}\right\rVert^{2},
\end{equation}
where the unnormalized state $\ket{\tilde{\phi}_{n,n'}}_{\vec{\alpha};f_{1}g_{1}p_{1}q_{1}}^{\theta_{\rm A},\theta_{\rm B}}$ has the form
\begin{equation}
\ket{\tilde{\phi}_{n,n'}}_{\vec{\alpha};f_{1}g_{1}p_{1}q_{1}}^{\theta_{\rm A},\theta_{\rm B}}={\braket{\vec{\alpha}}{\phi_{n,n'}}}_{a_{2}b_{2}c_{2}d_{2};f_{1}g_{1}p_{1}q_{1}}^{\theta_{\rm A},\theta_{\rm B}},
\end{equation}
being $\ket{\vec{\alpha}}=\ket{\alpha,\beta,\gamma,\delta,\mu,\nu,\tau,\lambda}$.

To compute $\ket{\tilde{\phi}_{n,n'}}_{\vec{\alpha};f_{1}g_{1}p_{1}q_{1}}^{\theta_{\rm A},\theta_{\rm B}}$, we make use of the following orthogonality relation:
\begin{align}\label{orthogonality}
&\bra{\vec{\alpha}}{{a_{2,\rm{h}}^{\dagger}}^{e}{a_{2,\rm{v}}^{\dagger}}^{l+j-e}{d_{2,\rm{h}}^{\dagger}}^{r}{d_{2,\rm{v}}^{\dagger}}^{z+w-r}{c_{2,\rm{h}}^{\dagger}}^{k+t-s}{c_{2,\rm{v}}^{\dagger}}^{m+y-o}}{b_{2,\rm{h}}^{\dagger}}^{s}\nonumber \\
&\times {b_{2,\rm{v}}^{\dagger}}^{o}\ket{0} =\left({\alpha!}{\beta!}{\gamma!}{\delta!}{\mu!}{\nu!}{\tau!}{\lambda!}\right)^{1/2}\delta_{\alpha}^{e}\delta_{\beta}^{l+j-e}\delta_{\gamma}^{r}\delta_{\delta}^{z+w-r} \nonumber \\ 
&\times\delta_{\mu}^{k+t-s}\delta_{\nu}^{m+y-o}\delta_{\tau}^{s}\delta_{\lambda}^{o},
\end{align} 
where $\delta_{i}^{j}$ stands for the Kronecker's delta symbol, {\it i.e.}, $\delta_{i}^{j}=1$ only if $i=j$, otherwise it is zero. Also, we recall that for finite range sums, $\sum_{a=A_{1}}^{A_{2}}f(a)\delta_{x}^{a}=f(x)\Theta_{x-A_{1}}\Theta_{A_{2}-x}$, where $\Theta_{j}$ is the ``discrete'' Heaviside function~\footnote{For all $j\in\mathbb{Z}$, $\Theta_{j}=1$ if $j\geq{0}$ and $\Theta_{j}=0$ otherwise.}. Then, in order to incorporate the effect of the different $\Theta$'s, one must modify the affected index ranges accordingly. This yields
\begin{widetext}
	\begin{align}\label{unnormalized}
	&\ket{\tilde{\phi}_{n,n'}}_{\vec{\alpha};f_{1}g_{1}p_{1}q_{1}}^{\theta_{\rm A},\theta_{\rm B}}=\frac{1}{n!n'!}\left[{\frac{{\alpha!}{\beta!}{\gamma!}{\delta!}{\mu!}{\nu!}{\tau!}{\lambda!}}{(n+1)(n'+1)2^{\mu+\nu+\tau+\lambda}}}\right]^{\frac{1}{2}}T_{\rm c,d}^{\alpha+\beta+\gamma+\delta+\mu+\tau}T_{\rm c,ch,d}^{\nu+\lambda}R_{\rm c,d}^{n+2n'-\alpha-\beta-\gamma-\delta-\mu-\tau}R_{\rm c,ch,d}^{n-\nu-\lambda}\cos{\theta_{\rm A}}^{\alpha} \nonumber \\
	&\times\sin{\theta_{\rm A}}^{\beta}\cos{\theta_{\rm B}}^{\gamma}\sin{\theta_{\rm B}}^{\delta}\sum_{i=0}^{n}\sum_{x=0}^{n'}\sum_{j=\max\{0,\alpha+\beta+i-n\}}^{\min\{i,\alpha+\beta\}}\sum_{y=\max\{0,\nu+\lambda+i-n\}}^{\min\{x,\nu+\lambda\}}\sum_{k=\max\{0,\mu+\tau+x-n'\}}^{\min\{i,\mu+\tau\}}\sum_{w=\max\{0,\gamma+\delta-x\}}^{\min\{n'-x,\gamma+\delta\}}\sum_{u=\max\{0,\tau-k\}}^{\min\{\tau,\mu+\tau-k\}} \nonumber \\
	&\times\sum_{v=\max\{0,y-\nu\}}^{\min\{\lambda,y\}}\sum_{h=\max\{0,j-\beta\}}^{\min\{\alpha,j\}}\sum_{\tilde{n}=\max\{0,w-\delta\}}^{\min\{\gamma,w\}}{{n}\choose{i}}{{n'}\choose{x}}{{i}\choose{j}}{{x}\choose{y}}{{i}\choose{k}}{{x}\choose{\gamma+\delta-w}}{{n-i}\choose{\alpha+\beta-j}}{{n'-x}\choose{\mu+\tau-k}}{{w}\choose{\tilde{n}}} \nonumber \\
	&\times{{n-i}\choose{\nu+\lambda-y}}{{n'-x}\choose{w}}{{k}\choose{\tau-u}}{{\nu+\lambda-y}\choose{\lambda-v}}{{\mu+\tau-k}\choose{u}}{{y}\choose{v}}{{\alpha+\beta-j}\choose{\alpha-h}}{{j}\choose{h}}{{\gamma+\delta-w}\choose{\gamma-\tilde{n}}}\left(\frac{T_{\rm c,d}R_{\rm c,ch,d}}{T_{\rm c,ch,d}R_{\rm c,d}}\right)^{y-k} \nonumber \\
	&\times(-1)^{i+x+u+v+h+\tilde{n}}\cos{\theta_{\rm A}}^{j-2h}\sin{\theta_{\rm A}}^{2h-j}\cos{\theta_{\rm B}}^{w-2\tilde{n}}\sin{\theta_{\rm B}}^{2\tilde{n}-w}{f_{1,\rm{h}}^{\dagger}}^{n+j-i-\alpha-\beta}{f_{1,\rm{v}}^{\dagger}}^{i-j}{g_{1,\rm{h}}^{\dagger}}^{i-k}{g_{1,\rm{v}}^{\dagger}}^{n+y-i-\nu-\lambda} \nonumber \\
	&\times{p_{1,\rm{h}}^{\dagger}}^{n'+k-x-\mu-\tau}{p_{1,\rm{v}}^{\dagger}}^{x-y}{q_{1,\rm{h}}^{\dagger}}^{x+w-\gamma-\delta}{q_{1,\rm{v}}^{\dagger}}^{n'-x-w}\ket{0},
	\end{align}
\end{widetext}
where some overall constant terms were factored from the sums. 

Then, by applying Eq.~(\ref{qwe}) on the state given by Eq.~(\ref{unnormalized}), we obtain
\begin{widetext}
	\begin{align}\label{distribution}
	&{{\rm{P}}\left(\vec{\alpha}|n,n'\right)}_{\theta_{\rm A},\theta_{\rm B}}={\frac{{\alpha!}{\beta!}{\gamma!}{\delta!}{\mu!}{\nu!}{\tau!}{\lambda!}{\eta}_{\rm B}^{\nu+\lambda}}{(n+1)(n'+1)2^{\mu+\nu+\tau+\lambda}}}{\eta}_{\rm A}^{\alpha+\beta+\gamma+\delta+\mu+\tau}(1-\zeta_{\rm c,d})^{n+2n'-\alpha-\beta-\gamma-\delta-\mu-\tau}(1-\zeta_{\rm c,ch,d})^{n-\nu-\lambda}\cos{\theta_{\rm A}}^{2\alpha} \nonumber \\
	&\times\sin{\theta_{\rm A}}^{2\beta}\cos{\theta_{\rm B}}^{2\gamma}\sin{\theta_{\rm B}}^{2\delta}\sum_{i=0}^{n}\sum_{\Delta=-i}^{n-i}\sum_{x=\max\{0,-\Delta\}}^{\min\{n',n'-\Delta\}}\sum_{j=\max\{0,\alpha+\beta+i-n,-\Delta\}}^{\min\{i,\alpha+\beta,\alpha+\beta-\Delta\}}\sum_{y=\max\{0,\nu+\lambda+i-n,-\Delta\}}^{\min\{x,\nu+\lambda,\nu+\lambda-\Delta\}}\sum_{k=\max\{0,\mu+\tau+x-n',-\Delta\}}^{\min\{i,\mu+\tau,\mu+\tau-\Delta\}} \nonumber \\
	&\times\sum_{w=\max\{0,\gamma+\delta-x,\Delta\}}^{\min\{n'-x,\gamma+\delta,\gamma+\delta+\Delta\}}\sum_{u=\max\{0,\tau-k\}}^{\min\{\tau,\mu+\tau-k\}}\sum_{U=\max\{0,\tau-k-\Delta\}}^{\min\{\tau,\mu+\tau-k-\Delta\}}\sum_{v=\max\{0,y-\nu\}}^{\min\{\lambda,y\}}\sum_{V=\max\{0,y+\Delta-\nu\}}^{\min\{\lambda,y+\Delta\}}\sum_{h=\max\{0,j-\beta\}}^{\min\{\alpha,j\}}\sum_{H=\max\{0,j+\Delta-\beta\}}^{\min\{\alpha,j+\Delta\}} \nonumber \\
	&\times\sum_{\tilde{n}=\max\{0,w-\delta\}}^{\min\{\gamma,w\}}\sum_{\tilde{N}=\max\{0,w-\Delta-\delta\}}^{\min\{\gamma,w-\Delta\}}\left[\frac{\zeta_{\rm c,d}(1-\zeta_{\rm c,ch,d})}{\zeta_{\rm c,ch,d}(1-\zeta_{\rm c,d})}\right]^{y-k}(-1)^{u+v+h+\tilde{n}+U+V+H+\tilde{N}}\left(\frac{\sin{\theta_{\rm A}}}{\cos{\theta_{\rm A}}}\right)^{2(h+H-j)-\Delta}\nonumber \\
	&\times\left(\frac{\sin{\theta_{\rm B}}}{\cos{\theta_{\rm B}}}\right)^{2(\tilde{n}+\tilde{N}-w)+\Delta}\Upsilon(n,n',i,x,j,y,k,w,u,U,v,V,h,H,\tilde{n},\tilde{N},\Delta,\alpha,\beta,\gamma,\delta,\mu,\nu,\tau,\lambda),
	\end{align}
\end{widetext}
where we have defined
\begin{widetext}	
	\begin{align}\label{upsilon} &\Upsilon(n,n',i,x,j,y,k,w,u,U,v,V,h,H,\tilde{n},\tilde{N},\Delta,\alpha,\beta,\gamma,\delta,\mu,\nu,\tau,\lambda)={} \nonumber \\
	&\times{i!(i+\Delta)!x!(x+\Delta)!(n-i)!(n-i-\Delta)!(n'-x)!(n'-x-\Delta)!(i-j)!(x-y)!(i-k)!} \nonumber \\
	& \times\frac{[(x+w-\gamma-\delta)!(n+j-i-\alpha-\beta)!(n'+k-x-\mu-\tau)!(n+y-i-\nu-\lambda)!(n'-x-w)!]^{-1}}{{h!H!(j-h)!(j+\Delta-H)!v!V!(y-v)!(y+\Delta-V)!(\tau-u)!(\tau-U)!(k+u-\tau)!(k+\Delta+U-\tau)!}}\nonumber \\
	&\times\frac{[(\gamma-\tilde{n})!(\gamma-\tilde{N})!(\delta+\tilde{n}-w)!(\delta+\tilde{N}+\Delta-w)!(\alpha-h)!(\alpha-H)!(\beta+h-j)!(\beta+H-j-\Delta)!u!U!]^{-1}}{(\mu+\tau-k-u)!(\mu+\tau-k-\Delta-U)!(\lambda-v)!(\lambda-V)!(\nu+v-y)!(\nu+V-y-\Delta)!\tilde{n}!\tilde{N}!(w-\tilde{n})!(w-\Delta-\tilde{N})!}. \nonumber \\
	\end{align}
\end{widetext}
Of course, the condition $\sum_{\vec{\alpha}}{{\rm{P}}\left(\vec{\alpha}|n,n'\right)}_{\theta_{\rm A},\theta_{\rm B}}=1$ holds, and only those click patterns $\vec{\alpha}$ such that $\alpha+\beta\leq{n}$, $\gamma+\delta\leq{n'}$ and $\mu+\nu+\tau+\lambda\leq{n+n'}$ give a non vanishing contribution due to the absence of noise, which we take into account next. 

In particular, if one considers the noise model introduced in the main text, we find that the resulting distribution in the noisy scenario is given by
\begin{eqnarray}\label{noisy_dist}
{\tilde{\rm{P}}\left(\vec{\alpha}|n,n'\right)}_{\theta_{\rm A},\theta_{\rm B}}&=&(1-{8}p_{\rm d}){{\rm{P}}\left(\vec{\alpha}|n,n'\right)}_{\theta_{\rm A},\theta_{\rm B}} \nonumber \\
&+&p_{\rm d}\sum_{\vec{\sigma}\in{\Gamma_{\vec{\alpha}}}}{{\rm{P}}\left(\vec{\sigma}|n,n'\right)}_{\theta_{\rm A},\theta_{\rm B}}+O(p_{\rm d}^{2}), \nonumber \\
\end{eqnarray}
where $\Gamma_{\vec{\alpha}}={\left\{\vec{\sigma}:|\vec{\alpha}|=|\vec{\sigma}|+1\right\}}$, $|\vec{\alpha}|$ being the overall number of photons corresponding to the pattern $\vec{\alpha}$, {\it i.e.},  $|\vec{\alpha}|=\alpha+\beta+\gamma+\delta+\mu+\nu+\tau+\lambda$. As a consequence, and up to first order in $p_{\rm d}$, we have that ${\tilde{\rm{P}}\left(\vec{\alpha}|n,n'\right)}_{\theta_{\rm A},\theta_{\rm B}}$ vanishes for any pattern $\vec{\alpha}$ that does not fulfill $\alpha+\beta\leq{n+1}$, $\gamma+\delta\leq{n'+1}$, $\mu+\nu+\tau+\lambda\leq{n+n'+1}$ and $\alpha+\beta+\gamma+\delta+\mu+\nu+\tau+\lambda\leq{2(n+n')+1}$. 

Alice's and Bob's outcomes are expected to be anti-correlated, as the entanglement sources we are considering emit singlet states. Therefore, we flip say Alice's outcomes in such a way that, whenever Alice and Bob select the same measurement settings, their outcomes are correlated. For this, we define the distribution 
\begin{equation}\label{permutation}
{\tilde{\rm{p}}\left(\vec{\alpha}|n,n'\right)}_{\theta_{\rm A},\theta_{\rm B}}={{\tilde{\rm{P}}_{\alpha\leftrightarrow\beta}\left(\vec{\alpha}|n,n'\right)}_{\theta_{\rm A},\theta_{\rm B}}}. 
\end{equation}
That is to say, ${\tilde{\rm{p}}\left(\alpha,\beta,\gamma,\delta,\mu,\nu,\tau,\lambda|n,n'\right)}_{\theta_{\rm A},\theta_{\rm B}}={\tilde{\rm{P}}\left(\beta,\alpha,\gamma,\delta,\mu,\nu,\tau,\lambda|n,n'\right)}_{\theta_{\rm A},\theta_{\rm B}}$.

Finally, it only remains to take into account the deterministic assignment performed by Alice and Bob whenever they observe an inconclusive event, required to have a distribution with binary outcomes on both sides. To be precise, the assignments read
\begin{equation}\label{assignment}
A_{\rm A} = \left\{
\begin{array}{ll}
0      & \mathrm{if\ } (\alpha,\beta)=(1,0) \\
1 & \rm{otherwise,} \\
\end{array}
\right. \\
A_{\rm B} = \left\{
\begin{array}{ll}
0      & \mathrm{if\ } (\gamma,\delta)=(1,0) \\
1 & \rm{otherwise.} \\
\end{array}
\right. \\
\end{equation}
In this way, every possible event $(\alpha,\beta,\gamma,\delta)$ regarding Alice's and Bob's detector outcomes is mapped to an element of the set of binary strings $\left\{(0,0),(0,1),(1,0),(1,1)\right\}$. 

Importantly, we remark that such a post-processing is not performed on the outcomes of the photodetectors inside the qubit amplifier, but only on the outcomes observed by the parties. In summary, the distribution we are finally interested in is that of the ``post-processed click pattern'' $A_{\vec{\alpha}}={(A_{\rm A},A_{\rm B},\mu,\nu,\tau,\lambda)}$, which we shall denote by $\mathbf{P}(A_{\vec{\alpha}}|n,n')_{\theta_{\rm A},\theta_{\rm B}}$. From Eqs.~($\ref{assignment}$), it is obvious that
\begin{align}\label{post-processed}
&\mathbf{P}(0,0,\mu,\nu,\tau,\lambda|n,n')_{\theta_{\rm A},\theta_{\rm B}}= \nonumber \\
&{\tilde{\rm{p}}\left(1,0,1,0,\mu,\nu,\tau,\lambda|n,n'\right)}_{\theta_{\rm A},\theta_{\rm B}}, \nonumber \\
&\nonumber \\
&\mathbf{P}(0,1,\mu,\nu,\tau,\lambda|n,n')_{\theta_{\rm A},\theta_{\rm B}}= \nonumber \\
&\sum_{(\gamma,\delta)\neq{(1,0)}}{\tilde{\rm{p}}\left(1,0,\gamma,\delta,\mu,\nu,\tau,\lambda|n,n'\right)}_{\theta_{\rm A},\theta_{\rm B}}, \nonumber \nonumber \\
&\nonumber \\
&\mathbf{P}(1,0,\mu,\nu,\tau,\lambda|n,n')_{\theta_{\rm A},\theta_{\rm B}}= \nonumber \\
&\sum_{(\alpha,\beta)\neq{(1,0)}}{\tilde{\rm{p}}\left(\alpha,\beta,1,0,\mu,\nu,\tau,\lambda|n,n'\right)}_{\theta_{\rm A},\theta_{\rm B}}\hspace{.2cm} \nonumber \nonumber \\
&\nonumber \\
&\mathbf{P}(1,1,\mu,\nu,\tau,\lambda|n,n')_{\theta_{\rm A},\theta_{\rm B}}= \nonumber \\
&\sum_{(\alpha,\beta)\neq{(1,0)}}\sum_{(\gamma,\delta)\neq{(1,0)}}{\tilde{\rm{p}}\left(\alpha,\beta,\gamma,\delta,\mu,\nu,\tau,\lambda|n,n'\right)}_{\theta_{\rm A},\theta_{\rm B}}. \nonumber \nonumber \\
\end{align}
\subsubsection{Parameters $P_{\rm SH}$, $Q|_{\rm SH}$ and $\omega|_{\rm SH}$}\label{ESR_2}

To obtain the value of these parameters we have to take into account that there are four different click patterns in the ESR which are considered to be successful heralding events, as explained in the main text. These are $(\mu,\nu,\tau,\lambda)=\{(1,1,0,0),(0,1,1,0),(1,0,0,1),(0,0,1,1)\}$. Due to the symmetries of the channel model, we can consider only one of these successful heralding events, say $\Omega={\left\{(\mu,\nu,\tau,\lambda)=(1,1,0,0)\right\}}$, and the next holds: $P_{\rm{SH}}=4P_{\Omega}$, $\omega|_{\rm{SH}}=\omega|_{\Omega}$ and $Q|_{\rm{SH}}=Q|_{\Omega}$. As a consequence, we can restrict ourselves to the calculation of $P_{\Omega}$, $Q|_{\Omega}$ and $\omega|_{\Omega}$. 

To begin with, we have that the probability $P_{\Omega}$ is simply given by
\begin{align}\label{success_prob}
&P_{\Omega}=\sum_{A_{\rm A},A_{\rm B}}\mathbf{P}(A_{\rm A},A_{\rm B},\Omega)_{\theta_{\rm A},\theta_{\rm B}}= \nonumber \\
&\sum_{n,n'}{p_{n}p'_{n'}}\sum_{A_{\rm A},A_{\rm B}}\mathbf{P}(A_{\rm A},A_{\rm B},\Omega|n,n')_{\theta_{\rm A},\theta_{\rm B}},
\end{align}
where $A_{\rm A},A_{\rm B}\in\{0,1\}$. Obviously, since we are summing over all possible measurement outcomes for Alice and Bob in Eq.~(\ref{success_prob}), $P_{\Omega}$ does not depend on the rotation angles $\theta_{\rm{A}}$ and $\theta_{\rm B}$. Therefore, one can simply set them both to zero for the numerical calculations.

Secondly, the conditional QBER is given by
\begin{align}\label{QBER}
&Q|_{\Omega}={\frac{1}{P_{\Omega}}\left[\mathbf{P}(0,1,\Omega)_{0,0}+\mathbf{P}(1,0,\Omega)_{0,0}\right]} \nonumber \\
&=\frac{1}{P_{\Omega}}\sum_{n,n'}{p_{n}p'_{n'}}\left[\mathbf{P}(0,1,\Omega|n,n')_{0,0}+\mathbf{P}(1,0,\Omega|n,n')_{0,0}\right]. \nonumber \\
\end{align}
We remark that this quantity is referred to the events in which both parties select the Z-basis, so that $\theta_{\rm{A}}=\theta_{\rm{B}}=0$.

Finally, the conditional winning probability at the CHSH game, $\omega|_{\Omega}$, can be defined in terms of the conditional CHSH violation, $S|_{\Omega}$, via 
\begin{equation}
\omega|_{\Omega}=\frac{1}{8}S|_{\Omega}+\frac{1}{2},
\end{equation}
where~\footnote{Actually, the summand that carries the minus sign in the definition of $S|_{\Omega}$ depends on the successful heralding event considered. Indeed, two out of the four successful heralding events require the definition of $S|_{\Omega}$ given by Eq.~(\ref{violation}), while the other two require the minus sign to be carried by the third summand in Eq.~(\ref{violation}). This is so because the quantum measurements that lead to a maximum violation of the CHSH inequality are different for each Bell pair, and one can equivalently account for this fact by changing the definition of $S|_{\Omega}$ depending on the Bell pair.}
\begin{align}\label{violation}
&S|_{\Omega}={E_{0,-\frac{\pi}{8}}|_{\Omega}+E_{0,\frac{\pi}{8}}|_{\Omega}+E_{\frac{\pi}{4},-\frac{\pi}{8}}|_{\Omega}-E_{\frac{\pi}{4},\frac{\pi}{8}}|_{\Omega}}, 
\end{align}
and the parameters $E_{\theta_{\rm A},\theta_{\rm B}}|_{\Omega}$ are given by
\begin{eqnarray}\label{violation_summand}
E_{\theta_{\rm A},\theta_{\rm B}}|_{\Omega}&=&
\frac{1}{P_{\Omega}}\left[\mathbf{P}(1,1,\Omega)_{\theta_{\rm A},\theta_{\rm B}}+\mathbf{P}(0,0,\Omega)_{\theta_{\rm A},\theta_{\rm B}}\right] \nonumber \\
&-&\frac{1}{P_{\Omega}}\left[\mathbf{P}(0,1,\Omega)_{\theta_{\rm A},\theta_{\rm B}}+\mathbf{P}(1,0,\Omega)_{\theta_{\rm A},\theta_{\rm B}}\right] \nonumber \\
&=&\frac{2}{P_{\Omega}}\sum_{n,n'}{p_{n}p'_{n'}}[\mathbf{P}(0,0,\Omega|n,n')_{\theta_{\rm A},\theta_{\rm B}} \nonumber \\
&+&\mathbf{P}(1,1,\Omega|n,n')_{\theta_{\rm A},\theta_{\rm B}}]-1.
\end{eqnarray}
We note that in the second equality of Eq.~(\ref{violation_summand}), we simply used Eq.~(\ref{success_prob}) and the law of total probability, conditioning on the photon numbers $n$ and $n'$. 

\subsection{Polarizing qubit amplifier}\label{PQA}
\begin{figure*}[!htbp]
	\includegraphics[width=\textwidth,height=12cm]{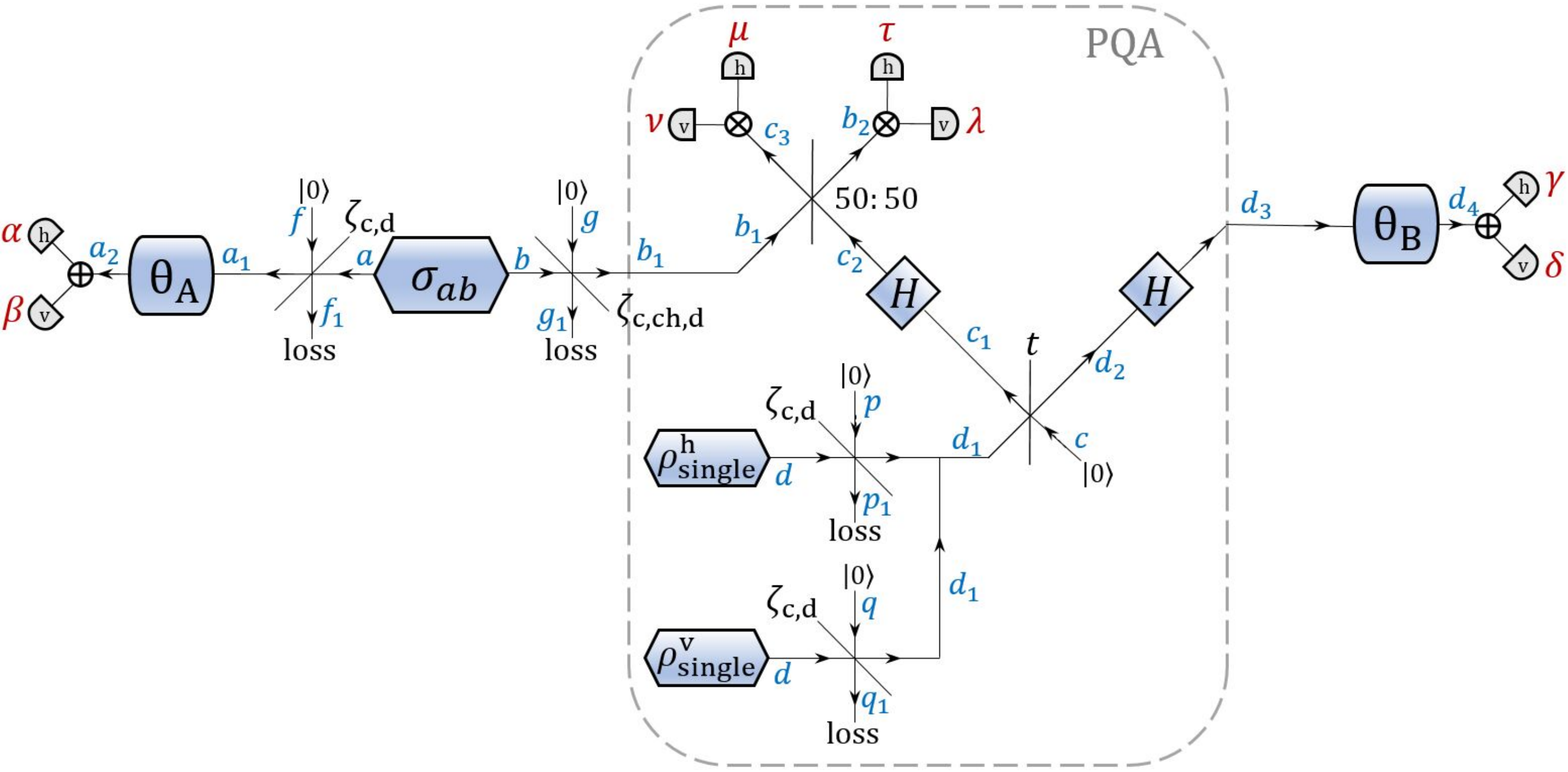} 
	\caption{Schematic of the PQA-based DIQKD setup matching the mathematical models presented in the main text. $\sigma_{ab}$ stands for the entanglement source held by Alice, while $\rho_{\rm single}^{\rm h}$ and $\rho_{\rm single}^{\rm v}$ denote the single-photon sources inside the PQA. $\theta_{\rm A}$ ($\theta_{\rm B}$) denotes the rotation angle of Alice's (Bob's) measurement settings and $\zeta_{\rm c,d}$ and $\zeta_{\rm c,ch,d}$ tag the effective efficiency parameters, $\zeta_{\rm c,d}=\eta_{\rm c}\eta_{\rm d}$ and $\zeta_{\rm c,ch,d}=\eta_{\rm c}\eta_{\rm ch}\eta_{\rm d}$, where $\eta_{\rm c}$, $\eta_{\rm d}$ and $\eta_{\rm ch}$ denote, respectively, the transmittance of the BSs modeling the coupling loss, the detection inefficiency and the channel loss. Again, the symbol ``$\oplus$'' is used to denote the PBSs that precede the photodetectors. Similarly, the latin letters in blue color indicate the different modes, while the greek letters in red color are used to tag the number of photons observed at each of the detectors. The output modes $f_{1}$, $g_{1}$, $p_{1}$ and $q_{1}$ correspond to the losses, $\ket{0}$ is the vacuum state, $t$ is the tunable transmittance of the BS within the PQA, two Hadamard gates are denoted by $H$, and a dashed grey rectangle identifies the PQA.}
	\label{fig:PQA_setup}
\end{figure*}
In this Appendix we now consider a PQA-based implementation of the DIQKD protocol. A schematic of the mathematical model that describes the optical 
setup is illustrated in Fig.~(\ref{fig:PQA_setup}).

In contrast to the ESR-based implementation, Alice and Bob only keep those detection events for which, first, both triggered single-photon sources at Bob's lab record one photon in their idler mode, and second, a successful BSM occurs. For this reason, we need to calculate the trigger probability $P_{\rm trigger}$ at the idler mode of a source of this kind, and the conditional quantum state $\rho_{\rm single}$ at its signal mode, given that a trigger occurred. This is what we do next.

\subsubsection{Triggered single-photon sources}\label{triggered_PDC}

As described in the main text, we shall consider triggered single-photon sources generated by measuring the idler mode of an entanglement state of the form $\ket{\Psi}=\sum_{n=0}^{\infty}\sqrt{p_{n}}\ket{n,n}$.

We have that $P_{\rm trigger}$ can be written as
\begin{equation}\label{p_trigger_1}
P_{\rm trigger}=(1-p_{\rm d})P_{{\rm trigger}|0\rm{d}}+p_{\rm d}P_{{\rm trigger}|1\rm{d}},
\end{equation}
where $P_{{\rm trigger}|0\rm{d}}$ ($P_{{\rm trigger}|1\rm{d}}$) denotes de trigger probability given that there is no dark count (one single dark count) in the detector. These conditional probabilities can be written as
\begin{align}\label{p_trigger_2}
&P_{{\rm trigger}|0\rm{d}}=\sum_{n=0}^{\infty}p_{n}P_{{\rm trigger}|n,0\rm{d}}, \nonumber \\
&P_{{\rm trigger}|1\rm{d}}=\sum_{n=0}^{\infty}p_{n}P_{{\rm trigger}|n,1\rm{d}}.
\end{align}
where $P_{{\rm trigger}|n,0\rm{d}}=n\zeta_{\rm c,d}(1-\zeta_{\rm c,d})^{n-1}$ and $P_{{\rm trigger}|n,1\rm{d}}=(1-\zeta_{\rm c,d})^{n}$. Here, like in Appendix~\ref{ESR}, $\zeta_{\rm c,d}={\eta_{\rm c}\eta_{\rm d}}$ with $\eta_{\rm c}$ and $\eta_{\rm d}$ being, respectively, the coupling and the detection efficiencies. 

Similarly, the conditional quantum state at the signal mode given that a trigger occurred has the form $\rho_{\rm single}=\sum_{n=0}^{\infty}r_{n}\ket{n}\bra{n}$, where the probability distribution $r_{n}$ can be written as:
\begin{align}\label{weights}
&r_{n}=\frac{p_{n}[(1-p_{\rm d})P_{{\rm trigger}|n,0\rm{d}}+p_{\rm d}P_{{\rm trigger}|n,1\rm{d}}]}{P_{\rm trigger}}.
\end{align}

For example, in the case of a triggered PDC source with $p_{n}=\mu^{n}(1+\mu)^{-n-1}$ we find that $P_{{\rm trigger}|0\rm{d}}=\mu\zeta_{\rm c,d}(1+\mu\zeta_{\rm c,d})^{-2}$, and $P_{{\rm trigger}|1\rm{d}}=(1+\mu\zeta_{\rm c,d})^{-1}$. Hence,
\begin{equation}\label{p_trigger_3}
P_{\rm trigger}=\frac{p_{\rm d}+\mu\zeta_{\rm c,d}}{(1+\mu\zeta_{\rm c,d})^{2}}.
\end{equation}
Also, we obtain
\begin{eqnarray}\label{weights2}
r_{n}&=&\frac{\mu^{n}(1-\zeta_{\rm c,d})^{n-1}(1+\mu\zeta_{\rm c,d})^{2}}{(1+\mu)^{n+1}(p_{\rm d}+\mu\zeta_{\rm c,d})} \nonumber \\
&\times&\left[(1-p_{\rm d})n\zeta_{\rm c,d}+p_{\rm d}(1-\zeta_{\rm c,d})\right]. 
\end{eqnarray}

In what follows, we will use the notation $\rho_{\rm single}^{\rm h}$ and $\rho_{\rm single}^{\rm v}$ in order to specify the polarization state of the generated photons.

\subsubsection{Click pattern distribution}\label{PQA_1}

Our starting point is the state of the whole system conditioned on the trigger of both triggered single-photon sources at Bob's PQA,
\begin{align}\label{input}
&\rho_{0}|_{\rm double\hspace{.05cm}trigger}=\sigma_{ab}\otimes\rho_{\rm single}^{\rm h}\otimes\rho_{\rm single}^{\rm v} \nonumber \\
&=\sum_{n=0}^{\infty}\sum_{n_{1}=0}^{\infty}\sum_{n_{2}=0}^{\infty}p_{n}r_{n_{1}}r_{n_{2}}\ket{\phi_{n}}_{ab}\bra{\phi_{n}}\otimes\ket{n_{1},n_{2}}_{d}\bra{n_{1},n_{2}}, \nonumber \\
\end{align}
where $\ket{n_{1},n_{2}}_{d}={d_{\rm h}^{\dagger}}^{n_{1}}{d_{\rm v}^{\dagger}}^{n_{2}}/\sqrt{n_{1}!n_{2}!}\ket{0}$. 

Next, we have that the quantum states $\ket{\phi_{n}}_{ab}$ undergo exactly the same transformation as in the case of the ESR-based setup (see Fig.~(\ref{fig:ESR_setup})), leading to the states $\ket{\phi_{n}}_{a_{1}b_{1};f_{1}g_{1}}$ given by Eq.~(\ref{first_branch}) right before the interference at the BSM. 

On the other hand, the states $\ket{n_{1},n_{2}}_{d}$ evolve to
\begin{align}\label{auxiliary_states_1}
&\ket{\psi_{n_{1}},\psi_{n_{2}}}_{d_{1};p_{1}q_{1}}=\frac{1}{\sqrt{n_{1}!n_{2}!}}\sum_{x=0}^{n_{1}}\sum_{y=0}^{n_{2}}{{n_{1}}\choose{x}}{{n_{2}}\choose{y}}T_{\rm c,d}^{x+y} \nonumber \\
&\times{}R_{\rm c,d}^{n_{1}+n_{2}-x-y}{d_{\rm h}^{\dagger}}^{x}{d_{\rm v}^{\dagger}}^{y}{p_{1,\rm h}^{\dagger}}^{n_{1}-x}{q_{1,\rm v}^{\dagger}}^{n_{2}-y}\ket{0},
\end{align}
where $T_{\rm c,d}={\zeta_{\rm c,d}}^{1/2}$ and $R_{\rm c,d}=(1-\zeta_{\rm c,d})^{1/2}$ as in Appendix~\ref{ESR}. Here, like in the ESR-based setup, we have modeled the effect of coupling and detection loss by means of a BS with transmittance $\zeta_{\rm c,d}=\eta_{\rm c}\eta_{\rm d}$. This means that for the remaining calculations we consider lossless PNR detectors.

Next, the quantum signals coming from both single-photon sources enter a BS of transmittance $t$. This BS transforms the quantum states $\ket{\psi_{n_{1}},\psi_{n_{2}}}_{d_{1};p_{1}q_{1}}$ into
\begin{widetext}
	\begin{eqnarray}\label{auxiliary_states_2}
	\ket{\psi_{n_{1}},\psi_{n_{2}}}_{c_{1}d_{2};p_{1}q_{1}}&=&\frac{1}{\sqrt{n_{1}!n_{2}!}}\sum_{x=0}^{n_{1}}\sum_{y=0}^{n_{2}}\sum_{z=0}^{x}\sum_{w=0}^{y}{{n_{1}}\choose{x}}{{n_{2}}\choose{y}}{{x}\choose{z}}{{y}\choose{w}}T_{\rm c,d}^{x+y}R_{\rm c,d}^{n_{1}+n_{2}-x-y}T_{\rm t}^{z+w}R_{\rm t}^{x+y-z-w}{d_{2,\rm h}^{\dagger}}^{z}{d_{2,\rm v}^{\dagger}}^{w} \nonumber \\
	&\times&{c_{1,\rm h}^{\dagger}}^{x-z}{c_{1,\rm v}^{\dagger}}^{y-w}{p_{1,\rm h}^{\dagger}}^{n_{1}-x}{q_{1,\rm v}^{\dagger}}^{n_{2}-y}\ket{0}, 
	\end{eqnarray}
\end{widetext}
where $T_{\rm t}=t^{1/2}$. 

Afterwards, we have that two Hadamard gates rotate the outgoing signals at each output port of the beamsplitter. These Hadamard gates transform the operator ${d_{2,\rm{h}}^{\dagger}}^{z}{d_{2,\rm{v}}^{\dagger}}^{w}{c_{1,\rm{h}}^{\dagger}}^{x-z}{c_{1,\rm{v}}^{\dagger}}^{y-w}$ as
\begin{align}\label{Hadamard}
&{d_{2,\rm{h}}^{\dagger}}^{z}{d_{2,\rm{v}}^{\dagger}}^{w}{c_{1,\rm{h}}^{\dagger}}^{x-z}{c_{1,\rm{v}}^{\dagger}}^{y-w}\xrightarrow{\rm{Hadamard}}\sum_{a=0}^{z}\sum_{b=0}^{w}\sum_{c=0}^{x-z}\sum_{d=0}^{y-w} \nonumber \\
&\frac{(-1)^{b+d}}{{\sqrt{2}}^{x+y}}{{z}\choose{a}}{{w}\choose{b}}{{x-z}\choose{c}}{{y-w}\choose{d}}{d_{3,\rm{h}}^{\dagger}}^{a+b}{d_{3,\rm{v}}^{\dagger}}^{z+w-a-b}  \nonumber \\
&\times{c_{2,\rm{h}}^{\dagger}}^{c+d}{c_{2,\rm{v}}^{\dagger}}^{x+y-z-w-c-d}. 
\end{align}
By using the same change of variables that we applied to Eq.~(\ref{interference}), we have that Eq.~(\ref{Hadamard}) can be rewritten as
\begin{align}\label{Hadamard_2}
&\sum_{u=0}^{z+w}\sum_{r=0}^{x+y-z-w}\sum_{v=\max\{0,u-z\}}^{\min\{w,u\}}\sum_{s=\max\{0,r+z-x\}}^{\min\{y-w,r\}}\frac{(-1)^{v+s}}{{\sqrt{2}}^{x+y}} \nonumber \\
&\times{{z}\choose{u-v}}{{w}\choose{v}}{{x-z}\choose{r-s}}{{y-w}\choose{s}}{d_{3,\rm{h}}^{\dagger}}^{u}{d_{3,\rm{v}}^{\dagger}}^{z+w-u}{c_{2,\rm{h}}^{\dagger}}^{r} \nonumber \\
&\times{c_{2,\rm{v}}^{\dagger}}^{x+y-z-w-r}.
\end{align}
By combining Eqs.~(\ref{auxiliary_states_2}) and~(\ref{Hadamard_2}), we obtain the quantum states that emerge from the PQA for the teleportation
\begin{align}\label{auxiliary_states_3}
&\ket{\psi_{n_{1},n_{2}}}_{c_{2}d_{3};p_{1}q_{1}}=\frac{1}{\sqrt{n_{1}!n_{2}!}}\sum_{x=0}^{n_{1}}\sum_{y=0}^{n_{2}}\sum_{z=0}^{x}\sum_{w=0}^{y}\sum_{u=0}^{z+w}\sum_{r=0}^{x+y-z-w} \nonumber \\
&\sum_{v=\max\{0,u-z\}}^{\min\{w,u\}}\sum_{s=\max\{0,r+z-x\}}^{\min\{y-w,r\}}\frac{(-1)^{v+s}}{{\sqrt{2}}^{x+y}}{{n_{1}}\choose{x}}{{n_{2}}\choose{y}}{{x}\choose{z}} \nonumber \\
&\times{{y}\choose{w}}{{z}\choose{u-v}}{{w}\choose{v}}{{x-z}\choose{r-s}}{{y-w}\choose{s}}T_{\rm c,d}^{x+y} \nonumber \\
&\times R_{\rm c,d}^{n_{1}+n_{2}-x-y}T_{\rm t}^{z+w}R_{\rm t}^{x+y-z-w}{d_{3,\rm{h}}^{\dagger}}^{u}{d_{3,\rm{v}}^{\dagger}}^{z+w-u}{c_{2,\rm{h}}^{\dagger}}^{r} \nonumber \\
&\times{c_{2,\rm{v}}^{\dagger}}^{x+y-z-w-r}{p_{1,\rm h}^{\dagger}}^{n_{1}-x}{q_{1,\rm v}^{\dagger}}^{n_{2}-y}\ket{0}. 
\end{align}

Then, by putting it all together, we have that the overall quantum state prior to the BSM is given by
\begin{align}\label{BSM_PQA}
&\rho_{\rm BSM}|_{\rm double\hspace{.05cm}trigger}=\sum_{n=0}^{\infty}\sum_{n_{1}=0}^{\infty}\sum_{n_{2}=0}^{\infty}{p_{n}r_{n_{1}}r_{n_{2}}} \nonumber \\
&\times{}\ket{\phi_{n}}_{a_{1}b_{1};f_{1}g_{1}}\bra{\phi_{n}}\otimes\ket{\psi_{n_{1},n_{2}}}_{c_{2}d_{3};p_{1}q_{1}}\bra{\psi_{n_{1},n_{2}}}, 
\end{align}
where the pure states $\ket{\phi_{n}}_{a_{1}b_{1};f_{1}g_{1}}\otimes\ket{\psi_{n_{1},n_{2}}}_{c_{2}d_{3};p_{1}q_{1}}$ can be written as
\begin{widetext}
	\begin{align}\label{input_PQA}
	&\ket{\phi_{n}}_{a_{1}b_{1};f_{1}g_{1}}\otimes\ket{\psi_{n_{1},n_{2}}}_{c_{2}d_{3};p_{1}q_{1}}=\frac{1}{n!\sqrt{(n+1)n_{1}!n_{2}!}}\sum_{i=0}^{n}\sum_{x=0}^{n_{1}}\sum_{y=0}^{n_{2}}\sum_{j=0}^{i}\sum_{k=0}^{i}\sum_{l=0}^{n-i}\sum_{m=0}^{n-i}\sum_{z=0}^{x}\sum_{w=0}^{y}\sum_{u=0}^{z+w}\sum_{r=0}^{x+y-z-w}\sum_{v=\max\{0,u-z\}}^{\min\{w,u\}} \nonumber \\
	&\sum_{s=\max\{0,r+z-x\}}^{\min\{y-w,r\}}\frac{(-1)^{i+v+s}}{{\sqrt{2}}^{x+y}}{{n}\choose{i}}{{n_{1}}\choose{x}}{{n_{2}}\choose{y}}{{i}\choose{j}}{{i}\choose{k}}{{n-i}\choose{l}}{{n-i}\choose{m}}{{x}\choose{z}}{{y}\choose{w}}{{z}\choose{u-v}}{{w}\choose{v}}{{x-z}\choose{r-s}}{{y-w}\choose{s}} \nonumber \\
	&\times{}T_{\rm c,d}^{j+l+x+y}R_{\rm c,d}^{n+n_{1}+n_{2}-j-l-x-y}T_{\rm c,ch,d}^{k+m}R_{\rm c,ch,d}^{n-k-m}T_{\rm t}^{z+w}R_{\rm t}^{x+y-z-w}{a_{1,\rm{h}}^{\dagger}}^{l}{a_{1,\rm{v}}^{\dagger}}^{j}{b_{1,\rm{h}}^{\dagger}}^{k}{b_{1,\rm{v}}^{\dagger}}^{m}{d_{3,\rm{h}}^{\dagger}}^{u}{d_{3,\rm{v}}^{\dagger}}^{z+w-u}{c_{2,\rm{h}}^{\dagger}}^{r}{c_{2,\rm{v}}^{\dagger}}^{x+y-z-w-r} \nonumber \\
	&\times{f_{1,\rm{h}}^{\dagger}}^{n-i-l}{f_{1,\rm{v}}^{\dagger}}^{i-j}{g_{1,\rm{h}}^{\dagger}}^{i-k}{g_{1,\rm{v}}^{\dagger}}^{n-i-m}{p_{1,\rm h}^{\dagger}}^{n_{1}-x}{q_{1,\rm v}^{\dagger}}^{n_{2}-y}\ket{0}. 
	\end{align}
\end{widetext}

Next, modes $b_{1}$ and $c_{2}$ interfere at the 50:50 BS within the PQA. This BS transforms the state $\ket{\phi_{n}}_{a_{1}b_{1};f_{1}g_{1}}\otimes\ket{\psi_{n_{1},n_{2}}}_{c_{2}d_{3};p_{1}q_{1}}$ into the state $\ket{\phi_{n,n_{1},n_{2}}}_{a_{1}b_{2}c_{3}d_{3};f_{1}g_{1}p_{1}q_{1}}$ given by
\begin{widetext}
	\begin{align}\label{interference_PQA}
	&\ket{\phi_{n,n_{1},n_{2}}}_{a_{1}b_{2}c_{3}d_{3};f_{1}g_{1}p_{1}q_{1}}=\frac{1}{n!\sqrt{(n+1)n_{1}!n_{2}!}}\sum_{i=0}^{n}\sum_{x=0}^{n_{1}}\sum_{y=0}^{n_{2}}\sum_{j=0}^{i}\sum_{k=0}^{i}\sum_{l=0}^{n-i}\sum_{m=0}^{n-i}\sum_{z=0}^{x}\sum_{w=0}^{y}\sum_{u=0}^{z+w}\sum_{r=0}^{x+y-z-w}\sum_{v=\max\{0,u-z\}}^{\min\{w,u\}} \nonumber \\
	&\sum_{s=\max\{0,r+z-x\}}^{\min\{y-w,r\}}\sum_{a=0}^{r}\sum_{b=0}^{x+y-z-w-r}\sum_{c=0}^{k}\sum_{d=0}^{m} \frac{(-1)^{i+v+s+a+b}}{{\sqrt{2}}^{2(x+y)+k+m-z-w}}{{n}\choose{i}}{{n_{1}}\choose{x}}{{n_{2}}\choose{y}}{{i}\choose{j}}{{i}\choose{k}}{{n-i}\choose{l}}{{n-i}\choose{m}}{{x}\choose{z}}{{y}\choose{w}} \nonumber \\
	&\times{{z}\choose{u-v}}{{w}\choose{v}}{{x-z}\choose{r-s}}{{y-w}\choose{s}}{{r}\choose{a}}{{x+y-z-w-r}\choose{b}}{{k}\choose{c}}{{m}\choose{d}}T_{\rm c,d}^{j+l+x+y}R_{\rm c,d}^{n+n_{1}+n_{2}-j-l-x-y}T_{\rm c,ch,d}^{k+m}R_{\rm c,ch,d}^{n-k-m}\nonumber \\
	&\times{}T_{\rm t}^{z+w}R_{\rm t}^{x+y-z-w}{a_{1,\rm{h}}^{\dagger}}^{l}{a_{1,\rm{v}}^{\dagger}}^{j}{d_{3,\rm{h}}^{\dagger}}^{u}{d_{3,\rm{v}}^{\dagger}}^{z+w-u}{b_{2,\rm{h}}^{\dagger}}^{a+c}{b_{2,\rm{v}}^{\dagger}}^{b+d}{c_{3,\rm{h}}^{\dagger}}^{k+r-a-c}{c_{3,\rm{v}}^{\dagger}}^{x+y+m-z-w-r-b-d}{f_{1,\rm{h}}^{\dagger}}^{n-i-l}{f_{1,\rm{v}}^{\dagger}}^{i-j}{g_{1,\rm{h}}^{\dagger}}^{i-k}\nonumber \\
	&\times{g_{1,\rm{v}}^{\dagger}}^{n-i-m}{p_{1,\rm h}^{\dagger}}^{n_{1}-x}{q_{1,\rm v}^{\dagger}}^{n_{2}-y}\ket{0}. 
	\end{align}
\end{widetext}

As in the case of the ESR-based setup, Alice's and Bob's measurement settings can be incorporated at this stage by rotating the pairs of operators $(a_{1,\rm{h}}^{\dagger},a_{1,\rm{v}}^{\dagger})$ and $(d_{3,\rm{h}}^{\dagger},d_{3,\rm{v}}^{\dagger})$ with angles $\theta_{\rm{A}}$ and $\theta_{\rm{B}}$, respectively. We denote the resulting pure state by $\ket{\phi_{n,n_{1},n_{2}}}_{a_{2}b_{2}c_{3}d_{4};f_{1}g_{1}p_{1}q_{1}}^{\theta_{\rm A},\theta_{\rm B}}$, and it is given by
\begin{widetext}
	\begin{align}\label{detectors_PQA}
	&\ket{\phi_{n,n_{1},n_{2}}}_{a_{2}b_{2}c_{3}d_{4};f_{1}g_{1}p_{1}q_{1}}^{\theta_{\rm A},\theta_{\rm B}}=\frac{1}{n!\sqrt{(n+1)n_{1}!n_{2}!}}\sum_{i=0}^{n}\sum_{x=0}^{n_{1}}\sum_{y=0}^{n_{2}}\sum_{j=0}^{i}\sum_{k=0}^{i}\sum_{l=0}^{n-i}\sum_{m=0}^{n-i}\sum_{z=0}^{x}\sum_{w=0}^{y}\sum_{u=0}^{z+w}\sum_{r=0}^{x+y-z-w}\sum_{v=\max\{0,u-z\}}^{\min\{w,u\}} \nonumber \\
	&\sum_{s=\max\{0,r+z-x\}}^{\min\{y-w,r\}}\sum_{a=0}^{r}\sum_{b=0}^{x+y-z-w-r}\sum_{c=0}^{k}\sum_{d=0}^{m}\sum_{e=0}^{l+j}\sum_{h=\max\{0,e-l\}}^{\min\{e,j\}}\sum_{o=0}^{z+w}\sum_{\tilde{n}=\max\{0,o-u\}}^{\min\{o,z+w-u\}}\frac{(-1)^{i+v+s+a+b+h+\tilde{n}}}{{\sqrt{2}}^{2(x+y)+k+m-z-w}}{{n}\choose{i}}{{n_{1}}\choose{x}}{{n_{2}}\choose{y}}{{i}\choose{j}} \nonumber \\
	&\times{{i}\choose{k}}{{n-i}\choose{l}}{{n-i}\choose{m}}{{x}\choose{z}}{{y}\choose{w}}{{z}\choose{u-v}}{{w}\choose{v}}{{x-z}\choose{r-s}}{{y-w}\choose{s}}{{r}\choose{a}}{{x+y-z-w-r}\choose{b}}{{k}\choose{c}}{{m}\choose{d}}{{l}\choose{e-h}}{{j}\choose{h}} \nonumber \\
	&\times{{u}\choose{o-\tilde{n}}}{{z+w-u}\choose{\tilde{n}}}T_{\rm c,d}^{j+l+x+y}R_{\rm c,d}^{n+n_{1}+n_{2}-j-l-x-y}T_{\rm c,ch,d}^{k+m}R_{\rm c,ch,d}^{n-k-m}T_{\rm t}^{z+w}R_{\rm t}^{x+y-z-w}\cos{\theta_{\rm A}}^{e+j-2h}\sin{\theta_{\rm A}}^{l+2h-e} \nonumber \\
	&\times\cos{\theta_{\rm B}}^{o+z+w-u-2\tilde{n}}\sin{\theta_{\rm B}}^{u+2\tilde{n}-o}{a_{2,\rm{h}}^{\dagger}}^{e}{a_{2,\rm{v}}^{\dagger}}^{l+j-e}{d_{4,\rm{h}}^{\dagger}}^{o}{d_{4,\rm{v}}^{\dagger}}^{z+w-o}{b_{2,\rm{h}}^{\dagger}}^{a+c}{b_{2,\rm{v}}^{\dagger}}^{b+d}{c_{3,\rm{h}}^{\dagger}}^{k+r-a-c}{c_{3,\rm{v}}^{\dagger}}^{x+y+m-z-w-r-b-d} \nonumber \\
	&\times{f_{1,\rm{h}}^{\dagger}}^{n-i-l}{f_{1,\rm{v}}^{\dagger}}^{i-j}{g_{1,\rm{h}}^{\dagger}}^{i-k}{g_{1,\rm{v}}^{\dagger}}^{n-i-m}{p_{1,\rm h}^{\dagger}}^{n_{1}-x}{q_{1,\rm v}^{\dagger}}^{n_{2}-y}\ket{0}.
	\end{align}
\end{widetext}

Then, in an identical way as in the ESR-based setup, the probability ${{\rm{P}}\left(\vec{\alpha}|n,n_{1},n_{2}\right)}_{\theta_{\rm A},\theta_{\rm B}}$ that a specific click pattern $\vec{\alpha}=(\alpha,\beta,\gamma,\delta,\mu,\nu,\tau,\lambda)$ occurs with the state written in Eq.~(\ref{detectors_PQA}) is given by 
\begin{equation}\label{qwev}
{{\rm{P}}\left(\vec{\alpha}|n,n_{1},n_{2}\right)}_{\theta_{\rm A},\theta_{\rm B}}=\left\lVert{\ket{\tilde{\phi}_{n,n_{1},n_{2}}}_{\vec{\alpha};f_{1}g_{1}p_{1}q_{1}}^{\theta_{\rm A},\theta_{\rm B}}}\right\rVert^{2},
\end{equation}
where the unnormalized state $\ket{\tilde{\phi}_{n,n_{1},n_{2}}}_{\vec{\alpha};f_{1}g_{1}p_{1}q_{1}}^{\theta_{\rm A},\theta_{\rm B}}$ has the form
\begin{equation}
\ket{\tilde{\phi}_{n,n_{1},n_{2}}}_{\vec{\alpha};f_{1}g_{1}p_{1}q_{1}}^{\theta_{\rm A},\theta_{\rm B}}={\braket{\vec{\alpha}}{\phi_{n,n_{1},n_{2}}}}_{a_{2}b_{2}c_{3}d_{4};f_{1}g_{1}p_{1}q_{1}}^{\theta_{\rm A},\theta_{\rm B}}, \quad\quad
\end{equation}
being $\ket{\vec{\alpha}}=\ket{\alpha,\beta,\gamma,\delta,\mu,\nu,\tau,\lambda}$.

After addressing the relevant orthogonality relation, given by
\begin{align}\label{orthogonality_PQA}
&\bra{\vec{\alpha}}{a_{2,\rm{h}}^{\dagger}}^{e}{a_{2,\rm{v}}^{\dagger}}^{l+j-e}{d_{4,\rm{h}}^{\dagger}}^{o}{d_{4,\rm{v}}^{\dagger}}^{z+w-o}{c_{3,\rm{h}}^{\dagger}}^{k+r-a-c} \nonumber \\
&\times{c_{3,\rm{v}}^{\dagger}}^{x+y+m-z-w-r-b-d}{b_{2,\rm{h}}^{\dagger}}^{a+c}{b_{\rm{v}}^{\dagger}}^{b+d}\ket{0} \nonumber \\
&=\left({\alpha!}{\beta!}{\gamma!}{\delta!}{\mu!}{\nu!}{\tau!}{\lambda!}\right)^{1/2}\delta_{\alpha}^{e}\delta_{\beta}^{l+j-e}\delta_{\gamma}^{o}\delta_{\delta}^{z+w-o}\delta_{\mu}^{k+r-a-c}\nonumber \\
&\times\delta_{\nu}^{x+y+m-z-w-r-b-d}\delta_{\tau}^{a+c}\delta_{\lambda}^{b+d}, 
\end{align}
one finds that the state $\ket{\tilde{\phi}_{n,n_{1},n_{2}}}_{\vec{\alpha};f_{1}g_{1}p_{1}q_{1}}^{\theta_{\rm A},\theta_{\rm B}}$ can be written as
\begin{widetext}
	\begin{align}\label{unnormalized_PQA}
	&\ket{\tilde{\phi}_{n,n_{1},n_{2}}}_{\vec{\alpha};f_{1}g_{1}p_{1}q_{1}}^{\theta_{\rm A},\theta_{\rm B}}=\frac{1}{n!}\left[\frac{{\alpha!}{\beta!}{\gamma!}{\delta!}{\mu!}{\nu!}{\tau!}{\lambda!}}{(n+1)n_{1}!n_{2}!2^{\mu+\nu+\tau+\lambda}}\right]^{\frac{1}{2}}T_{\rm c,d}^{\alpha+\beta}R_{\rm c,d}^{n+n_{1}+n_{2}-\alpha-\beta}T_{\rm c,ch,d}^{\gamma+\delta+\mu+\nu+\tau+\lambda}R_{\rm c,ch,d}^{n-\gamma-\delta-\mu-\nu-\tau-\lambda}T_{\rm t}^{\gamma+\delta} \nonumber \\
	&\times{}R_{\rm t}^{-\gamma-\delta}\cos{\theta_{\rm A}}^{\alpha}\sin{\theta_{\rm A}}^{\beta}\cos{\theta_{\rm B}}^{2\gamma+\delta}\sin{\theta_{\rm B}}^{-\gamma}\sum_{i=0}^{n}\sum_{x=0}^{n_{1}}\sum_{y=0}^{n_{2}}\sum_{k=\max\{0,\gamma+\delta+\mu+\tau-x-y,\gamma+\delta+\mu+\nu+\tau+\lambda+i-n-x-y\}}^{\min\{i,\mu+\tau,\gamma+\delta+\mu+\nu+\tau+\lambda-x-y\}} \nonumber \\
	&\times\sum_{j=\max\{0,\alpha+\beta+i-n\}}^{\min\{i,\alpha+\beta\}}\sum_{w=\max\{0,\gamma+\delta-x\}}^{\min\{y,\gamma+\delta\}}\sum_{u=0}^{\gamma+\delta}\sum_{v=\max\{0,u+w-\gamma-\delta\}}^{\min\{w,u\}}\sum_{s=\max\{0,\mu+\tau+\gamma+\delta-x-k-w\}}^{\min\{y-w,\mu+\tau-k\}}\sum_{a=\max\{0,\tau-k\}}^{\min\{\tau,\mu+\tau-k\}} \nonumber \\
	&\times\sum_{b=\max\{0,x+y+k-\gamma-\delta-\mu-\nu-\tau\}}^{\min\{\lambda,x+y+k-\gamma-\delta-\mu-\tau\}}\sum_{h=\max\{0,j-\beta\}}^{\min\{\alpha,j\}}\sum_{\tilde{n}=\max\{0,\gamma-u\}}^{\min\{\gamma,\gamma+\delta-u\}}\frac{(-1)^{i+v+s+a+b+h+\tilde{n}}}{{\sqrt{2}}^{x+y}}{{n}\choose{i}}{{n_{1}}\choose{x}}{{n_{2}}\choose{y}}{{i}\choose{j}}{{i}\choose{k}}{{n-i}\choose{\alpha+\beta-j}} \nonumber \\
	&\times{{x}\choose{\gamma+\delta-w}}{{y}\choose{w}}{{y-w}\choose{s}}{{n-i}\choose{\gamma+\delta+\mu+\nu+\tau+\lambda-x-y-k}}{{\gamma+\delta-w}\choose{u-v}}{{w}\choose{v}}{{x+w-\gamma-\delta}\choose{\mu+\tau-k-s}}{{\mu+\tau-k}\choose{a}} \nonumber \\
	&\times{{j}\choose{h}}{{u}\choose{\gamma-\tilde{n}}}{{\gamma+\delta-u}\choose{\tilde{n}}}{{k}\choose{\tau-a}}{{\gamma+\delta+\mu+\nu+\tau+\lambda-x-y-k}\choose{\lambda-b}}{{\alpha+\beta-j}\choose{\alpha-h}}{{x+y+k-\gamma-\delta-\mu-\tau}\choose{b}} \nonumber \\
	&\times\left(\frac{T_{\rm c,d}R_{\rm c,ch,d}R_{\rm t}}{T_{\rm c,ch,d}R_{\rm c,d}}\right)^{x+y}\left(\frac{\sin{\theta_{\rm A}}}{\cos{\theta_{\rm A}}}\right)^{2h-j}\left(\frac{\sin{\theta_{\rm B}}}{\cos{\theta_{\rm B}}}\right)^{2\tilde{n}+u}{f_{1,\rm{h}}^{\dagger}}^{n+j-i-\alpha-\beta}{f_{1,\rm{v}}^{\dagger}}^{i-j}{g_{1,\rm{h}}^{\dagger}}^{i-k}{g_{1,\rm{v}}^{\dagger}}^{n+x+y+k-i-\gamma-\delta-\mu-\nu-\tau-\lambda} \nonumber \\
	&\times{p_{1,\rm h}^{\dagger}}^{n_{1}-x}{q_{1,\rm v}^{\dagger}}^{n_{2}-y}\ket{0}. 
	\end{align}
\end{widetext}

Finally, by taking the squared norm of the previous state as decribed in Eq.~(\ref{qwev}), we obtain that
\begin{widetext}
	\begin{align}\label{distribution_PQA}
	&{{\rm{P}}\left(\vec{\alpha}|n,n_{1},n_{2}\right)}_{\theta_{\rm A},\theta_{\rm B}}=\frac{{\alpha!}{\beta!}{\gamma!}{\delta!}{\mu!}{\nu!}{\tau!}{\lambda!}}{(n+1)2^{\mu+\nu+\tau+\lambda}}\zeta_{\rm c,d}^{\alpha+\beta}(1-\zeta_{\rm c,d})^{n+n_{1}+n_{2}-\alpha-\beta}\zeta_{\rm c,ch,d}^{\gamma+\delta+\mu+\nu+\tau+\lambda}(1-\zeta_{\rm c,ch,d})^{n-\gamma-\delta-\mu-\nu-\tau-\lambda} \nonumber \\
	&\times\left(\frac{t}{1-t}\right)^{\gamma+\delta}\cos{\theta_{\rm A}}^{2\alpha}\sin{\theta_{\rm A}}^{2\beta}\cos{\theta_{\rm B}}^{2(2\gamma+\delta)}\sin{\theta_{\rm B}}^{-2\gamma}\sum_{i=0}^{n}\sum_{\Delta=-i}^{n-i}\sum_{x=0}^{n_{1}}\sum_{y=0}^{n_{2}}\sum_{j=\max\{0,\alpha+\beta+i-n,-\Delta\}}^{\min\{i,\alpha+\beta,\alpha+\beta-\Delta\}} \nonumber \\
	&\times\sum_{k=\max\{0,\gamma+\delta+\mu+\tau-x-y,\gamma+\delta+\mu+\nu+\tau+\lambda+i-n-x-y,-\Delta,\gamma+\delta+\mu+\tau-x-y-\Delta\}}^{\min\{i,\mu+\tau,\gamma+\delta+\mu+\nu+\tau+\lambda-x-y,\mu+\tau-\Delta,\gamma+\delta+\mu+\nu+\tau+\lambda-x-y-\Delta\}}\sum_{w=\max\{0,\gamma+\delta-x\}}^{\min\{y,\gamma+\delta\}}\sum_{W=\max\{0,\gamma+\delta-x\}}^{\min\{y,\gamma+\delta\}}\sum_{u=0}^{\gamma+\delta}\sum_{U=0}^{\gamma+\delta} \nonumber \\
	&\times\sum_{v=\max\{0,u+w-\gamma-\delta\}}^{\min\{w,u\}}\sum_{V=\max\{0,U+W-\gamma-\delta\}}^{\min\{W,U\}}\sum_{s=\max\{0,\mu+\tau+\gamma+\delta-x-k-w\}}^{\min\{y-w,\mu+\tau-k\}}\sum_{S=\max\{0,\mu+\tau+\gamma+\delta-x-k-\Delta-W\}}^{\min\{y-W,\mu+\tau-k-\Delta\}}\sum_{a=\max\{0,\tau-k\}}^{\min\{\tau,\mu+\tau-k\}} \nonumber \\
	&\times\sum_{A=\max\{0,\tau-k-\Delta\}}^{\min\{\tau,\mu+\tau-k-\Delta\}}\sum_{b=\max\{0,x+y+k-\gamma-\delta-\mu-\nu-\tau\}}^{\min\{\lambda,x+y+k-\gamma-\delta-\mu-\tau\}}\sum_{B=\max\{0,x+y+k+\Delta-\gamma-\delta-\mu-\nu-\tau\}}^{\min\{\lambda,x+y+k+\Delta-\gamma-\delta-\mu-\tau\}}\sum_{h=\max\{0,j-\beta\}}^{\min\{\alpha,j\}}\sum_{H=\max\{0,j+\Delta-\beta\}}^{\min\{\alpha,j+\Delta\}} \nonumber \\
	&\times\sum_{\tilde{n}=\max\{0,\gamma-u\}}^{\min\{\gamma,\gamma+\delta-u\}}\sum_{\tilde{N}=\max\{0,\gamma-U\}}^{\min\{\gamma,\gamma+\delta-U\}}(-1)^{v+s+a+b+h+\tilde{n}+V+S+A+B+H+\tilde{N}+\Delta}\left(\frac{\zeta_{\rm c,d}(1-\zeta_{\rm c,ch,d})(1-t)}{2\zeta_{\rm c,ch,d}(1-\zeta_{\rm c,d})}\right)^{x+y} \nonumber \\
	&\times\left(\frac{\sin{\theta_{\rm A}}}{\cos{\theta_{\rm A}}}\right)^{2(h+H-j)-\Delta}\left(\frac{\sin{\theta_{\rm B}}}{\cos{\theta_{\rm B}}}\right)^{2(\tilde{n}+\tilde{N})+u+U} \nonumber \\
	&\times\Upsilon(n,n_{1},n_{2},i,x,y,j,k,w,W,u,U,v,V,s,S,a,A,b,B,h,H,\tilde{n},\tilde{N},\Delta,\alpha,\beta,\gamma,\delta,\mu,\nu,\tau,\lambda), 
	\end{align}
where
\begin{align}\label{upsilon_PQA}
	&\Upsilon(n,n_{1},n_{2},i,x,y,j,k,w,W,u,U,v,V,s,S,a,A,b,B,h,H,\tilde{n},\tilde{N},\Delta,\alpha,\beta,\gamma,\delta,\mu,\nu,\tau,\lambda)={n_{1}}!{n_{2}}!{i}!(i+\Delta)!(n-i)! \nonumber \\
	&\times(n-i-\Delta)!{{\mu+\tau-k}\choose{a}}{{\mu+\tau-k-\Delta}\choose{A}}{{x+y+k-\gamma-\delta-\mu-\tau}\choose{b}}{{x+y+k+\Delta-\gamma-\delta-\mu-\tau}\choose{B}}{{u}\choose{\gamma-\tilde{n}}} \nonumber \\
	&\times{{U}\choose{\gamma-\tilde{N}}}{{\gamma+\delta-u}\choose{\tilde{n}}}{{\gamma+\delta-U}\choose{\tilde{N}}}\left[(\alpha-h)!(\alpha-H)!(\beta+h-j)!(\beta+H-j-\Delta)!h!H!(j-h)!(j+\Delta-H)!\right]^{-1} \nonumber \\
	&\times\frac{\left[(n_{1}-x)!(n_{2}-y)!(i-j)!(i-k)!(n+j-i-\alpha-\beta)!(n+x+y+k-i-\gamma-\delta-\mu-\nu-\tau-\lambda)!(u-v)!\right]^{-1}}{(\gamma+\delta+v-u-w)!(\gamma+\delta+V-U-W)!v!V!(w-v)!(W-V)!(\mu+\tau-k-s)!(\mu+\tau-k-\Delta-S)!} \nonumber \\
	&\times\frac{\left[(x+w+k+s-\gamma-\delta-\mu-\tau)!(x+W+k+\Delta+S-\gamma-\delta-\mu-\tau)!(y-w-s)!(y-W-S)!s!S!(\tau-a)!\right]^{-1}}{(k+a-\tau)!(k+\Delta+A-\tau)!(b+\gamma+\delta+\mu+\nu+\tau-x-y-k)!(B+\gamma+\delta+\mu+\nu+\tau-x-y-k-\Delta)!} \nonumber \\
	&\times\frac{\left[(U-V)!(\tau-A)!\right]^{-1}}{(\lambda-b)!(\lambda-B)!}. 
	\end{align}
\end{widetext}
The normalization condition $\sum_{\vec{\alpha}}{{\rm{P}}\left(\vec{\alpha}|n,n_{1},n_{2}\right)}_{\theta_{\rm A},\theta_{\rm B}}=1$ holds for any set of physical parameters $\theta_{\rm A}$, $\theta_{\rm B}$, $\zeta_{\rm c,d}$, $\zeta_{\rm c,ch,d}$ and $t$, and only those click patterns $\vec{\alpha}$ such that $\alpha+\beta\leq{n}$, $\gamma+\delta\leq{n_{1}+n_{2}}$ and $\gamma+\delta+\mu+\nu+\tau+\lambda\leq{n+n_{1}+n_{2}}$ have a nonzero probability to happen due to the fact that for the moment we have disregarded dark counts.

As a final step, we need to define the click pattern distribution in the noisy scenario, ${\tilde{\rm{p}}\left(\vec{\alpha}|n,n_{1},n_{2}\right)}_{\theta_{\rm A},\theta_{\rm B}}$, as well as the post-processed click pattern distribution, $\mathbf{P}(A_{\vec{\alpha}}|n,n_{1},n_{2})_{\theta_{\rm A},\theta_{\rm B}}$ (with $A_{\vec{\alpha}}={(A_{\rm A},A_{\rm B},\mu,\nu,\tau,\lambda)}$). This is exactly analogous to what we did for the  ESR-based setup in Appendix~\ref{ESR_1}, and we omit the details here for simplicity. The only difference is that no permutation step $\alpha\leftrightarrow\beta$ is performed in this case. This is so because, conditioned on a successful heralding event, Alice's and Bob's outcomes are expected to be directly correlated in this case.

\subsubsection{Parameters $P_{\rm SH}$, $Q|_{\rm SH}$ and $\omega|_{\rm SH}$}\label{PQA_2}

To determine $P_{\rm SH}$, we note that any given detection event is discarded unless a trigger is observed at the idler mode of both single photon sources and, at the same time, a success occurs at the PQA. Therefore, and due to the symmetries of the channel model under consideration, we can define, say the event $\Omega={\left\{(\mu,\nu,\tau,\lambda)=(1,1,0,0)\right\}}$ (as we did in Appendix~\ref{ESR_2}) and thus the overall success probability is given by
\begin{equation}
P_{\rm SH}=4P_{\Omega,\rm double \hspace{.05cm} trigger}=4P_{\rm trigger}^{2}P_{\Omega|\rm double \hspace{.05cm} trigger}, 
\end{equation}
where the conditional probability $P_{\Omega|\rm double \hspace{.05cm} trigger}$ has the form
\begin{align}\label{success_prob_PQA}
&P_{\Omega|\rm double \hspace{.05cm} trigger}=\sum_{A_{\rm A},A_{\rm B}}\mathbf{P}(A_{\rm A},A_{\rm B},\Omega|\rm double \hspace{.05cm} trigger)_{\theta_{\rm A},\theta_{\rm B}} \nonumber \\
&=\sum_{n,n_{1},n_{2}}{p_{n}r_{n_{1}}r_{n_{2}}}\sum_{A_{\rm A},A_{\rm B}}\mathbf{P}(A_{\rm A},A_{\rm B},\Omega|n,n_{1},n_{2})_{\theta_{\rm A},\theta_{\rm B}}, \nonumber \\
\end{align}
with $A_{\rm A},A_{\rm B}\in\{0,1\}$. On the other hand, we have that the conditional QBER reads
\begin{align}\label{QBER_PQA}
&Q|_{\Omega,\rm double \hspace{.05cm} trigger}=\frac{\mathbf{P}(0,1,\Omega)_{0,0}+\mathbf{P}(1,0,\Omega)_{0,0}}{P_{\Omega|\rm double \hspace{.05cm} trigger}} \nonumber \\
&=\frac{1}{P_{\Omega|\rm double \hspace{.05cm} trigger}}\sum_{n,n_{1},n_{2}}{p_{n}r_{n_{1}}r_{n_{2}}}\big[\mathbf{P}(0,1,\Omega|n,n_{1},n_{2})_{0,0} \nonumber \\
&+\mathbf{P}(1,0,\Omega|n,n_{1},n_{2})_{0,0}\big]. 
\end{align}

Similarly, the conditional winning probability reads $\omega|_{\Omega,\rm double \hspace{.05cm} trigger}=S|_{\Omega,\rm double \hspace{.05cm} trigger}/8+1/2$, where
\begin{eqnarray}\label{violation_PQA}
S|_{\Omega,\rm double \hspace{.05cm} trigger}&=&E_{0,-\frac{\pi}{8}}|_{\Omega}+E_{0,\frac{\pi}{8}}|_{\Omega}-E_{\frac{\pi}{4},-\frac{\pi}{8}}|_{\Omega}, \nonumber \\
&+&E_{\frac{\pi}{4},\frac{\pi}{8}}|_{\Omega}
\end{eqnarray}
and the quantities $E_{\theta_{\rm A},\theta_{\rm B}}|_{\Omega}$ are given by
\begin{align}\label{violation_summand_PQA}
&E_{\theta_{\rm A},\theta_{\rm B}}|_{\Omega}=\frac{2}{P_{\Omega|\rm double \hspace{.05cm} trigger}}\sum_{n,n_{1},n_{2}}{p_{n}r_{n_{1}}r_{n_{2}}}\nonumber \\
&\times[\mathbf{P}(0,0,\Omega|n,n_{1},n_{2})_{\theta_{\rm A},\theta_{\rm B}}+\mathbf{P}(1,1,\Omega|n,n_{1},n_{2})_{\theta_{\rm A},\theta_{\rm B}}] \nonumber \\
&-1. 
\end{align}
Note that in Eq.~(\ref{violation_PQA}) the summand that carries the minus sign is different from that of Eq.~(\ref{violation}). As already explained in Appendix~\ref{ESR_2}, the definition of the conditional CHSH violation depends on the particular Bell pair shared by the parties after a successful BSM~\footnote{Again, also here two successful heralding events require the given definition of $S|_{\Omega,\rm double \hspace{.05cm} trigger}$, while the other two require the minus sign to be carried by the fourth summand in Eq.~(\ref{violation_PQA}).}.

\subsection{DIQKD without heralding mechanism}\label{unassisted_scenario}
\begin{figure}[!htbp]
	\includegraphics[width=7.5cm,height=4.5cm]{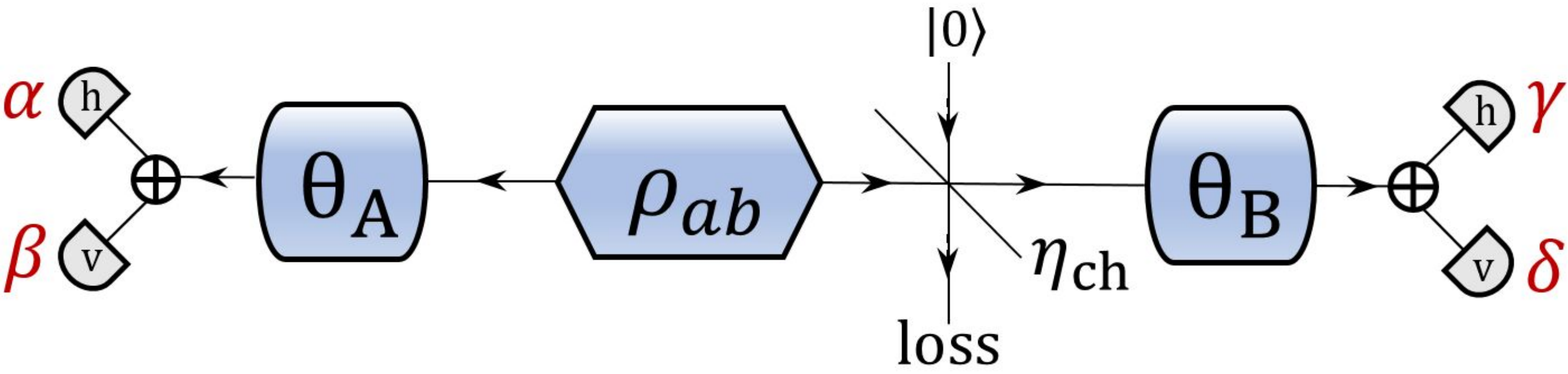} 
	\caption{Schematic of the DIQKD setup without a qubit amplifier. Since we are interested in determining the maximum achievable distance, detection and coupling efficiencies are assumed to be perfect, {\it i.e.}, $\eta_{\rm c}=\eta_{\rm d}=1$, and detector noise is set to zero, {\it i.e.}, $p_{\rm d}=0$. Again, $\eta_{\rm ch}$ tags the transmission efficiency of the channel, $\eta_{\rm ch}=10^{-\Lambda/10}$, the symbol ``$\oplus$'' represents the PBSs, the greek letters in red color identify the number of photons observed at each of the detectors and $\ket{0}$ denotes the vacuum state.}
	\label{fig:unassisted}
\end{figure}

In this Appendix, we calculate the maximum channel loss that a photonic DIQKD implementation can tolerate in the absence of an heralding mechanism. For that purpose, we consider the setup where Bob does not hold a qubit amplifier in his lab, as shown by Fig.~\ref{fig:unassisted}. Also, since we are interested in the maximum achievable distance, we further assume perfect coupling and detection efficiencies, {\it i.e.}, $\eta_{\rm c}=\eta_{\rm d}=1$, and no detector noise, {\it i.e.}, $p_{\rm d}=0$. In this way, channel loss is the only source of loss that we contemplate, modeled by a transmission efficiency $\eta_{\rm ch}=10^{-\Lambda/10}$ as usual. 

In the case of a perfect entanglement source, $\rho_{ab}=\ket{\phi_{1}}_{ab}\bra{\phi_{1}}$ (see Eq.~(\ref{2})) we have that the parameters of the honest implementation are simply given by $Q=(1-\eta_{\rm ch})/2$ and $\omega=(\sqrt{2}\eta_{\rm ch}+2)/4$, so that the maximum tolerated channel loss is given by $\Lambda_{\rm max}=\max\{\Lambda\in\mathbb{R}^{+}|K_{\infty}\geq{0}\}\approx{0.7}\rm dB$, where $K_{\infty}$ is the asymptotic secret key rate in the absence of a qubit amplifier,
\begin{align}\label{asymptotic_key}
K_{\infty}=1-h\left[\frac{1}{2}+\frac{1}{2}\sqrt{16\omega(\omega-1)+3}\right]-h(Q).
\end{align}
If we assume, for instance, an attenuation coefficient of, say, $\alpha=0.2$ dB/km, which corresponds to the typical value for single-mode fibers in the telecom wavelength, then $\Lambda_{\rm max}=0.7$ dB means that the maximum achievable distance between the parties in the DIQKD link is $L_{\rm max}=3.5$ km.

Let us now consider the case where $\rho_{ab}$ is a more general entanglement source, $\rho_{ab}=\ket{\psi}_{ab}\bra{\psi}$, with $\ket{\psi}_{ab}$ given by Eq.~(\ref{1}) and some arbitrary photon number statistics $p_{n}$. Then, by using the same techniques employed in Appendix~\ref{ESR} and in Appendix~\ref{PQA}, it is possible to derive the click pattern distribution ${{\rm{P}}\left(\alpha,\beta,\gamma,\delta|n\right)}_{\theta_{\rm A},\theta_{\rm B}}$ that matches the setup of Fig.~(\ref{fig:unassisted}), conditioned on the number of photon pairs $n$ emitted by the source. Precisely, one finds 
\begin{equation}
{{\rm{P}}\left(\alpha,\beta,\gamma,\delta|n\right)}_{\theta_{\rm A},\theta_{\rm B}}=\delta_{\alpha+\beta}^{n}{\tilde{\rm{P}}\left(\alpha,\beta,\gamma,\delta|n\right)}_{\theta_{\rm A},\theta_{\rm B}}, 
\end{equation}
where
\begin{widetext}
	\begin{align}\label{unassisted}
	&{\tilde{\rm{P}}\left(\alpha,\beta,\gamma,\delta|n\right)}_{\theta_{\rm A},\theta_{\rm B}}=\frac{{\alpha!}{\beta!}{\gamma!}{\delta!}}{n+1}\eta_{\rm ch}^{\gamma+\delta}(1-\eta_{\rm ch})^{n-\gamma-\delta}\cos{\theta_{\rm A}}^{2\alpha}\sin{\theta_{\rm A}}^{2(n-\alpha)}\cos{\theta_{\rm B}}^{2\gamma}\sin{\theta_{\rm B}}^{2\delta}\sum_{j=0}^{n}\sum_{\Delta=-j}^{n-j}\sum_{l=\max\{0,\gamma+\delta-j,\Delta\}}^{\min\{n-j,\gamma+\delta,\gamma+\delta+\Delta\}} \nonumber \\
	&\times\sum_{h=\max\{0,\alpha+j-n\}}^{\min\{\alpha,j\}}\sum_{H=\max\{0,\alpha+j+\Delta-n\}}^{\min\{\alpha,j+\Delta\}}\sum_{s=\max\{0,l-\delta\}}^{\min\{\gamma,l\}}\sum_{S=\max\{0,l-\delta-\Delta\}}^{\min\{\gamma,l-\Delta\}}(-1)^{h+H+s+S+\Delta}\left(\frac{\sin{\theta_{\rm A}}}{\cos{\theta_{\rm A}}}\right)^{2(h+H-j)-\Delta} \nonumber \\
	&\times\left(\frac{\sin{\theta_{\rm B}}}{\cos{\theta_{\rm B}}}\right)^{2(s+S-l)+\Delta}\Upsilon(n,j,l,h,H,s,S,\alpha,\beta,\gamma,\delta,\Delta),
\end{align}
\end{widetext}
with the parameter $\Upsilon(n,j,l,h,H,s,S,\alpha,\beta,\gamma,\delta,\Delta)$ being of the form
\begin{widetext}
	\begin{align}\label{upsilon_unassisted}
	&\Upsilon(n,j,l,h,H,s,S,\alpha,\beta,\gamma,\delta,\Delta)={{n-j}\choose{\alpha-h}}{{n-j-\Delta}\choose{\alpha-H}}{{j}\choose{h}}{{j+\Delta}\choose{H}}\left[(j+l-\gamma-\delta)!(n-j-l)!(\gamma-s)!(\gamma-S)!\right]^{-1} \nonumber \\
	&\times\left[(s+\delta-l)!(S+\delta+\Delta-l)!s!S!(l-s)!(l-\Delta-S)!\right]^{-1}.
	\end{align}
\end{widetext}
Then, summing over $n$ and taking into account the $\delta_{\alpha+\beta}^{n}$ factor, we obtain that the overall distribution reads
\begin{align}\label{overall}
&{\rm{P}}\left(\alpha,\beta,\gamma,\delta\right)_{\theta_{\rm A},\theta_{\rm B}}=\sum_{n=0}^{\infty}p_{n}{{\rm{P}}\left(\alpha,\beta,\gamma,\delta|n\right)}_{\theta_{\rm A},\theta_{\rm B}} \nonumber \\
&=p_{\alpha+\beta}{\tilde{\rm{P}}\left(\alpha,\beta,\gamma,\delta|\alpha+\beta\right)}_{\theta_{\rm A},\theta_{\rm B}}.
\end{align}
As in Appendix~(\ref{ESR_1}), a permutation step is required to enforce correlation between Alice's and Bob's outcomes. Therefore, we define
\begin{equation}\label{permutation_unassisted}
{\tilde{\rm{p}}\left(\alpha,\beta,\gamma,\delta\right)}_{\theta_{\rm A},\theta_{\rm B}}={\rm{P}}\left(\beta,\alpha,\gamma,\delta\right)_{\theta_{\rm A},\theta_{\rm B}}. 
\end{equation}
Given ${\tilde{\rm{p}}\left(\alpha,\beta,\gamma,\delta\right)}_{\theta_{\rm A},\theta_{\rm B}}$, one can readily define the post-processed click pattern distribution $\mathbf{P}(A_{\rm A},A_{\rm B})_{\theta_{\rm A},\theta_{\rm B}}$ by simply summing over all click patterns $(\alpha,\beta,\gamma,\delta)$ that are mapped to a specific pair of deterministic assignments, $(A_{\rm A},A_{\rm B})$, as we are assuming here that $p_{\rm d}=0$. Once this is done, it is straightforward to define the parameters of the honest implementation. 

In particular, we have that
\begin{align}\label{QBER_unassisted}
&Q=\mathbf{P}(0,1)_{0,0}+\mathbf{P}(1,0)_{0,0}.
\end{align}
On the other hand, $\omega=S/8+1/2$, where in this case the CHSH violation reads $S=E_{0,-\frac{\pi}{8}}+E_{0,\frac{\pi}{8}}-E_{\frac{\pi}{4},-\frac{\pi}{8}}+E_{\frac{\pi}{4},\frac{\pi}{8}}$ and
\begin{align}\label{violation_summand_unassisted}
&E_{\theta_{\rm A},\theta_{\rm B}}=2\left[\mathbf{P}(0,0)_{\theta_{\rm A},\theta_{\rm B}}+\mathbf{P}(1,1)_{\theta_{\rm A},\theta_{\rm B}}\right]-1.
\end{align}

In doing so, one can numerically compute the maximum tolerated channel loss for any given photon-number statistics $p_{n}$. For instance, in the case of PDC sources, the statistics read $p_{n}=(n+1){\lambda^{n}}(1+\lambda)^{-n-2}$, so that $\Lambda_{\rm max}=\max\{\Lambda\in\mathbb{R}^{+}|K_{\infty}\geq{0}\}$. In this scenario, however, the definition of $K_{\infty}$ includes a maximization over the free parameter $\lambda$ characterizing the intensity of the PDC source, {\it i.e.},
\begin{align}\label{asymptotic_key_PDC}
K_{\infty}=\max_{\lambda\in\mathbb{R}^{+}}\bigg\{1-h\left[\frac{1}{2}+\frac{1}{2}\sqrt{16\omega(\omega-1)+3}\right]-h(Q)\bigg\}.
\end{align}
In this way, we find $\Lambda_{\rm max}\approx{0.4}$ dB, which results in a maximum transmission distance $L_{\rm max}\approx{2}$ km for an attenuation coefficient $\alpha=0.2$ dB/km.
\subsection{DIQKD with two qubit amplifiers}\label{two_QA}

Finally, in this Appendix we consider a different setup from that presented in Fig.~\ref{fig:2}, {\it i.e.}, we now assume that the entanglement source $\rho_{ab}$ is located in the middle of the channel, equidistant from Alice's and Bob's labs. Also, we suppose that both parties hold an ESR-based qubit amplifier to palliate the effect of the channel loss (the case with two PQAs is briefly discussed afterwards also below). The goal is to investigate whether or not such setup could increase the maximum transmission distance before the secret key rate sharply drops to zero. 

Intuitively speaking, the cutoff point where the secret key rate drops down to zero happens at the range of distances for which a significant fraction of the successful heralding events at the qubit amplifier are triggered by the dark counts of the detectors. In this scenario, the conditional quantum state shared by Alice and Bob after a successful heralding takes place is a separable state, thus leading to a vanishing conditional secret key rate. 

In the setup with a central source $\rho_{ab}$ and two qubit amplifiers, the input signal to each qubit amplifier has only traveled a half of the overall transmission distance, so the probability of still carrying a photon that hits a detector within the amplifier may still be large compared to that of a dark count, and this could lead to an enhancement of the transmission distance. Of course, one also expects that the overall secret key rate decreases, as the probability to have a simultaneous successful heralding event at both qubit amplifiers is lower than that of a single success in a unique qubit amplifier, as required in the setup given by Fig.~\ref{fig:2}. Nevertheless, it takes longer distances for the conditional secret key rate to sharply drop down to zero, since it comes from an entangled state with a higher probability than in the setup of Fig.~\ref{fig:2}. As a result, the cutoff point is shifted to further distances.

\begin{figure}
	\centering 
	\includegraphics[width=10.5cm,height=12cm]{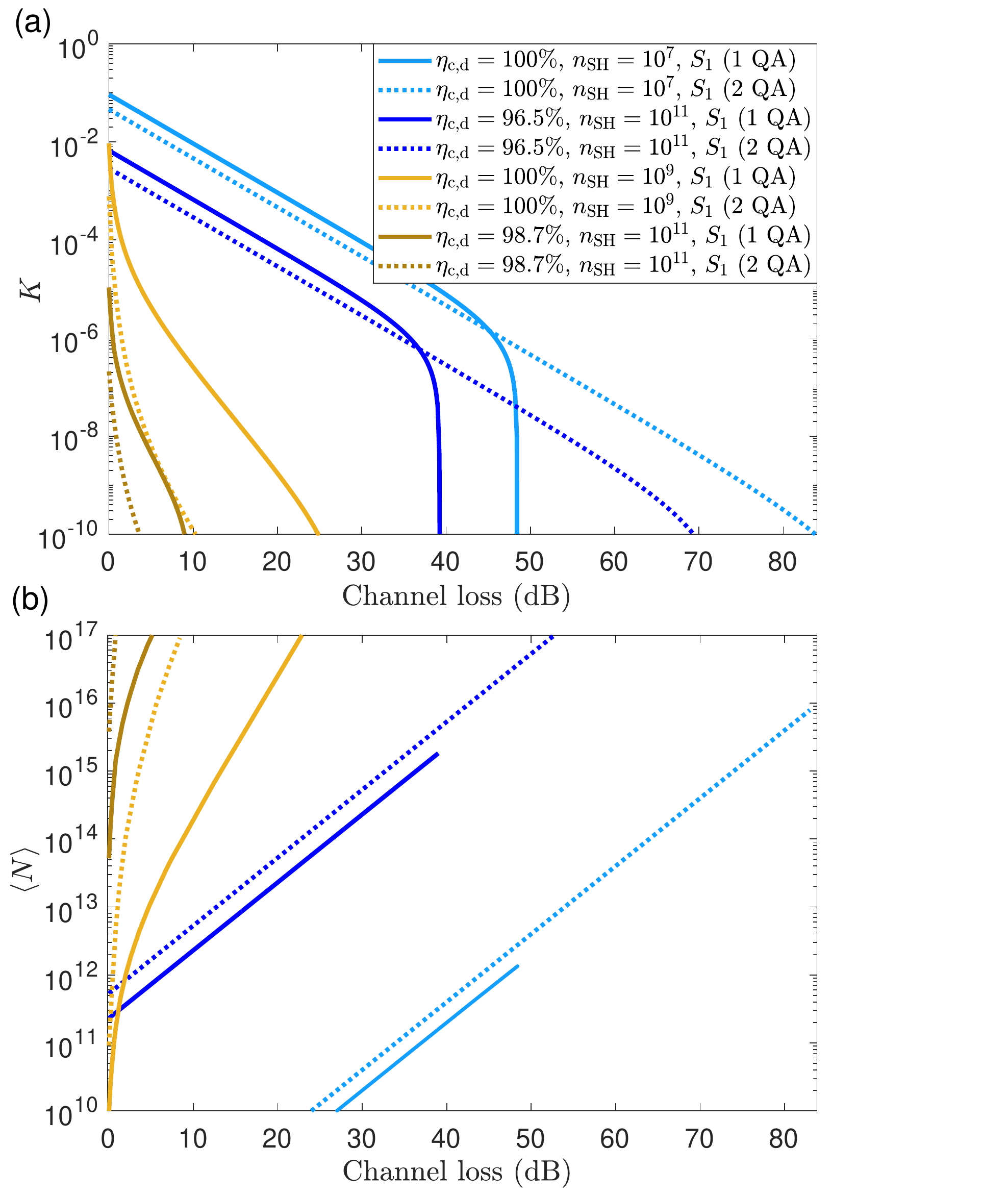} 
	\caption{Comparison between the performance of the DIQKD scheme with one and two ESR-based qubit amplifiers for ideal sources (bluish lines) and for PDC sources (yellowish lines). In both cases, solid (dotted) lines are used for the case with one (two) qubit amplifiers. For illustration purposes, here we consider the same combinations of parameters $\eta_{\rm c,d}$ and $n_{\rm SH}$ like in Sec.~\ref{Section5}. Also, we suppose the less demanding set of security requirements given by $S_{1}$.
		(a) Secret key rate as a function of the channel loss.  
		(b) Average number of transmitted signals as a function of the channel loss.}
	\label{fig:13}
\end{figure}
For instance, in the case where $\rho_{ab}$ is an ideal entanglement source, the setup with two qubit amplifiers roughly doubles the maximum distance without significantly affecting the secret key rate. This is illustrated by the bluish lines in Fig.~\ref{fig:13}, where we compare the secret key rate with one and two qubit amplifiers as a function of the channel loss. Here, as in the previous examples, we set the dark count rate to $p_{\rm d}=10^{-7}$. Also, we assume the security settings given by the set $S_{1}$, and we evaluate the same two examples used in Sec.~\ref{Section5}. From Fig.~\ref{fig:13} we have that, if ideal entanglement sources were available, it would actually be beneficial to use two ESRs for long-distance transmissions. Similar conclusions would be obtained if two PQAs were used instead.

However, this might not be the case if one considers practical light sources with a nonzero probability of emitting multi-photon pulses. For example, in the case of PDC sources, it turns out that the use of two qubit amplifiers does not seem to improve the performance of the system in a practical regime. This is exemplified by the yellowish lines in Fig.~\ref{fig:13}, which again correspond to the same examples as in Sec.~\ref{Section5} with the security settings $S_{1}$. Precisely, Fig.~\ref{fig:13}(a) shows that no advantage is obtained with two qubit amplifiers in a practical key rate regime in this scenario. Arguably, one would expect to see an advantage by considering lower detection and coupling efficiencies or higher dark count rates, as in both situations the cutoff point where the key rate sharply drops to zero is shifted to lower distances. Nevertheless, these two scenarios are notably restricted by time considerations. To see this, in Fig.~\ref{fig:13}(b) we plot the average number of transmitted signals $\left\langle{N}\right\rangle$ required to achieve the secret key rates of Fig.~\ref{fig:13}(a), comparing again the cases of one and two ESRs. As expected, using two qubit amplifiers instead of one demands longer DIQKD sessions in order to harvest a specific block size, due to the requirement of having simultaneous successful heralding events.    In fact, even for $\eta_{\rm c,d}=100\%$ and $n_{\rm SH}=10^{9}$, the necessary value of $\left\langle{N}\right\rangle$ is larger than $10^{15}$ for a channel loss as low as $\Lambda\approx{4}$ dB, and any lower value of the detection and coupling efficiencies would result in smaller values of $\Lambda$. This being the case, even if one considers higher values of the dark count rate $p_{\rm d}$, no advantage is expected from the setup with two qubit amplifiers within such a short channel loss interval (before the duration of the DIQKD session becomes impractical).

Indeed, if one considers two PQAs instead of two ESRs and compares again this setup with the one that uses a single PQA, the time constraint with PDC sources becomes even more strict, as each PQA additionally includes two triggered single-photon sources that must yield a success in their idler mode in order not to dismiss a detection event. That is, a setup with two PQAs requires the trigger of four single-photon sources and two simultaneous successful BSMs afterwards. As a result, this overall decrease of the success probability (which is particularly relevant for PDC sources) translates into larger values of $\left\langle{N}\right\rangle$ which seem to render this solution impractical.

In short, the potential advantage of using two qubit amplifiers to enhance further the distance covered with DIQKD strongly depends on the photon-number statistics of the entanglement sources under consideration. In this sense, although Fig.~(\ref{fig:13}) is restricted to ideal sources and PDC sources, we include below all the necessary calculations to evaluate the case of general photon-number statistics, maintaining the form of the states given in Eqs.~(\ref{1}) and (\ref{2}). In this regard, we remark that the results are similar to those illustrated in Sec.~\ref{general}, {i.e.}, by reducing the probability to emit multiple photon pulses with respect to that of PDC sources, one could approach the behaviour of ideal sources shown in Fig.~(\ref{fig:13}).  

\subsubsection{Click pattern distribution}\label{2ESR_1}
\begin{figure*}[!htbp]
	\includegraphics[width=\textwidth,height=12cm]{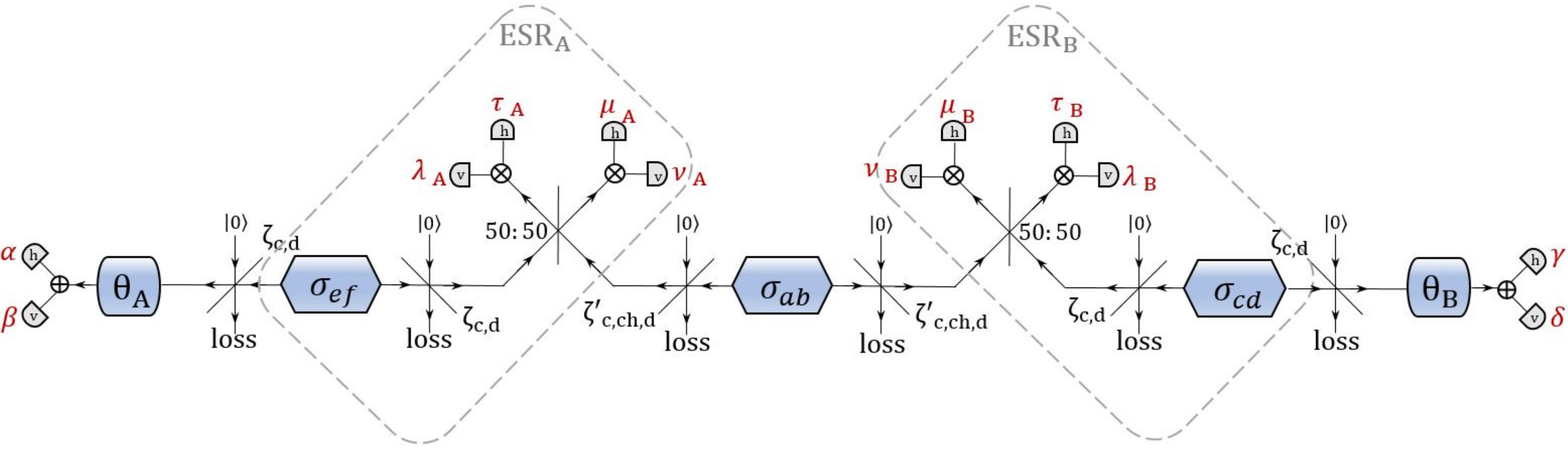} 
	\caption{Schematic of the DIQKD protocol with a central source and two ESR-based qubit amplifiers. The central source is denoted by $\sigma_{ab}$, while $\sigma_{ef}$ ($\sigma_{cd}$) stands for the source within the ESR held by Alice (Bob). As usual, $\theta_{\rm A}$ ($\theta_{\rm B}$) denotes the rotation angle of Alice's (Bob's) measurement settings and $\zeta_{\rm c,d}$ and $\zeta_{\rm c,ch,d}$ tag the effective efficiency parameters, $\zeta_{\rm c,d}=\eta_{\rm c}\eta_{\rm d}$ and $\zeta_{\rm c,ch,d}=\eta_{\rm c}\eta_{\rm ch}\eta_{\rm d}$, where $\eta_{\rm c}$, $\eta_{\rm d}$ and $\eta_{\rm ch}$ denote the transmittance of the BSs modeling the coupling loss, the detection loss and the channel loss, respectively. The symbol ``$\oplus$'' is used again to indicate the PBSs, the greek letters in red color tag the number of photons observed at each of the detectors and $\ket{0}$ stands for the vacuum state. Two dashed grey rectangles identify the two ESRs, ${\rm ESR}_{\rm A}$ and ${\rm ESR}_{\rm B}$.}
	\label{fig:2ESR}
\end{figure*}
A schematic of the DIQKD setup with a central source and assisted by two ESR-based qubit amplifiers is shown in Fig.~\ref{fig:2ESR}. Using the same techniques employed in Appendixes~\ref{ESR} and~\ref{PQA}, it is possible to derive the click pattern distribution that matches the setup of Fig.~\ref{fig:2ESR}, conditioned on the numbers of photon pairs $n_{1}$, $n_{2}$ and $n_{3}$ emitted by the entanglement sources $\sigma_{ab}$, $\sigma_{ef}$ and $\sigma_{cd}$, respectively. Precisely, let us define the click pattern (in vector notation) as $\vec{\alpha}=(\alpha,\beta,\gamma,\delta,\mu_{\rm A},\nu_{\rm A},\tau_{\rm A},\lambda_{\rm A},\mu_{\rm B},\nu_{\rm B},\tau_{\rm B},\lambda_{\rm B})$, where $\mu_{\rm A}$, $\nu_{\rm A}$, $\tau_{\rm A}$ and $\lambda_{\rm A}$ ($\mu_{\rm B}$, $\nu_{\rm B}$, $\tau_{\rm B}$ and $\lambda_{\rm B}$) are the numbers of photons recorded within Alice's (Bob's) ESR in any given detection event, while $\alpha$ and $\beta$ ($\gamma$ and $\delta$) denote again the numbers of photons she (he) observes with the detectors in her (his) lab when performing a measurement with rotation angle $\theta_{\rm A}$ ($\theta_{\rm B}$). In particular, one finds that ${{\rm{P}}\left(\vec{\alpha}|n_{1},n_{2},n_{3}\right)}_{\theta_{\rm A},\theta_{\rm B}}$ reads
\begin{widetext}
	\begin{align}\label{distribution_2QA}
	&{{\rm{P}}\left(\vec{\alpha}|n_{1},n_{2},n_{3}\right)}_{\theta_{\rm A},\theta_{\rm B}}=\frac{{\alpha!}{\beta!}{\gamma!}{\delta!}{\mu_{\rm A}!}{\nu_{\rm A}!}{\tau_{\rm A}!}{\lambda_{\rm A}!}{\mu_{\rm B}!}{\nu_{\rm B}!}{\tau_{\rm B}!}{\lambda_{\rm B}!}}{(n_{1}+1)(n_{2}+1)(n_{3}+1)2^{\mu_{\rm A}+\nu_{\rm A}+\tau_{\rm A}+\lambda_{\rm A}+\mu_{\rm B}+\nu_{\rm B}+\tau_{\rm B}+\lambda_{\rm B}}}\zeta_{\rm c,d}^{\alpha+\beta+\gamma+\delta+\mu_{\rm A}+\nu_{\rm A}+\tau_{\rm A}+\lambda_{\rm A}+\mu_{\rm B}+\nu_{\rm B}+\tau_{\rm B}+\lambda_{\rm B}} \nonumber \\
	&\times(1-\zeta_{\rm c,d})^{2(n_{2}+n_{3})-\alpha-\beta-\gamma-\delta-\mu_{\rm A}-\nu_{\rm A}-\tau_{\rm A}-\lambda_{\rm A}-\mu_{\rm B}-\nu_{\rm B}-\tau_{\rm B}-\lambda_{\rm B}}(1-\zeta_{\rm c,ch,d})^{2n_{1}}\cos{\theta_{\rm A}}^{-2\beta}\sin{\theta_{\rm A}}^{2(\alpha+2\beta)}\cos{\theta_{\rm B}}^{-2\delta}\sin{\theta_{\rm B}}^{2(\gamma+2\delta)} \nonumber \\
	&\times\sum_{i_{1}=0}^{n_{1}}\sum_{i_{2}=0}^{n_{2}}\sum_{i_{3}=0}^{n_{3}}\sum_{\Delta=\max\{-i_1,i_2-n_2,-i_3\}}^{\min\{n_1-i_1,i_2,n_3-i_3\}}\sum_{j_{1}=\max\{0,-\Delta,\nu_{\rm A}+\lambda_{\rm A}-i_2\}}^{\min\{i_1,\nu_{\rm A}+\lambda_{\rm A},\nu_{\rm A}+\lambda_{\rm A}-\Delta\}}\sum_{j'_{1}=\max\{0,-\Delta,\mu_{\rm B}+\tau_{\rm B}+i_3-n_3\}}^{\min\{i_1,\mu_{\rm B}+\tau_{\rm B},\mu_{\rm B}+\tau_{\rm B}-\Delta\}}\sum_{j'_{2}=\max\{0,\Delta,\alpha+\beta+i_2-n_2\}}^{\min\{i_2,\alpha+\beta,\alpha+\beta+\Delta\}} \nonumber \\
	&\times\sum_{j'_{3}=\max\{0,-\Delta,\gamma+\delta+i_3-n_3\}}^{\min\{i_3,\gamma+\delta,\gamma+\delta-\Delta\}}\sum_{l=\max\{0,\Delta,\mu_{\rm A}+\tau_{\rm A}+i_2-n_2\}}^{\min\{n_1-i_1,\mu_{\rm A}+\tau_{\rm A},\mu_{\rm A}+\tau_{\rm A}+\Delta\}}\sum_{l'=\max\{0,\Delta,\nu_{\rm B}+\lambda_{\rm B}-i_3\}}^{\min\{n_1-i_1,\nu_{\rm B}+\lambda_{\rm B},\nu_{\rm B}+\lambda_{\rm B}+\Delta\}}\sum_{k=\max\{0,l-\tau_{\rm A}\}}^{\min\{l,\mu_{\rm A}\}}\sum_{K=\max\{0,l-\tau_{\rm A}-\Delta\}}^{\min\{l-\Delta,\mu_{\rm A}\}} \nonumber \\
	&\times\sum_{m=\max\{0,j_1-\lambda_{\rm A}\}}^{\min\{j_1,\nu_{\rm A}\}}\sum_{M=\max\{0,j_1+\Delta-\lambda_{\rm A}\}}^{\min\{j_1+\Delta,\nu_{\rm A}\}}\sum_{p=\max\{0,j'_1-\tau_{\rm B}\}}^{\min\{j'_1,\mu_{\rm B}\}}\sum_{P=\max\{0,j'_1+\Delta-\tau_{\rm B}\}}^{\min\{j'_1+\Delta,\mu_{\rm B}\}}\sum_{q=\max\{0,l'-\lambda_{\rm B}\}}^{\min\{l',\nu_{\rm B}\}}\sum_{Q=\max\{0,l'-\Delta-\lambda_{\rm B}\}}^{\min\{l'-\Delta,\nu_{\rm B}\}}\sum_{s=\max\{0,\beta-j'_{2}\}}^{\min\{\beta,\alpha+\beta-j'_{2}\}} \nonumber \\
	&\times\sum_{S=\max\{0,\beta-j'_2+\Delta\}}^{\min\{\beta,\alpha+\beta-j'_2+\Delta\}}\sum_{z=\max\{0,\delta-j'_3\}}^{\min\{\delta,\gamma+\delta-j'_3\}}\sum_{Z=\max\{0,\delta-j'_3-\Delta\}}^{\min\{\delta,\gamma+\delta-j'_3-\Delta\}}(-1)^{k+K+m+M+p+P+q+Q+s+S+z+Z+\Delta}\left[{\frac{\zeta_{\rm c,ch,d}(1-\zeta_{\rm c,d})}{\zeta_{\rm c,d}(1-\zeta_{\rm c,ch,d})}}\right]^{j_1+l+j'_1+l'} \nonumber \\
	&\times\left(\frac{\sin{\theta_{\rm A}}}{\cos{\theta_{\rm A}}}\right)^{\Delta-2(j'_2+s+S)}\left(\frac{\sin{\theta_{\rm B}}}{\cos{\theta_{\rm B}}}\right)^{-2(j'_3+z+Z)-\Delta} \nonumber \\
	&\times\Upsilon(n_1,n_2,n_3,i_1,i_2,i_3,j_1,j'_1,j'_2,j'_3,l,l',k,K,m,M,p,P,q,Q,s,S,z,Z,\alpha,\beta,\gamma,\delta,\mu_{\rm A},\nu_{\rm A},\tau_{\rm A},\lambda_{\rm A},\mu_{\rm B},\nu_{\rm B},\tau_{\rm B},\lambda_{\rm B},\Delta),	
	\end{align}
\end{widetext}
	with 
\begin{widetext}
	\begin{align}\label{upsilon_2QA}
	&\Upsilon(n_1,n_2,n_3,i_1,i_2,i_3,j_1,j'_1,j'_2,j'_3,l,l',k,K,m,M,p,P,q,Q,s,S,z,Z,\alpha,\beta,\gamma,\delta,\mu_{\rm A},\nu_{\rm A},\tau_{\rm A},\lambda_{\rm A},\mu_{\rm B},\nu_{\rm B},\tau_{\rm B},\lambda_{\rm B},\Delta)= \nonumber \\
	&\frac{i_1!(n_1-i_1)!(i_1+\Delta)!(n_1-i_1-\Delta)!i_2!(n_2-i_2)!(i_2-\Delta)!(n_2-i_2+\Delta)!i_3!(n_3-i_3)!(i_3+\Delta)!(n_3-i_3-\Delta)!}{(i_1-j'_1)!(i_1-j_1)!(n_1-i_1-l)!(n_1-i_1-l')!(i_2-\nu_{\rm A}-\lambda_{\rm A}+j_1)!(n_2-i_2-\mu_{\rm A}-\tau_{\rm A}+l)!(n_2-i_2-\alpha-\beta+j'_2)!} \nonumber \\
	&\times\frac{[(i_3-j'_3)!(n_3-i_3-\mu_{\rm B}-\tau_{\rm B}+j'_1)!(n_3-i_3-\gamma-\delta+j'_3)!(i_3-\nu_{\rm B}-\lambda_{\rm B}+l')!(\delta-z)!(\delta-Z)!z!Z!(j'_3-\delta+z)!S!]^{-1}}{(i_2-j'_2)!(j'_3+\Delta-\delta+Z)!(\gamma+\delta-j'_3-z)!(\gamma+\delta-j'_3-\Delta-Z)!(\beta-s)!(\beta-S)!(j'_2-\beta+s)!(j'_2-\Delta-\beta+S)!s!} \nonumber \\
	&\times\frac{[(\alpha+\beta-j'_2-s)!(\alpha+\beta-j'_2+\Delta-S)!k!K!(l-k)!(l-\Delta-K)!(\mu_{\rm A}-k)!(\mu_{\rm A}-K)!(\tau_{\rm A}-l+k)!]^{-1}}{m!M!(j_1-m)!(j_1+\Delta-M)!(\nu_{\rm A}-m)!(\nu_{\rm A}-M)!(\lambda_{\rm A}-j_1+m)!(\lambda_{\rm A}-j_1-\Delta+M)!p!P!(j'_1-p)!(j'_1+\Delta-P)!} \nonumber \\
	&\times[(\tau_{\rm A}-l+\Delta+K)!(\mu_{\rm B}-P)!(\tau_{\rm B}-j'_1+p)!(\tau_{\rm B}-j'_1-\Delta+P)!q!Q!(l'-q)!(l'-\Delta-Q)!(\nu_{\rm B}-q)!(\nu_{\rm B}-Q)!]^{-1} \nonumber \\
	&\times[(\mu_{\rm B}-p)!(\lambda_{\rm B}-l'+q)!(\lambda_{\rm B}-l'+\Delta+Q)!]^{-1}.
	\end{align}
\end{widetext}
The normalization condition $\sum_{\vec{\alpha}}{{\rm{P}}\left(\vec{\alpha}|n_{1},n_{2},n_{3}\right)}_{\theta_{\rm A},\theta_{\rm B}}=1$ holds for any set of physical parameters $\theta_{\rm A}$, $\theta_{\rm B}$, $\zeta_{\rm c,d}$, and $\zeta_{\rm c,ch,d}$, and only those click patterns $\vec{\alpha}$ such that $\alpha+\beta\leq{n_{2}}$, $\gamma+\delta\leq{n_{3}}$, $\mu_{\rm A}+\nu_{\rm A}+\tau_{\rm A}+\lambda_{\rm A}\leq{n_{1}+n_{2}}$ and $\mu_{\rm B}+\nu_{\rm B}+\tau_{\rm B}+\lambda_{\rm B}\leq{n_{1}+n_{3}}$ have a nonzero probability to happen, since we have disregarded dark counts so far. If we incorporate now the noise model introduced in the main text, the click pattern distribution in the noisy scenario becomes
\begin{eqnarray}\label{noisy_dist_2ESR}
&{\tilde{\rm{P}}\left(\vec{\alpha}|n_{1},n_{2},n_{3}\right)}_{\theta_{\rm A},\theta_{\rm B}}=(1-{12}p_{\rm d}){{\rm{P}}\left(\vec{\alpha}|n_{1},n_{2},n_{3}\right)}_{\theta_{\rm A},\theta_{\rm B}} \nonumber \\
&+p_{\rm d}\sum_{\vec{\sigma}\in{\Gamma_{\vec{\alpha}}}}{{\rm{P}}\left(\vec{\sigma}|n_{1},n_{2},n_{3}\right)}_{\theta_{\rm A},\theta_{\rm B}}+O(p_{\rm d}^{2}), 
\end{eqnarray}
where, again, $\Gamma_{\vec{\alpha}}={\left\{\vec{\sigma}:|\vec{\alpha}|=|\vec{\sigma}|+1\right\}}$ and, for an arbitrary $\vec{\alpha}$, $|\vec{\alpha}|=\alpha+\beta+\gamma+\delta+\mu_{\rm A}+\nu_{\rm A}+\tau_{\rm A}+\lambda_{\rm A}+\mu_{\rm B}+\nu_{\rm B}+\tau_{\rm B}+\lambda_{\rm B}$. Coming next, by defining
\begin{equation}\label{permutation_2ESR}
{\tilde{\rm{p}}\left(\vec{\alpha}|n_1,n_2,n_3\right)}_{\theta_{\rm A},\theta_{\rm B}}={{\tilde{\rm{P}}_{\alpha\leftrightarrow\beta}\left(\vec{\alpha}|n_1,n_2,n_3\right)}_{\theta_{\rm A},\theta_{\rm B}}},
\end{equation}
we flip Alice's outcomes to enforce direct correlation with Bob's, and by summing over all click patterns $\vec{\alpha}$ that are mapped to the pair of deterministic assignments $(A_{\rm A},A_{\rm B})$, we finally obtain the post-processed click pattern distribution, $\mathbf{P}(A_{\vec{\alpha}}|n_{1},n_{2},n_{3})_{\theta_{\rm A},\theta_{\rm B}}$. In accordance with Appendixes~\ref{ESR_1} and~\ref{PQA_1}, $A_{\vec{\alpha}}$ is defined as $A_{\vec{\alpha}}={(A_{\rm A},A_{\rm B},\mu_{\rm A},\nu_{\rm A},\tau_{\rm A},\lambda_{\rm A},\mu_{\rm B},\nu_{\rm B},\tau_{\rm B},\lambda_{\rm B})}$. Note that if we sum over all possible photon numbers $n_{1}$, $n_{2}$ and $n_{3}$ and assume arbitrary statistics for the entanglement sources, say $p_{n_1}$, $p'_{n_2}$ and $p''_{n_3}$, we obtain the overall distribution
\begin{equation}\label{overall_2ESR}
\mathbf{P}(A_{\vec{\alpha}})_{\theta_{\rm A},\theta_{\rm B}}=\sum_{n_1,n_2,n_3}p_{n_1}p'_{n_2}p''_{n_3}\mathbf{P}(A_{\vec{\alpha}}|n_{1},n_{2},n_{3})_{\theta_{\rm A},\theta_{\rm B}}.
\end{equation}
\subsubsection{Parameters $P_{\rm SH}$, $Q|_{\rm SH}$ and $\omega|_{\rm SH}$}\label{2ESR_2}

In contrast to the setup with a single qubit amplifier, two BSMs are performed in the current scenario, one per ESR (see Fig.~(\ref{fig:2ESR})). Therefore, given that four different click patterns are considered to be successful heralding events in each qubit amplifier, there exist sixteen events that are not discarded by either Alice or Bob, corresponding to those cases for which both $(\mu_{\rm A},\nu_{\rm A},\tau_{\rm A},\lambda_{\rm A})$ and $(\mu_{\rm B},\nu_{\rm B},\tau_{\rm B},\lambda_{\rm B})$ belong to the set $\{(1,1,0,0),(0,1,1,0),(1,0,0,1),(0,0,1,1)\}$. Let $\Omega$ be one of them, say, 
\begin{eqnarray}
\Omega=&&\big\{(\mu_{\rm A},\nu_{\rm A},\tau_{\rm A},\lambda_{\rm A},\mu_{\rm B},\nu_{\rm B},\tau_{\rm B},\lambda_{\rm B}) \nonumber \\
&&=(1,1,0,0,1,1,0,0)\big\}. 
\end{eqnarray}
Then, due to the symmetries of the channel model, it turns out that $P_{\rm{SH}}=16P_{\Omega}$, $\omega|_{\rm{SH}}=\omega|_{\Omega}$ and $Q|_{\rm{SH}}=Q|_{\Omega}$. As in Appendix~\ref{ESR_2}, $P_{\Omega}$ can be computed as
\begin{align}\label{success_prob_2ESR}
&P_{\Omega}=\sum_{A_{\rm A},A_{\rm B}}\mathbf{P}(A_{\rm A},A_{\rm B},\Omega)_{0,0},
\end{align}
with $A_{\rm A},A_{\rm B}\in\{0,1\}$. Analogously,
\begin{align}\label{QBER_2ESR}
&Q|_{\Omega}={\frac{1}{P_{\Omega}}\left[\mathbf{P}(0,1,\Omega)_{0,0}+\mathbf{P}(1,0,\Omega)_{0,0}\right]}, 
\end{align}
and $\omega|_{\Omega}=S|_{\Omega}/8+1/2$ with the conditional CHSH violation given by
\begin{align}\label{violation_2ESR}
&S|_{\Omega}={E_{0,-\frac{\pi}{8}}|_{\Omega}+E_{0,\frac{\pi}{8}}|_{\Omega}-E_{\frac{\pi}{4},-\frac{\pi}{8}}|_{\Omega}+E_{\frac{\pi}{4},\frac{\pi}{8}}|_{\Omega}},
\end{align}
being
\begin{eqnarray}\label{violation_summand_2ESR}
E_{\theta_{\rm A},\theta_{\rm B}}|_{\Omega}=\frac{2}{P_{\Omega}}[\mathbf{P}(0,0,\Omega)_{\theta_{\rm A},\theta_{\rm B}}+\mathbf{P}(1,1,\Omega)_{\theta_{\rm A},\theta_{\rm B}}]-1. \nonumber \\
\end{eqnarray}
Note that, with our choice of $\Omega$, it is the third summand that carries the minus sign in Eq.~(\ref{violation_2ESR}).

\section{Entropy rate function and error correction leakage}\label{ap2}

In this Appendix we provide the explicit expressions for the functions $\eta_{\rm{opt}}$ and $leak_{\rm{IR}}$ which appear in the secret key length formula given by Eq.~(\ref{18}). In particular, we have that~\cite{Rotem}
\begin{equation}
\begin{split}
&\eta_{\rm{opt}}(\omega|_{\rm{SH}},n_{\rm{SH}},\gamma,\delta_{\rm{est}},\epsilon_{\rm{s}}/4,\epsilon_{\rm{EA}}+\epsilon_{\rm{IR}})= \\
&\max\limits_{\frac{3}{4}<p_{\rm{t}}<\frac{2+\sqrt{2}}{4}}\eta\bigg(\frac{\omega|_{\rm{SH}}\gamma-\delta_{\rm{est}}}{\gamma},p_{\rm{t}},n_{\rm{SH}},\gamma,\epsilon_{\rm{s}}/4,\epsilon_{\rm{EA}}+\epsilon_{\rm{IR}}\bigg),
\end{split}
\end{equation}
where the function $\eta(p,p_{\rm{t}},n_{\rm{SH}},\gamma,\epsilon_{1},\epsilon_{2})$ has the form
\begin{eqnarray}
\eta(p,p_{\rm{t}},n_{\rm{SH}},\gamma,\epsilon_{1},\epsilon_{2})&=&f_{\rm{min}}(p,p_{\rm{t}})-\frac{2}{\sqrt{n_{\rm{SH}}}}\nonumber\\
&\times&\left[\log13+\frac{1}{\gamma}\frac{dg(p)}{dp}\bigg\rvert_{p_{\rm{t}}}\right]\nonumber \\
&\times&\sqrt{1-2\log(\epsilon_{1}\epsilon_{2})}.
\end{eqnarray}
In this equation, the function $f_{\rm{min}}(p,p_{\rm{t}})$ is given by
\begin{equation}
f_{\rm{min}}(p,p_{\rm{t}}) = \left\{ \begin{array}{ll}
 g(p) & p<p_{\rm{t}}\\
g(p_{\rm{t}})+\frac{dg(p)}{dp}\bigg\rvert_{p_{\rm{t}}}(p-p_{\rm{t}}) & p\geq{p_{\rm{t}}},
  \end{array} \right.
\end{equation}
and the function $g(p)$ has the form
\begin{equation}
g(p)=1-h\left[\frac{1}{2}+\frac{1}{2}\sqrt{16p(p-1)+3}\right],
\end{equation}
where the winning probability $p$ lies in the interval $0\leq{p}\leq(2+\sqrt{2})/4$, and $h(x)$ is the binary entropy function introduced in the main text.

The function $leak_{\rm{IR}}$, on the other hand, can be written as
\begin{align}
&leak_{\rm{IR}}(Q|_{\rm{SH}},\omega|_{\rm{SH}},n_{\rm{SH}},\gamma,\epsilon_{\rm{IR}},\epsilon_{\rm{rob}}^{\rm{IR}})= \nonumber\\
&n_{\rm{SH}}[(1-\gamma)h(Q|_{\rm{SH}})+\gamma{h(\omega|_{\rm{SH}})}] \nonumber\\
&+4\sqrt{n_{\rm{SH}}}\log\left(2\sqrt{2}+1\right)\sqrt{2\log\left[\frac{8}{{\epsilon'_{\rm{IR}}}^{2}}\right]} \nonumber\\
&+\log\left[\frac{8}{{\epsilon'_{\rm{IR}}}^{2}}+\frac{2}{2-{\epsilon'_{\rm{IR}}}}\right]+\log\left(\frac{1}{\epsilon_{\rm{IR}}}\right),
\end{align}
where $\epsilon_{\rm{rob}}^{\rm{IR}}=\epsilon'_{\rm{IR}}+\epsilon_{\rm{IR}}$. 

\bibliographystyle{apsrev}

\begin{thebibliography}{99}
\bibitem{comment0}
We note that an honest implementation of the CHSH game violates the CHSH inequality whenever its winning probability exceeds $3/4$. Therefore, $\omega|_{\rm{SH}}>3/4$ is a necessary condition to achieve a non vanishing secret key rate. The maximum quantum-mechanical winning probability of the game is $(2+\sqrt{2})/4$, achievable, for instance, by performing adequate projective measurements on a singlet state.
\end{thebibliography}

\end{document}